%% file: main.tex
\documentclass[showpacs,preprintnumbers,nofootinbib,amsmath,amssymb,aps,prd,floatfix]{revtex4-1}
\pdfoutput=1
\usepackage{graphicx}
\graphicspath{{plots/}}
\usepackage{bm}
\usepackage{subfigure}
\usepackage{url}
\usepackage[hyperindex]{hyperref}
\usepackage{color}
\usepackage{bigdelim}
\usepackage{booktabs}
\usepackage{multirow}
\usepackage{cancel}
\usepackage{stackrel}
\usepackage{paralist}
\usepackage{xspace}
\usepackage{catchfile}
\usepackage{enumitem}
\usepackage[normalem]{ulem}
\usepackage[utf8]{inputenc}

\newcommand{\nua}[1]{\ensuremath{\rlap{\kern-2.5pt\ensuremath{\overset{\scriptscriptstyle(-)}{\phantom{\nu}}}}{\ensuremath{{\nu}_{#1}}}}}
\newcommand{\neff}{\ensuremath{N_{\rm eff}}}

\newcommand{\e}[1]{\ensuremath{\times10^{#1}}}
\newcommand{\mlight}{\ensuremath{m_{\rm lightest}}}
\newcommand{\deltacp}{\ensuremath{\delta_{\rm CP}}}
\newcommand{\dmsq}[1]{\ensuremath{\Delta m^2_{#1}}}
\newcommand{\mnu}{\ensuremath{\sum m_\nu}}
\newcommand{\mbb}{\ensuremath{m_{\beta\beta}}}
\newcommand{\Tbb}{\ensuremath{T^{0\nu}_{1/2}}}
\newcommand{\doublebeta}{\ensuremath{0\nu\beta\beta}}

\newcommand{\Bnoio}{\ensuremath{B_{\rm NO,IO}}}
\newcommand{\lnBsim}[1]{\ensuremath{\log(\Bnoio)\simeq#1}}
\newcommand{\Zno}{\ensuremath{Z_{\rm NO}}}
\newcommand{\Zio}{\ensuremath{Z_{\rm IO}}}
\newcommand{\eVq}{\ensuremath{\rm{eV}^2}}
\newcommand{\nme}[1]{\ensuremath{\mathcal{M}^{0\nu}_{#1}}}
\newcommand{\logA}{\ensuremath{\log(10^{10}A_s)}}

\newcommand{\pla}{\texttt{Planck}}
\newcommand{\plaSZ}{\texttt{Planck SZ}}
\newcommand{\plaT}{\texttt{Planck TT}}
\newcommand{\plaTlowP}{\texttt{Planck TT + lowP}}
\newcommand{\plaTE}{\texttt{Planck TT,TE,EE}}
\newcommand{\plaTElowP}{\texttt{Planck TT,TE,EE + lowP}}
\newcommand{\plaTsl}{\texttt{Planck TT + SimLow}}
\newcommand{\plaTEsl}{\texttt{Planck TT,TE,EE + SimLow}}
\newcommand{\lens}{\texttt{lensing}}

\makeatletter
\def\l@subsubsection#1#2{}
\makeatother

\begin{document}

\title{Neutrino Mass Ordering from Oscillations and Beyond: 2018 Status and Future
Prospects}

\author{P.F.\ de Salas}\email{pabferde@ific.uv.es}
\affiliation{Instituto de F\'isica Corpuscular (CSIC-Universitat de Val\`encia), Paterna (Valencia), Spain}

\author{S.\ Gariazzo}\email{gariazzo@ific.uv.es}
\affiliation{Instituto de F\'isica Corpuscular (CSIC-Universitat de Val\`encia), Paterna (Valencia), Spain}

\author{O.\ Mena}\email{omena@ific.uv.es}
\affiliation{Instituto de F\'isica Corpuscular (CSIC-Universitat de Val\`encia), Paterna (Valencia), Spain}

\author{C.A.\ Ternes}\email{chternes@ific.uv.es}
\affiliation{Instituto de F\'isica Corpuscular (CSIC-Universitat de Val\`encia), Paterna (Valencia), Spain}

\author{M.\ T\'ortola}\email{mariam@ific.uv.es}
\affiliation{Instituto de F\'isica Corpuscular (CSIC-Universitat de Val\`encia), Paterna (Valencia), Spain}

\begin{abstract}

The ordering of the neutrino masses is a crucial input for a deep
understanding of flavor physics,
and its determination may provide the key to
establish the relationship among the lepton masses and
mixings and their analogous properties in the quark sector.
The extraction of the neutrino mass ordering is a data-driven field
expected to evolve very rapidly in the next decade.
In this review,
we both analyze the present status and describe the physics of subsequent prospects.
Firstly, the different current available tools
to measure the neutrino mass ordering are described.
Namely, reactor, long-baseline (accelerator and atmospheric)
neutrino beams, laboratory searches for beta and neutrinoless
double beta decays and observations of the cosmic background
radiation and the large scale structure of the universe
are carefully reviewed.
Secondly, the results from an up-to-date comprehensive
global fit are reported:
the Bayesian analysis to the 2018 publicly available oscillation and
cosmological data sets provides
\emph{strong} evidence for the normal neutrino mass ordering versus
the inverted scenario,
with a significance of $3.5$ standard deviations.
This preference for the normal neutrino mass ordering is
mostly due to neutrino oscillation measurements.
Finally, we shall also emphasize the future perspectives for unveiling
the neutrino mass ordering.
In this regard, apart from describing the
expectations from the aforementioned probes,
we also focus on those arising from alternative and novel methods,
as 21~cm cosmology, core-collapse supernova neutrinos
and the direct detection of relic neutrinos.

\end{abstract}

%\pacs{}

\maketitle

\tableofcontents
\newpage
\section{Introduction}
\label{sec:intro}
\input{texs/intro.tex}

\section{Neutrino oscillations}
\label{sec:osc-current}
\input{texs/osc.tex}

\section{Mass ordering and decay experiments}
\label{sec:beta}
\input{texs/doublebeta.tex}

\section{Results from cosmology}
\label{sec:cosmo}
\input{texs/cosmo.tex}

\section{Global 2018 data analysis}
\label{sec:global}
\input{texs/analysis.tex}

\section{Future Prospects}
\label{sec:future}
In this last section, we will explore the future prospects for the detection
of the neutrino mass ordering.
Let us clarify that many of the proposed methods are much less robust
than the ones involving neutrino oscillations through matter (see section~\ref{sec:futureosc}),
and will likely give their first results much after
the first experimental 5$\sigma$ determinations which are likely to be reached in the next $5-10$ years.
Many of the discussed methods, indeed, will give constraints on the neutrino mass ordering
only as a secondary product of their operation and not as a main result,
hence they are not optimized nor mainly focused on the mass ordering determination.
Nevertheless, it is interesting to discuss these additional methods for different reasons.
First of all, independent tests of the neutrino mass ordering from different methods
are surely welcome to have more robust results.
Secondly, the different methods can provide complementary information:
if some inconsistencies or anomalies will appear, we will have new hints for our quest towards new physics beyond our current knowledge.
In conclusion, even if the question regarding the neutrino mass ordering
will be solved within the next few years by the currently running experiments or their immediate extensions,
its study through the other methods we discuss here will be useful to shed more light on the topic
and provide more interesting information on neutrino physics and beyond.
This is why we do not focus only on neutrino oscillation experiments (section~\ref{sec:futureosc}),
which will probably provide the first and strongest results,
but also on more exotic cases as
determinations from decay experiments (sections~\ref{sec:futbeta} and \ref{sec:futdoublebeta})
cosmological constraints (section~\ref{sec:futcosmo}),
measurements from the 21~cm surveys (section~\ref{sec:cosmo21}),
and probes which involve neutrinos emitted by core-collapse supernova explosions (section~\ref{sec:supernova})
or relic neutrinos from the early Universe (section~\ref{sec:relic}).

\subsection{Prospects from oscillations}
\label{sec:futureosc}
\input{texs/fut_osc.tex}

\input{texs/fut_dec.tex}

\subsection{Prospects from cosmology}
\label{sec:futcosmo}
\input{texs/fut_cosmo.tex}
\subsection{Prospects from $21$~cm surveys}
\label{sec:cosmo21}
\input{texs/cosmo21.tex}
\subsection{Prospects from core-collapse supernova}
\label{sec:supernova}
\input{texs/sn.tex}
\subsection{Prospects from relic neutrino direct detection}
\label{sec:relic}
\input{texs/relic.tex}

\section{Summary}
\label{sec:summary}
\input{texs/summary.tex}

\begin{acknowledgments}

We thank C.~Giunti for providing us the two panels constituting Figure~\ref{fig:mbb_vs_mlightest_3p1}
and M.~Hirsch for his suggestions on how to improve the first version of the manuscript.
Work supported by the Spanish grants
FPA2015-68783-REDT, %RENATA
FPA2017-90566-REDC (Red Consolider MultiDark),
FPA2017-85216-P, %AHEP
FPA2017-85985-P %SOM
and
SEV-2014-0398 %IFIC
(AEI/FEDER, UE, MINECO),
and PROMETEOII/2014/050, %SOM
PROMETEOII/2014/084 %AHEP
and GV2016-142 (Generalitat Valenciana).
SG receives support from the European Union's Horizon 2020 research and innovation programme under the Marie Sk{\l}odowska-Curie individual grant agreement No.\ 796941.
OM is also supported by the European Union's Horizon 2020 research and innovation program under the Marie
Sk\l odowska-Curie grant agreements No.\ 690575 and 674896 and
acknowledges the hospitality of the Fermilab Theoretical Physics Department.
PFdS and CAT are also supported by the MINECO fellowships FPU13/03729 and BES-2015-073593, respectively.
MT acknowledges financial support from MINECO through the Ram\'{o}n y Cajal contract RYC-2013-12438 as well as from the L'Or\'eal-UNESCO \textit{For Women in Science} initiative.

\end{acknowledgments}

\bibliography{bibliography}

\end{document}

%% file: texs/intro.tex
The Royal Swedish Academy of Sciences decided to award the 2015 Nobel Prize in Physics to
Takaaki Kajita and
Arthur B.\ McDonald
\emph{``for the discovery of neutrino oscillations, which shows that neutrinos have mass.
[...] New discoveries about the deepest neutrino secrets are expected to change our current understanding of the history, structure and future fate of the Universe''},
see
\cite{Fukuda:1998mi,Ahmad:2002jz,Ahmad:2001an,Eguchi:2002dm,An:2012eh,Abe:2011sj}
for essential publications.
These discoveries robustly established that neutrinos are massive particles.
However, neutrinos are massless particles in the Standard Model (SM)
of particle physics: in the absence of any direct indication for
their mass available at the time,
they were introduced as fermions for which no gauge
invariant renormalizable mass term can be constructed.
As a consequence, in the SM there is neither
mixing nor CP violation in the lepton sector.
Therefore, neutrino
oscillations and masses \textit{imply the first known departure from
the SM of particle physics}.

Despite the good precision that neutrino experiments
have reached in the recent years,
still many neutrino properties remain unknown.
Among them, the neutrino character, Dirac versus Majorana,
the existence of CP violation in the leptonic sector,
the absolute scale of neutrino masses,
and the
type of the neutrino mass spectrum.
Future laboratory, accelerator and reactor, astrophysical and
cosmological probes will address all these open questions,
that may further reinforce the evidence for physics beyond the SM.
The main focus of this review is, however, the last of the aforementioned unknowns.
We will discuss what we know and how we could improve
our current knowledge of the neutrino mass ordering.

Neutrino oscillation physics
is only sensitive to the squared mass differences
($\dmsq{ij}=m_i^2-m_j^2$).
Current oscillation data can be remarkably well-fitted in terms of two
squared mass differences, dubbed as the solar mass splitting
($\dmsq{21}\simeq 7.6\e{-5}$~eV$^2$)
and the atmospheric mass splitting
($|\dmsq{31}|\simeq 2.5\e{-3}$~eV$^2$)~\cite{deSalas:2017kay,globalfit}.
Thanks to matter effects in the Sun,
we know that $\dmsq{21}>0$~\footnote{Note that
the observation of matter effects in the Sun constrains
the product $\dmsq{21}\cos 2\theta_{12}$ to be positive.
Therefore,
depending on the convention chosen to describe solar neutrino
oscillations, matter effects either fix the sign of the solar mass
splitting $\dmsq{21}$ or the octant of the solar angle
$\theta_{12}$, with $\dmsq{21}$ positive by definition.}.
Since the atmospheric mass splitting $\Delta m_{31}^2$ is essentially 
measured only via neutrino oscillations in vacuum, which exclusively
depend on its absolute value, its sign is unknown at the moment.
As a consequence, we have two possibilities for the ordering of neutrino masses:
\emph{normal ordering} (NO, $\dmsq{31}>0$) or
\emph{inverted ordering} (IO, $\dmsq{31}<0$).

The situation for the mass ordering has changed a lot in the last few months.
The 2017 analyses dealing with global oscillation neutrino data
have only shown a mild preference for the normal ordering.
Namely, the authors of Ref.~\cite{Capozzi:2017ipn}, by means of a frequentist analysis, found
$\chi^2_{\rm{IO}}-\chi^2_{\rm{NO}} =3.6$ from all the oscillation data considered in their analyses.
Very similar results were reported in the first version of \cite{deSalas:2017kay}~%
\footnote{See the ``July 2017'' version in \cite{globalfit}.},
where a value of $\chi^2_{\rm{IO}}-\chi^2_{\rm{NO}}=4.3$ was quoted~%
\footnote{A somewhat milder preference in favor of normal mass ordering was obtained in the corresponding version of the analysis in Refs.~\cite{Esteban:2016qun,nufit}.}.
Furthermore, in Ref.~\cite{Gariazzo:2018pei}, the authors verified that the use of a Bayesian approach
and the introduction of cosmological or neutrinoless double beta decay
data did not alter the main result,
which was a weak-to-moderate evidence for the normal
neutrino mass ordering according to the Jeffreys' scale (see Table~\ref{tab:jeffreys}).
The most recent global fit to neutrino oscillation data, however,
reported a strengthened preference for normal ordering that is mainly
due to the new data from the
\texttt{Super-Kamiokande}~\cite{Abe:2017aap},
\texttt{T2K}~\cite{t2k-hartz}
and \texttt{NO$\nu$A}~\cite{nova-radovic} experiments.
The inclusion of these new data in both the analyses of Ref.~\cite{Capozzi:2018ubv} and
the 2018 update of Refs.~\cite{deSalas:2017kay,globalfit}
increases the preference for normal ordering, which now lies mildly above the
$3\sigma$ level.
In this review we will comment these new results (see section~\ref{sec:osc-current})
and use them to perform an updated global analysis, following the method of Ref.~\cite{Gariazzo:2018pei}
(see section~\ref{sec:global}).

\begin{figure}
\centering
\includegraphics[width=0.7\textwidth]{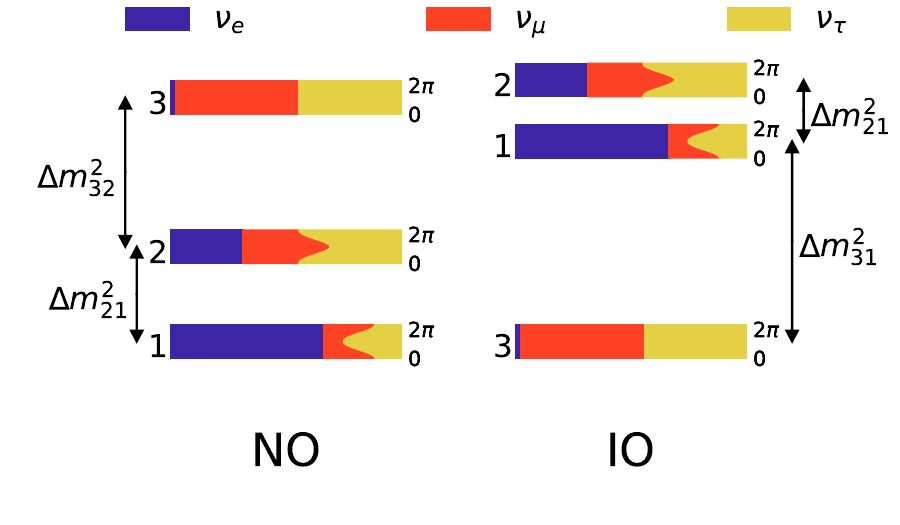}
\caption{\label{fig:compo}
Probability of finding the $\alpha$ neutrino flavor in the $i$-th
neutrino mass eigenstate as the CP-violating phase,
$\deltacp$, is varied. Inspired by Ref.~\cite{Mena:2003ug}.
}
\end{figure}

The two possible hierarchical%
\footnote{%
A clarification about the use of ``hierarchy'' and ``ordering'' is mandatory.
One talks about ``hierarchy'' when referring to the absolute scales
of neutrino masses,
in the sense that neutrino masses can be distinguished and
ranked from lower to higher.
This does not include the possibility that the lightest neutrino mass
is much larger than the mass splittings obtained by neutrino oscillation measurements,
since in this case the neutrino masses are degenerate.
On the other hand, the mass ``ordering''
is basically defined by the sign of \dmsq{31},
or by the fact that the lightest neutrino is the most (least) coupled
to the electron neutrino flavor in the normal (inverted) case.
}
neutrino mass scenarios are shown in
Figure~\ref{fig:compo},
inspired by Ref.~\cite{Mena:2003ug},
which provides a graphical representation of
the neutrino flavor content of each of the neutrino mass eigenstates
given the current preferred values of
the oscillation parameters~\cite{deSalas:2017kay},
see section~\ref{sec:osc-current}.
At present, even if the current preferred value of
$\deltacp$ for both normal and inverted mass orderings lies close to
$3\pi/2$~\cite{deSalas:2017kay},
the precise value of the CP violating phase in
the leptonic sector remains unknown.
Consequently, in Figure~\ref{fig:compo}, we have varied
$\deltacp$ within its entire range, ranging from $0$ to $2\pi$.

\begin{figure}
\centering
\includegraphics[width=0.7\textwidth]{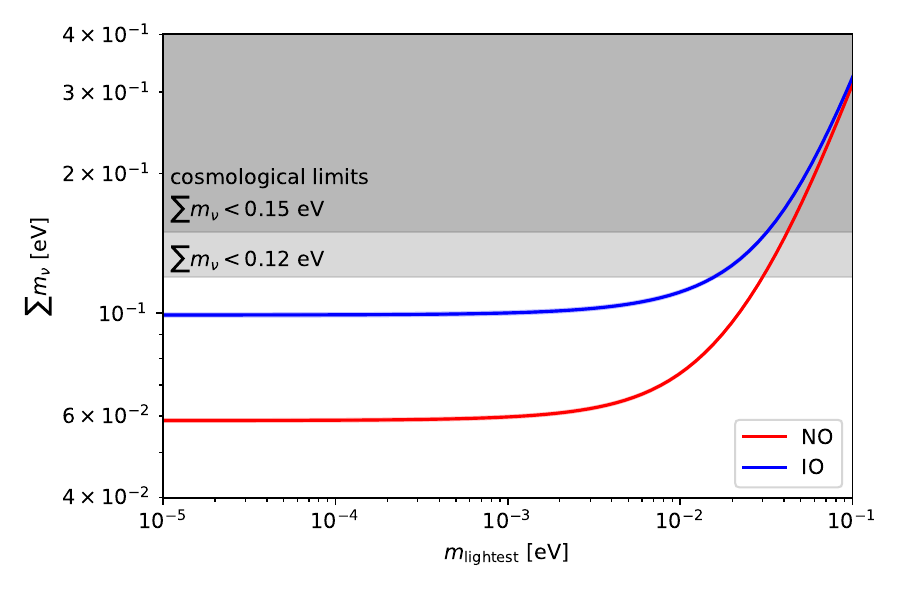}
 \caption{
 \label{fig:mnu_mlightest}
 The sum of the neutrino masses $\mnu$
 as a function of the mass of the lightest neutrino,
 $m_1$ ($m_3$) for the normal (inverted) ordering,
 in red (blue) respectively.
 The (indistinguishable) width of the lines
 represents the present 3$\sigma$ uncertainties
 in the neutrino mass splittings
 from the global fit to neutrino oscillation data~\cite{deSalas:2017kay}.
 The horizontal bands illustrate
 two distinct $95\%$~Confidence Level (CL) limits
 on \mnu\ from cosmology, see the text for details.}
\end{figure}

Given the two known mass splittings that oscillation experiments
provide us, we are sure that at least two neutrinos have a mass above
$\sqrt{\dmsq{21}}\simeq8$~meV and that at least one of these two neutrinos
has a mass larger than $\sqrt{|\dmsq{31}|}\simeq50$~meV.
For the same reason,
we also know that there exists a
lower bound on the sum of the three active neutrino masses
($\mnu=m_1+m_2+m_3$):
\begin{eqnarray}
\mnu^{\rm{NO}}
&=&
m_1
+ \sqrt{m_1^2 + \dmsq{21}}
+ \sqrt{m_1^2 + \dmsq{31}}\, ,\\ \nonumber
\mnu^{\rm{IO}}
&=&
m_3
+ \sqrt{m_3^2+ |\dmsq{31}| }
+ \sqrt{m_3^2+ |\dmsq{31}| + \dmsq{21}}~,
\end{eqnarray}
where the lightest neutrino mass eigenstate
corresponds to $m_1$ ($m_3$) in the normal (inverted) ordering.
Using the best-fit values for the neutrino mass splittings
in Table~\ref{tab:oscillation_summary} one finds that
$\mnu \gtrsim 0.06$~eV in normal ordering,
while $\mnu \gtrsim 0.10$~eV in inverted ordering.
Figure~\ref{fig:mnu_mlightest} illustrates the values
of $\mnu$ as a function of the lightest neutrino mass
for the two possible ordering schemes.
We also show the two representative bounds on the
sum of the neutrino masses from cosmology (discussed later in section~\ref{sec:cosmo})
which is currently providing the strongest limits
on \mnu\ thanks to the fact that
neutrinos affect both the evolution of
the cosmological background and perturbation quantities
(see e.g.\ the excellent detailed reviews of
Refs.~\cite{Lesgourgues:2006nd,Lesgourgues:2012uu,Lesgourgues-Mangano-Miele-Pastor-2013,Lesgourgues:2014zoa,Lattanzi:2017ubx}).

The state-of-knowledge of cosmological observations~\cite{Ade:2015xua}
points to a flat Universe whose mass-energy density includes
$5\%$ of ordinary matter (baryons),
$22\%$ non-baryonic \emph{dark matter},
and that is dominated by the \emph{dark energy},
identified as the motor for the accelerated expansion.
This is the so-called $\Lambda$CDM Universe,
which fits extremely well the Cosmic Microwave Background (CMB) fluctuations, distant Supernovae Ia and galaxy
clustering data.

Using the known neutrino oscillation parameters
and the standard cosmological evolution,
it is possible to compute the thermalization and the decoupling
of neutrinos in the early universe
(see e.g.~\cite{Mangano:2005cc,deSalas:2016ztq}).
While neutrinos decoupled as ultra-relativistic particles, currently
at least two out of the three neutrino mass eigenstates
are non-relativistic.
Neutrinos constitute
the first and only known form of dark matter so far.
Indeed, neutrinos behave as
\emph{hot} dark matter particles,
possessing large thermal velocities,
clustering only at scales
below their free streaming scale,
modifying the evolution of matter overdensities and
suppressing structure formation at small scales.
The CMB is also affected by the
presence of massive neutrinos, as these particles may turn
non-relativistic around the decoupling period.
However, the strong degeneracy between the Hubble constant and the total neutrino
mass requires additional constraints
(from Baryon Acoustic Oscillations,
Supernovae Ia luminosity distance data
and/or direct measurements of the Hubble constant)
to be added in the global analyses.
In this regard, CMB lensing is also helpful and
improves the CMB temperature and polarization constraints,
as the presence of massive neutrinos modify the matter
distribution along the line of sight through their free streaming
nature, reducing clustering and, consequently, CMB lensing.
The most constraining cosmological upper bounds to date
on $\mnu$ can be obtained combining CMB
with different large scale structure observations
and range from
$\mnu<0.12$~eV to $\mnu<0.15$~eV at $95\%$~CL
~\cite{Vagnozzi:2018jhn,Lattanzi:2017ubx,Vagnozzi:2017ovm,
Giusarma:2016phn,Cuesta:2015iho,Palanque-Delabrouille:2015pga,DiValentino:2015sam},
as illustrated in Figure~\ref{fig:mnu_mlightest}.

If the massive neutrino spectrum does not lie in the degenerate
region, the three distinct neutrino masses affect the cosmological
observables in a different way.
For instance, the transition to the
non-relativistic period takes place at different cosmic times, and
the associated free-streaming scale is different for each of the
neutrino mass eigenstates.
However, the effect on the power spectrum
is very small (permille level) and therefore an extraction of the
neutrino mass hierarchy via singling out each of the massive neutrino
states seems a very futuristic challenge.
This will be possibly attainable only via
huge effective volume surveys,
as those tracing the $21$~cm spin-flip
transition in neutral hydrogen,
see sections~\ref{sec:futcosmo} and \ref{sec:cosmo21}.
On the other hand,
should the cosmological measurements of $\mnu$ be strong
enough to rule out the $\mnu$ parameter space corresponding
to the inverted ordering
(i.e.\ strong enough to establish in a very significant way
that $\mnu< 0.1$ ~eV),
we would know that the neutrino mass ordering must be normal.
A word of caution is needed here
when dealing with Bayesian analyses,
usually performed when dealing with cosmological data:
a detection of
the neutrino mass ordering could be driven by volume effects in the marginalization,
and therefore the prior choice can make a huge difference,
if data are not powerful enough~\cite{Schwetz:2017fey}.

\begin{figure}
\centering
\includegraphics[width=0.49\textwidth]{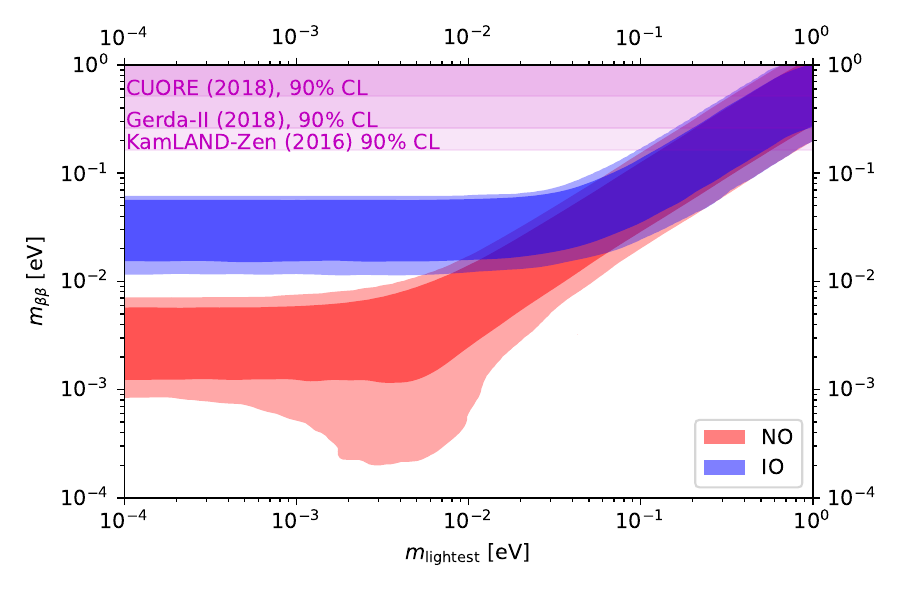}
\includegraphics[width=0.49\textwidth]{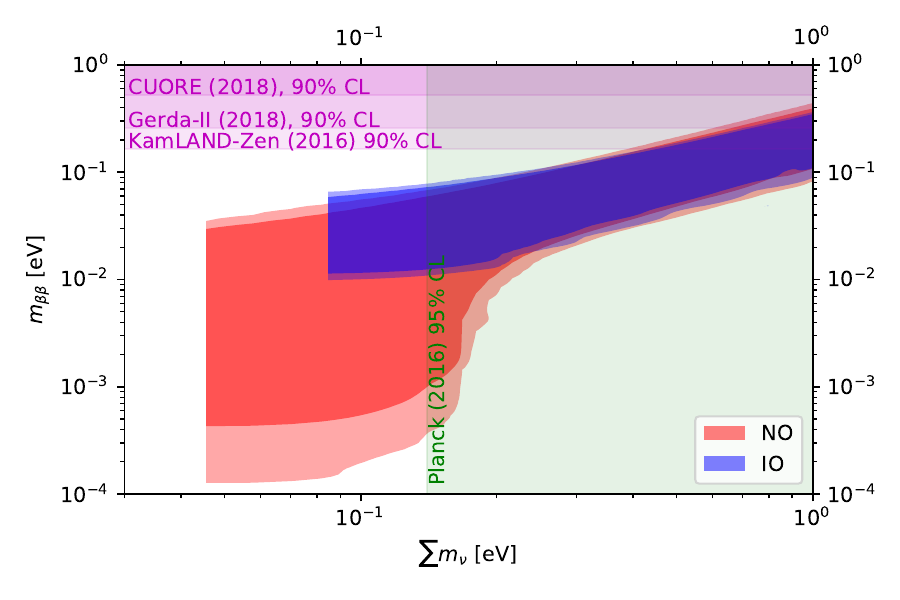}
\caption{
\label{fig:mbb_vs_mlightest} 95.5\% and 99.7\% Bayesian credible intervals
for the effective Majorana mass,
\mbb, as a function of the lightest neutrino mass (left panel)
or of the sum of the neutrino masses \mnu\ (right panel),
taking into account the current uncertainties
on the neutrino mixing parameters (angles and phases),
when three neutrinos are considered.
The horizontal bands indicate the most conservative version
(obtained by each collaboration
when assuming a disfavorable value
for the nuclear matrix element of the process)
of some of the most competitive upper bounds,
as those reported by KamLAND-Zen \cite{KamLAND-Zen:2016pfg},
GERDA Phase II \cite{Agostini:2018tnm} and
CUORE \cite{Alduino:2017ehq}.
The vertical band in the right panel indicates the strongest limit reported by \pla~\cite{Aghanim:2016yuo},
using the \plaTEsl\ + \lens\ data combination.
}
\end{figure}

Another way to probe the neutrino mass ordering,
apart from direct determinations of the sign of the atmospheric
mass splitting $\Delta m^2_{31}$ in neutrino oscillation experiments
and, indirectly, from cosmological bounds on the sum of
the neutrino masses,
is \emph{neutrinoless double $\beta$ decay}~%
\cite{GomezCadenas:2011it,DellOro:2016tmg,Vergados:2012xy,Rodejohann:2011mu}.
This process is a spontaneous nuclear transition in which the charge
of two isobaric nuclei would change by two units with
the simultaneous emission of two electrons and without the emission
of neutrinos.
This process is only possible if the neutrino is a Majorana particle and an experimental signal of the existence of this process
would constitute evidence of the
putative Majorana neutrino character.
The non-observation of the process provides bounds on the so-called
\emph{effective Majorana mass} $\mbb$,
which is a combination of the (Majorana) neutrino masses
weighted by the leptonic flavor mixing effects
(see section~\ref{sec:beta}).
Figure~\ref{fig:mbb_vs_mlightest} illustrates the
(Bayesian) 95.5\% and 99.7\% credible intervals
for $\mbb$ as a function of the lightest neutrino mass
in the case of three neutrino mixing,
considering a logarithmic prior on the lightest neutrino mass.
The picture differs from the plot that is usually shown, which features
an open band towards increasingly smaller values of \mbb\
for $\mlight\simeq5$~meV, due to cancellations
which depend on the values of the Majorana phases $\alpha_i$
(see section~\ref{sec:beta}).
In the Bayesian sense of credible intervals,
the values of $\alpha_i$ which produce such a suppression of
\mbb\ represent an extremely small fraction of the parameter space,
which is therefore not relevant when computing
the 95.5\% and 99.7\% credible intervals.
In other words,
given our knowledge of the neutrino mixing parameters,
having $\mbb\lesssim 2\e{-4}$~eV would require some amount of fine tuning
in the Majorana phases.
This figure is in perfect agreement with the results shown in Figure~1 of Ref.~\cite{Agostini:2017jim},
which shows that most of the allowed parameter space is not concentrated at small \mbb\
if one considers a linear prior on the lightest neutrino mass.
We also show the most conservative version of some of the
most competitive current limits,
as those from \texttt{KamLAND-Zen}
($\mbb <61-165$~meV at $90\%$~CL)~\cite{KamLAND-Zen:2016pfg},
\texttt{GERDA} Phase II
($\mbb <120-260$~meV at $90\%$~CL)~\cite{Agostini:2018tnm}
and \texttt{CUORE}
($\mbb <110-520$~meV at $90\%$~CL)~\cite{Alduino:2017ehq}.
Please note that a detection of the effective Majorana mass
will not be sufficient to determine the mass ordering
if the lightest neutrino mass is above $\sim40$~meV:
in this case, indeed, the normal and the inverted ordering
become indistinguishable from the point of view of
neutrinoless double beta decay.
Similarly to the case of the
cosmological bounds on the neutrino mass \mnu,
in which only constraining \mnu\ to be below 0.1~eV
could be used to disfavor the inverted mass ordering,
only a limit on \mbb\ below $\sim10$~meV
could be used to rule out the inverted ordering scheme,
and only assuming that neutrinos are Majorana particles.

Since neutrino oscillation measurements, cosmological observations and
neutrinoless double beta decay experiments are cornering the inverted
mass ordering region, it makes sense to combine their present results.
Indeed, plenty of works have been recently
devoted to test whether a preference for one mass ordering over the other exists,
given current oscillation, neutrinoless double beta decay and cosmological data.
A number of studies on the subject~\cite{Hannestad:2016fog,Gerbino:2016ehw,Capozzi:2017ipn,Wang:2017htc,Caldwell:2017mqu}
found that the preference for the normal versus the inverted mass scenario
is rather mild with current data,
regardless the frequentist versus Bayesian approach.
In the latter case, however, the results may be subject-dependent, as a consequence
of different possible choices of priors and parameterizations when describing the theoretical model,
for example in the case of sampling over the three individual neutrino mass states.
Therefore, one must be careful when playing with different priors,
as recently shown in Ref.~\cite{Gariazzo:2018pei}.
The current status of the preference of normal versus inverted ordering
will be further investigated carefully throughout this review.
Furthermore, as it will be carefully detailed in section~\ref{sec:global}, the Bayesian global fit to
the 2018 publicly available oscillation and cosmological data
points to a strong preference ($\input{results/s_osc_0n2b_cmb_H0_novsio.tex}$ standard
deviations) for the normal neutrino mass ordering versus the inverted one.

To summarize and conclude this introductory part,
we resume that the current available methods to determine the neutrino mass ordering can be grouped as:
\begin{itemize}
 \item[a)] neutrino oscillation facilities;
 \item[b)] neutrinoless double beta decay experiments,
with the caveat that the results will only apply in case neutrinos are
Majorana fermions;
 \item[c)] CMB and large scale structure surveys.
\end{itemize}
For each of these three categories we will review the current status and also analyse the future prospects,
with a particular focus on the existing experiments which will be improved in the future
and on new facilities which aim at determining
the neutrino mass ordering in the next ten-to-twenty years~%
\footnote{See also the review~\cite{Qian:2015waa},
focused mostly on neutrino oscillation perspectives.}
In the second part of this review
we will also focus on possible novel methods that in the future
will enable us to determine the neutrino mass ordering,
as for example
future cosmological observations of the 21~cm line,
the detection of neutrinos emitted by core-collapse supernovae,
measurements of the electron spectrum of $\beta$-decaying nuclei and
the direct detection of relic neutrinos.

We shall exploit the complementarity of both cosmology and particle physics approaches,
profiting from the highly multidisciplinary character of the topic.
We dedicate sections \ref{sec:osc-current}, \ref{sec:beta} and \ref{sec:cosmo}
to explain the extraction of the neutrino mass ordering
via neutrino oscillations,
$\beta$ and neutrinoless double $\beta$ decays
and cosmological observations,
which will be combined in section~\ref{sec:global} where we present the
analysis of current data related to these three data sets.
Future perspectives are described throughout section~\ref{sec:future} and its subsections,
while the final remarks will be outlined in section~\ref{sec:summary}.

%% file: results/s_osc_0n2b_cmb_H0_novsio.tex
3.5

%% file: texs/osc.tex
Our current knowledge on the neutrino mass ordering comes mainly from
the analysis of the available neutrino oscillation data.
The sensitivity to the neutrino mass spectrum at oscillation experiments
is mostly due to the presence of matter effects in the neutrino propagation.
Therefore, one can expect that this sensitivity will
increase with the size of matter effects, being larger for atmospheric
neutrino experiments, where a fraction of neutrinos travel through the Earth.
For long-baseline accelerator experiments, matter effects will increase with the baseline,
while these effects will be negligible at short-baseline and medium-baseline experiments.

When neutrinos travel through the Earth, the effective matter
potential due to the electron (anti)neutrino charged-current elastic
scatterings with the electrons in the medium will modify the three-flavor mixing processes.
The effect will strongly depend on the neutrino mass ordering:
in the normal (inverted) mass ordering scenario, the neutrino
flavor transition probabilities will get enhanced (suppressed).
In the case of antineutrino propagation, instead, the flavor transition probabilities
will get suppressed (enhanced) in the normal (inverted) mass ordering scenario.
This is the Wolfenstein effect~\cite{Wolfenstein:1977ue},
later expanded by Mikheev and Smirnov~\cite{Mikheev:1986gs,Mikheev:1986wj},
and commonly named as the Mikheev-Smirnov-Wolfenstein (MSW) effect
(see e.g.\ Ref.~\cite{Blennow:2013rca} for a detailed description of neutrino oscillations in matter).

Matter effects in long-baseline accelerator or atmospheric neutrino oscillation experiments depend on the size of the
effective mixing angle $\theta_{13}$ in matter, which leads the transitions
$\nu_e \leftrightarrow \nu_{\mu,\tau}$
governed by the atmospheric mass-squared difference $\Delta_{31}=\Delta m^2_{31}/2E$.
Within the simple two-flavor mixing framework,
the effective $\theta_{13}$ angle in matter reads as
\begin{equation}
\sin^2 2 \theta^{\textrm{m}}_{13}
=
\frac{\sin^2 2 \theta_{13}}%
{\sin^2 2 \theta_{13} +
\left(\cos 2 \theta_{13} \mp
\frac{\sqrt{2} G_{F} N_{e}}{\Delta_{31}}
\right)^2}~,
\label{eq:mixmatter}
\end{equation}
where the minus (plus) sign refers to neutrinos (antineutrinos) and $N_e$
is the electron number density in the Earth interior.
The neutrino mass ordering fixes the sign of $\Delta_{31}$,
that is positive (negative) for normal (inverted) ordering:
notice that, in the presence of matter effects,
the neutrino (antineutrino) oscillation probability
$P (\nu_\mu\to \nu_e)$
[$P (\bar{\nu}_\mu\to \bar{\nu}_e)$]
gets enhanced if the ordering is normal (inverted).
Exploiting the different matter effects for neutrinos and antineutrinos provides
therefore the ideal tool to unravel the mass ordering.

Matter effects are expected to be particularly relevant when the following resonance condition is satisfied:
\begin{equation}
\Delta m^2_{31} \cos 2 \theta_{13}
=
2 \sqrt{2} G_{F} N_{e} E~.
\label{eq:res}
\end{equation}
The precise location of the resonance will depend on
both the neutrino path and the neutrino energy.
For instance, for
$\Delta m^2_{31} \sim 2.5 \times 10^{-3}$~eV$^2$
and distances of several thousand kilometers,
as it is the case of atmospheric neutrinos,
the resonance effect is expected to happen for neutrino energies $\sim 3-8$~GeV.

\begin{figure}
\centering
\includegraphics[width=0.49\textwidth]{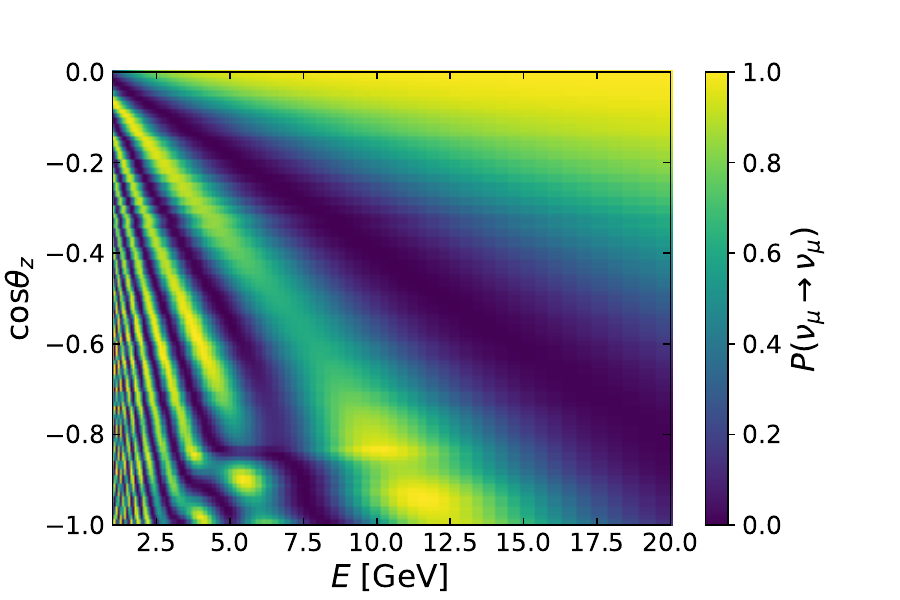}
\includegraphics[width=0.49\textwidth]{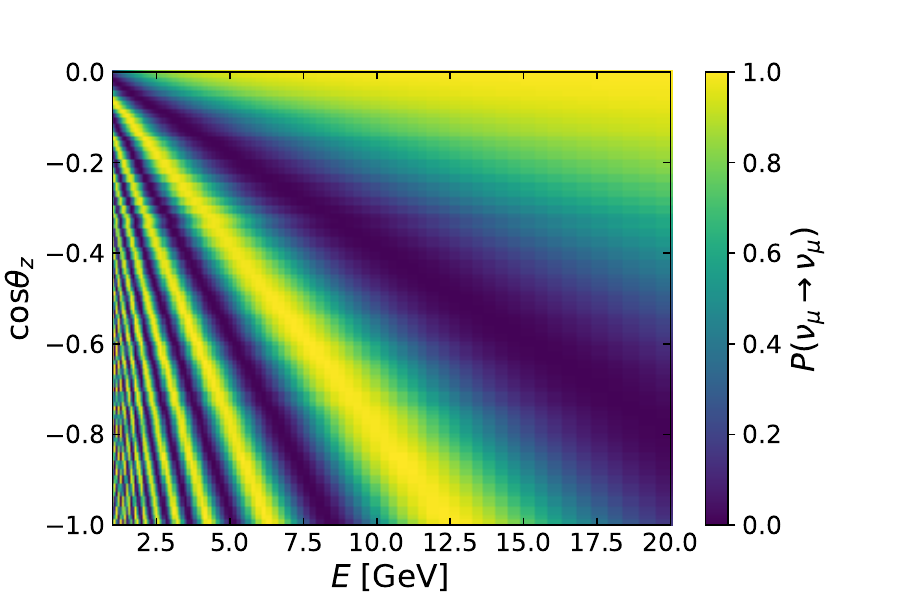}
\caption{\label{fig:Pmumu}
Survival probability $P_{\mu \mu}$,
as a function of the neutrino energy $E$ and
the cosine of the zenith angle $\cos\theta_z$,
for normal (inverted) ordering in the left (right) panel.
}
\end{figure}

In the case of muon disappearance experiments,
in the $\sim$ GeV energy range relevant for long-baseline and atmospheric neutrino beams,
the $P_{\mu \mu}$ survival probabilities
are suppressed (enhanced) due to matter effects
if the ordering is normal (inverted).
If the matter density is constant,
the $P_{\mu \mu}$ survival probability at terrestrial baselines~%
\footnote{For an expansion including also the solar
mixing parameters, see Ref.\cite{Akhmedov:2004ny}.} is given by
\begin{eqnarray}
P_{\mu \mu}
&=&
1-\cos^2 \theta^{\textrm{m}}_{13}
\sin^2 2 \theta_{23}
\times
\sin^2\left[1.27\left(\frac{\Delta m^2_{31} + A +
  (\Delta m^2_{31})^{\textrm{m}}}{2}\right)
  \frac{L}{E}\right]
\label{eq:mixmatter2}
\\
&&
-\sin^2 \theta^{\textrm{m}}_{13}
\sin^2 2 \theta_{23}
\times
\sin^2\left[1.27\left(\frac{\Delta m^2_{31} + A -
  (\Delta m^2_{31})^{\textrm{m}}}{2}\right)
  \frac{L}{E}\right]
  - \sin^4\theta_{23}\sin^2 2 \theta^{\textrm{m}}_{13}
\sin^2\left[1.27 (\Delta m^2_{31})^{\textrm{m}}\frac{L}{E}\right]
~ \nonumber,
\end{eqnarray}
where
$A=2 \sqrt{2} G_{F} N_{e} E$,
$\theta^{\textrm{m}}_{13}$ is that of Equation~\eqref{eq:mixmatter}
and
\begin{equation}
(\Delta m^2_{31})^{\textrm{m}}
=
\Delta m^2_{31}
\sqrt{\sin^2 2 \theta_{13}
+ \left( \cos 2 \theta_{13} \mp
\frac{2 \sqrt{2} G_{F} N_{e} E}{\Delta m^2_{31}}
\right)^2}~.
\label{eq:masseff}
\end{equation}
The dependence of the survival probability $P_{\mu \mu}$
on the neutrino energy $E$ and
the cosine of the zenith angle $\cos\theta_z$,
related to the distance the atmospheric neutrinos travel inside the Earth
before being detected at the experiments,
is shown in Figure~\ref{fig:Pmumu} for normal (left panel)
and inverted (right panel) ordering.
There, we can see that reconstructing the oscillation pattern
at different distances and energies allows to determine
the neutrino mass ordering (see also section~\ref{sec:futureosc}).

Until very recently, oscillation experiments were not showing
a particular preference for any of the mass orderings,
not even when combined in a global analysis,
see for instance Ref.~\cite{Forero:2014bxa}.
Lately, however, the most recent data releases
from some of the experiments have become more sensitive
to the  ordering of the neutrino mass spectrum.
In particular, the long-baseline experiments
\texttt{T2K} and \texttt{NO$\nu$A} on their own
obtain a slight preference in favor of normal mass ordering,
with $\Delta\chi^2\approx 4$ each~\cite{t2k-hartz,nova-radovic}.
Note that these results have been obtained imposing a prior
on the mixing angle $\theta_{13}$,
according to its most recent determination at reactor experiments.
Relaxing the prior on the reactor angle results
in a milder preference for normal over inverted mass ordering.
The latest atmospheric neutrino results from
\texttt{Super-Kamiokande} also show some sensitivity
to the neutrino mass ordering.
In this case, the collaboration obtains a preference
for normal ordering with $\Delta\chi^2\approx 3.5$,
without any prior on the reactor angle.
Constraining the value of $\theta_{13}$,
the preference for normal mass ordering increases up to
$\Delta\chi^2\approx 4.5$~\cite{Abe:2017aap}.

The full sensitivity to the ordering of the neutrino mass spectrum
from oscillations is obtained after combining the data samples
described above with all the available experimental results
in a global fit~\cite{deSalas:2017kay}.
This type of analysis exploits the complementarity
among the different results as well as the correlations
among the oscillation parameters
to obtain improved sensitivities on them.
In the global analysis to neutrino oscillations,
the parameters $\sin^2\theta_{12}$ and $\dmsq{21}$
are rather well measured by the solar experiments~\cite{Cleveland:1998nv,Kaether:2010ag,Abdurashitov:2009tn,Hosaka:2005um,Cravens:2008aa,Abe:2010hy,Nakano:PhD,Aharmim:2008kc,Aharmim:2009gd,Bellini:2013lnn}
and the long-baseline reactor experiment \texttt{KamLAND}~\cite{Gando:2010aa}.
The short-baseline reactor neutrino experiments
\texttt{Daya Bay}~\cite{An:2016ses}, \texttt{RENO}~\cite{Pac:2018scx} and
\texttt{Double Chooz}~\cite{Abe:2014bwa}
are the most efficient ones in measuring the reactor angle $\theta_{13}$
and also measure very well the atmospheric mass splitting, $\dmsq{31}$.
Notice however that the atmospheric mass splitting is
best measured by the combined data from MINOS (beam and atmospheric) and MINOS+, as shown in Ref.~\cite{aurisano_adam_2018_1286760}.
This mass splitting is also measured,
together with the atmospheric angle $\theta_{23}$,
by the atmospheric experiments
\texttt{IceCube-DeepCore}~\cite{Aartsen:2014yll},
\texttt{ANTARES}~\cite{AdrianMartinez:2012ph} and
\texttt{Super-Kamiokande}~\cite{Abe:2017aap},
where the latter shows some sensitivity to $\theta_{13}$ and $\deltacp$, too.
The long-baseline accelerator experiments
are also sensitive to these four parameters
through their appearance and disappearance neutrino channels.
Apart from the already mentioned
\texttt{T2K}~\cite{t2k-hartz} and
\texttt{NO$\nu$A}~\cite{nova-radovic}, the global fit also includes the previous experiments
\texttt{K2K}~\cite{Ahn:2006zza} and
\texttt{MINOS}~\cite{Adamson:2014vgd}.

\begin{table}[t]\centering
\catcode`?=\active \def?{\hphantom{0}}
\begin{tabular}{|l|c|c|c|}
\hline
parameter
& best-fit $\pm$ $1\sigma$
& \hphantom{x} 2$\sigma$ range \hphantom{x}
& \hphantom{x} 3$\sigma$ range \hphantom{x}
\\
\hline
$\dmsq{21}\: [10^{-5}\eVq]$
& 7.55$^{+0.20}_{-0.16}$  & 7.20--7.94 & 7.05--8.14 \\
\hline
$|\dmsq{31}|\: [10^{-3}\eVq]$ (NO)
&  2.50$\pm$0.03 &  2.44--2.57 &  2.41--2.60\\
$|\dmsq{31}|\: [10^{-3}\eVq]$ (IO)
&  2.42$^{+0.03}_{-0.04}$ &  2.34--2.47 &  2.31--2.51 \\
\hline
$\sin^2\theta_{12} / 10^{-1}$
& 3.20$^{+0.20}_{-0.16}$ & 2.89--3.59 & 2.73--3.79\\
\hline
$\sin^2\theta_{23} / 10^{-1}$ (NO)
& 5.47$^{+0.20}_{-0.30}$ & 4.67--5.83 & 4.45--5.99 \\
$\sin^2\theta_{23} / 10^{-1}$ (IO)
& 5.51$^{+0.18}_{-0.30}$ & 4.91--5.84 & 4.53--5.98\\
\hline
$\sin^2\theta_{13} / 10^{-2}$ (NO)
& 2.160$^{+0.083}_{-0.069}$ &  2.03--2.34 & 1.96--2.41 \\
$\sin^2\theta_{13} / 10^{-2}$ (IO)
& 2.220$^{+0.074}_{-0.076}$ & 2.07--2.36 & 1.99--2.44 \\
\hline
$\deltacp/\pi$ (NO)
& 1.32$^{+0.21}_{-0.15}$ & 1.01--1.75 & 0.87--1.94 \\
$\deltacp/\pi$ (IO)
& 1.56$^{+0.13}_{-0.15}$ & 1.27--1.82 & 1.12--1.94 \\
\hline
\end{tabular}
\caption{\label{tab:oscillation_summary}
Neutrino oscillation parameters summary determined from the global analysis.
The results for inverted mass ordering were calculated with respect to this mass ordering.}
\end{table}

\begin{figure}[ht]
\centering
\includegraphics[width=0.9\textwidth]{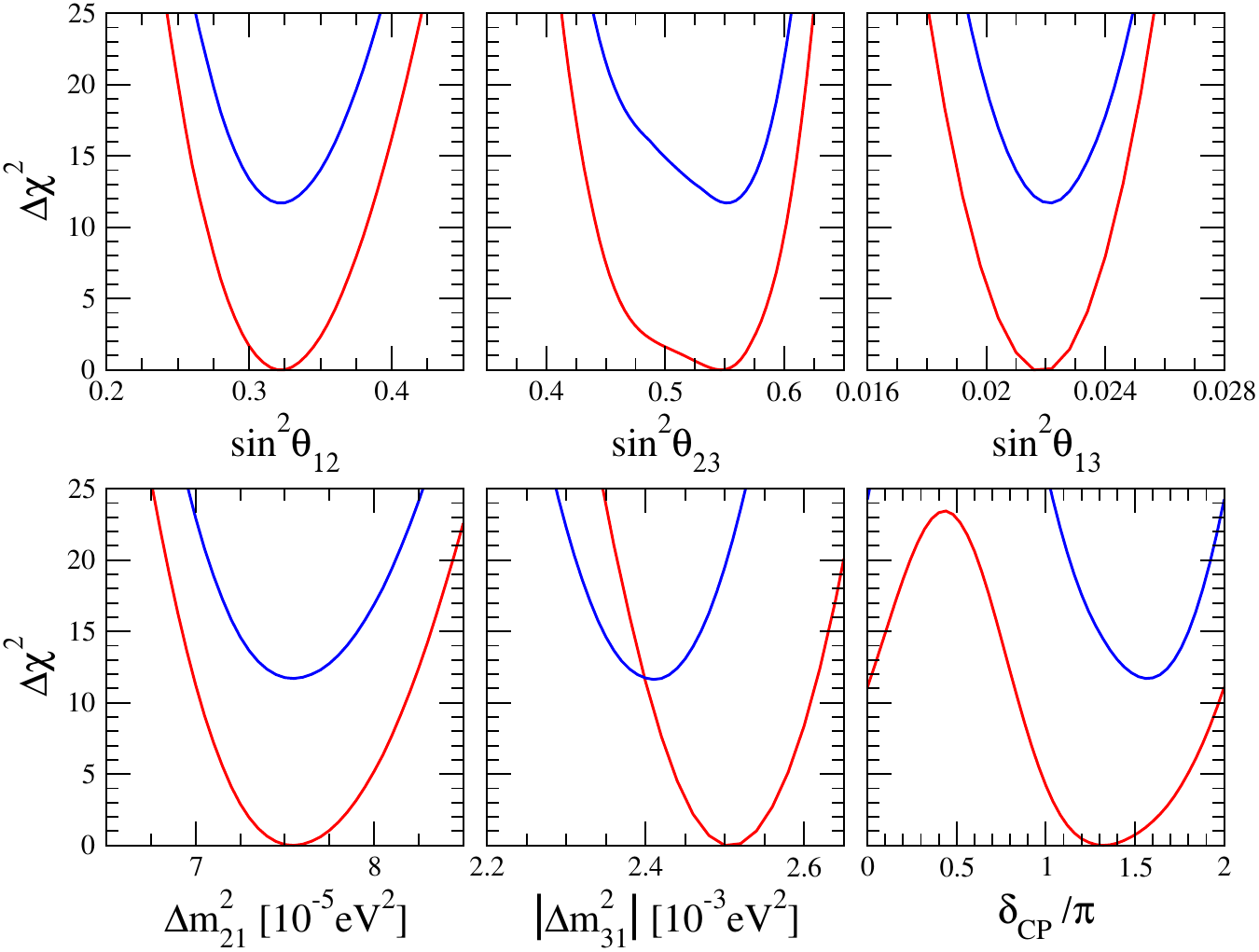}
\caption{
\label{fig:panel-dchi2}
Summary of neutrino oscillation parameters, 2018.
Red (blue) lines correspond to normal ordering (inverted ordering).
Notice that the $\Delta\chi^2$ profiles
for inverted ordering are plotted with respect to the minimum for
normal neutrino mass ordering.}
\end{figure}

The result of the global analysis is summarized
in Table~\ref{tab:oscillation_summary} and Figure~\ref{fig:panel-dchi2}.
Before discussing the sensitivity to the neutrino mass ordering,
we shall briefly discuss some other features of this global fit.
Notice first that now the best-fit value
for the atmospheric mixing angle $\theta_{23}$ lies in the second octant,
although values in the first octant are still allowed
with $\Delta\chi^2 = 1.6\ (3.2)$ for normal (inverted) ordering.
Therefore, the octant problem remains unsolved so far.
Note also that, for the first time,
the CP violating phase $\deltacp$ is determined
with rather good accuracy.
The best-fit values for this parameter lie
close to maximal CP violation,
being $\deltacp = 1.32\pi$ for normal ordering
and $\deltacp = 1.56\pi$ for inverted ordering.
As can be seen from the $\Delta\chi^2$ profile in Figure~\ref{fig:panel-dchi2},
values around $\deltacp\approx 0.5\pi$ are now highly disfavored by data.
Indeed, only around 50\% of the parameter space remains allowed at the 3$\sigma$ level,
roughly the interval $[0.9\pi,1.9\pi]$ for normal and
$[1.1\pi,1.9\pi]$ for inverted ordering.
In the case of normal ordering,
CP conservation remains allowed at 2$\sigma$,
while it is slightly more disfavored for inverted ordering.
For the remaining oscillation parameters,
one clearly sees that neutrino oscillations are entering the precision era,
with relative uncertainties on their determination of 5\% or below.
For a more detailed discussion about these parameters we refer the reader to Refs.~\cite{deSalas:2017kay,globalfit}.

Concerning the neutrino mass ordering,
we obtain a global preference of
$3.4\sigma$ ($\Delta\chi^2=11.7$) in favor of normal ordering.
This result emerges from the combination of all
the neutrino oscillation experiments,
as we explain in the following.
Starting with long-baseline data alone,
the inverted mass ordering is disfavored with $\Delta\chi^2=2.0$,
when no prior is considered on the value of $\theta_{13}$.
However, as explained above, the separate analysis
of the latest \texttt{T2K} and \texttt{NO$\nu$A} data
independently report a $\Delta\chi^2\approx 4$
among the two possible mass orderings
when a prior on the reactor angle is imposed.
This comes from the mismatch between the value of $\theta_{13}$
preferred by short-baseline reactor and long-baseline accelerator experiments,
which is more important for inverted ordering.
Besides that, the combination of \texttt{T2K} and reactor data
results in an additional tension relative to the preferred value
of the atmospheric mass splitting $\Delta m^2_{31}$,
which is again larger for the inverted mass ordering.
This further discrepancy results in a preference
for normal ordering with $\Delta\chi^2 = 5.3$
for the combination of ``\texttt{T2K} plus reactors''
and $\Delta\chi^2 = 3.7$ for the combination of ``\texttt{NO$\nu$A} plus reactors''.
From the combined analysis of all long-baseline accelerator
and short-baseline reactor data
one obtains a $\Delta\chi^2 = 7.5$ between normal and inverted ordering,
which corresponds to a preference of 2.7$\sigma$ in favor of normal mass ordering.
By adding the atmospheric data to the neutrino oscillations fit,
we finally obtain $\Delta\chi^2=11.7$~%
\footnote{Note that this extra sensitivity comes essentially
from \texttt{Super-Kamiokande}, since the effect of
\texttt{IceCube DeepCore} and \texttt{ANTARES}
is negligible in comparison.},
indicating a global preference for normal ordering
at the level of 3.4$\sigma$.

%% file: texs/doublebeta.tex
\subsection{Mass ordering through $\beta$-decay experiments}
The most reliable method to determine the absolute neutrino masses
in a completely model independent way is to measure the spectrum of $\beta$-decay
near the endpoint of the electron spectrum.
The reason for this is related to the fact that, if neutrinos are massive, part of the
released energy must go into the neutrino mass and the electron spectrum endpoint shifts to lower energies.
When there are more than one massive neutrino, each of the separate mass eigenstates
contributes to the suppression of the electron energy spectrum
and it becomes possible to study the pattern of the neutrino masses.
Nowadays none of the $\beta$-decay experiments can reach the energy
resolution required to be able to determine the mass
hierarchy~\footnote{In the case of quasi-degenerate spectrum, the distortion of the
spectrum will consist of just a bending and a shift of the end point,
similar to the case of an electron neutrino with a given mass without
mixing~\cite{Farzan:2001cj},
and the ordering cannot be measured.
Therefore, for future $\beta$-decay searches, measuring the
neutrino mass \emph{ordering} will be practically the same as measuring the
neutrino mass \emph{hierarchy}.},
but we will explain in the following how, in principle,
future experiments may aim at such result.

The best way to depict the effects of the separate mass eigenstates is to compute
the Kurie function for $\beta$-decay.
The complete expression can be written as (see e.g.\ Ref.~\cite{Giunti:2007ry}):
\begin{equation}
 \label{eq:kurie_beta}
 K(T)
 =
 \left[
 (Q_\beta-T)
 \sum^N_{i=1}
  |U_{ei}|^2
  \sqrt{(Q_\beta-T)^2-m_i^2}
  \,\,\Theta(Q_\beta-T-m_i)
 \right]^{1/2}\,,
\end{equation}
where $Q_\beta$ is the Q-value of the considered $\beta$-decay,
$T$ is the electron kinetic energy,
$\Theta$ is the Heaviside step function and
$|U_{ei}|^2$ is the mixing matrix element that defines the mixing between the
electron neutrino flavor and the $i$-th mass eigenstate with a mass $m_i$.
The standard scenario features $N=3$,
but the formula is valid also if a larger number of neutrinos exists
(i.e.\ if there are sterile neutrinos,
as explained for example in Ref.~\cite{Gariazzo:2015rra}).

\begin{figure}
 \centering
 \includegraphics[width=0.6\textwidth]{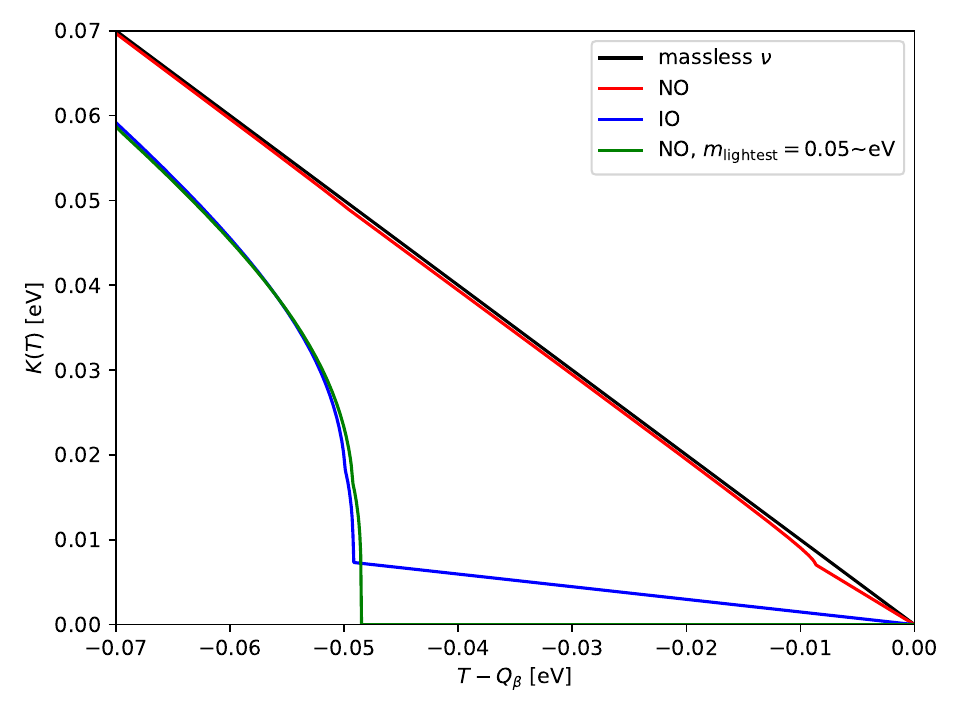}
 \caption{\label{fig:kurie_beta}
 Kurie function in $\beta$-decays.
 The red (blue) line indicates the normal (inverted) ordering case
 for a massless lightest neutrino,
 while the black line is for the case of three massless neutrinos.
 The green curve shows how the Kurie function of a normal ordering scenario with
 $\mlight \simeq \sqrt{\dmsq{31}}$ can mimic the inverted ordering case.}
\end{figure}

The Kurie function of Equation~\eqref{eq:kurie_beta} is depicted in Figure~\ref{fig:kurie_beta},
where we show in red (blue) the result obtained using a massless lightest neutrino
and the current best-fit mixing angles and mass splittings
for normal (inverted) ordering, as described in the previous section.
As a reference, we also plot $K(T)$ for a case with massless neutrinos
only (in black).
Should we consider higher values for the lightest neutrino mass,
the detection of the mass ordering would be increasingly more difficult,
since the separation of the mass eigenstates would decrease,
eventually becoming negligible in the degenerate case.
For this reason we will only discuss the case of a massless lightest
neutrino from the perspective of the $\beta$-decay experiments.

Given the unitarity of the mixing matrix
($\sum^N_{i=1} |U_{ei}|^2 =1$), the normalization of the Kurie function
is the same at $Q_\beta-T\gg m_i$.
Since we are interested in the small differences which appear near the endpoint,
the plot only focuses on the very end of the electron spectrum and
the common normalization is not visible for the inverted ordering case.
In the considered range, however, the effect of the different correspondence between the
mass eigenstates and the mixing matrix elements introduces a difference which
in principle would allow to determine the mass ordering through the observation
of the $\beta$-decay spectrum.
The observation of the kinks in the electron spectrum is very challenging,
especially in the case of normal ordering,
for which even the more pronounced kink
(at $Q_\beta-T\simeq \sqrt{\dmsq{21}}\simeq8$~meV)
is barely visible in Figure~\ref{fig:kurie_beta}.
In the case of inverted ordering, since the mass difference between
the two lightest mass eigenstates is the largest possible one ($\sqrt{\dmsq{31}}\simeq50$~meV),
and the lightest neutrino is the one with the smallest mixing with
the electron neutrino, the amplitude of the kink is much larger.
As a consequence,
an experiment with enough energy resolution to measure
the spectrum in the relevant energy range
can directly probe the mass ordering
observing the presence of a kink.
Note that this measurement can be obtained even without
a detection of the lightest neutrino mass.
As we show in Figure~\ref{fig:kurie_beta}, however,
it is crucial to have a non-zero observation of the electron spectrum
between $Q_\beta$ and $Q_\beta-\sqrt{\dmsq{31}}$,
otherwise one could confuse the inverted ordering spectrum
with a normal ordering spectrum obtained with a larger lightest neutrino mass
$\mlight \simeq \sqrt{\dmsq{31}}$ (green curve).

Another consideration is due.
One could think to probe the neutrino mass ordering just
using the fact that the expected number of events is smaller
in the inverted ordering than in the normal ordering case.
As we discussed above, this could be possible only if some
independent experiment would be able to determine the mass of the lightest neutrino,
in order to break the possible degeneracy between \mlight\ and the mass ordering
depicted by the blue and green curves in Figure~\ref{fig:kurie_beta},
otherwise the conditions required to observe the electron spectrum
between $Q_\beta$ and $Q_\beta-\sqrt{\dmsq{31}}$ would
be probably sufficient to guarantee
a direct observation or exclusion of the kink.
The best way to determine the neutrino mass ordering, however,
may be to use an estimator which compares the binned spectra
in the normal and inverted ordering cases,
as proposed for example in Ref.~\cite{Stanco:2017bpq}
in the context of reactor neutrino experiments.
The authors of the study, indeed, find that a dedicated estimator
can enhance the detection significance with respect to a standard
$\chi^2$ comparison.

To conclude, today the status of $\beta$-decay experiments
is far from the level of determining the mass ordering,
since the energy resolution achieved in past and current measurements
is not sufficient to guarantee a precise probe of the
interesting part of the spectrum.
\texttt{KATRIN}, for example, aims at a sensitivity of 0.2~eV
on the effective electron neutrino mass
\cite{Angrik:2005ep,SejersenRiis:2011sj}, only sufficient to probe
the fully degenerate region of the neutrino mass spectrum.

\subsection{Mass ordering from neutrinoless double beta decay}
\label{ssec:doublebeta_pres}
In the second part of this section we shall discuss instead
the perspectives from the neutrinoless double beta decay
(see e.g.\ the reviews \cite{GomezCadenas:2011it,DellOro:2016tmg} and
also Ref.~\cite{Pascoli:2002xq}),
a process allowed only if neutrinos
are Majorana particles \cite{Schechter:1981bd},
since it requires the lepton number to be violated by two units.
Neutrinoless double beta decay experiments therefore aim at measuring
the life time $\Tbb$ of the decay, which can be written as:
\begin{equation}\label{eq:0n2b_lifetime}
 \frac{1}{\Tbb(\mathcal{N})}
 =
 G_{0\nu}^\mathcal{N}|\nme{\mathcal{N}}|^2
 \left(\frac{|\mbb|}{m_e}\right)^2
 \,,
\end{equation}
where
$m_e$ is the electron mass,
$G_{0\nu}^\mathcal{N}$ is the phase space factor,
$\nme{\mathcal{N}}$ is the nuclear matrix element (NME)
of the neutrinoless double beta decay process, 
$\mathcal{N}$ indicates the chemical element which is adopted to build
the experiment and \mbb \ is the effective Majorana mass, see below. 
In case no events are observed, a lower bound on $\Tbb$
can be derived.
Recent constraints on the neutrinoless double beta decay half-life
come from the
\texttt{EXO-200} \cite{Albert:2014awa},
\texttt{KamLAND-Zen} \cite{KamLAND-Zen:2016pfg},
\texttt{CUORE} \cite{Alduino:2017ehq},
\texttt{Majorana} \cite{Aalseth:2017btx},
\texttt{CUPID-0} \cite{Azzolini:2018dyb},
\texttt{Gerda} \cite{Agostini:2018tnm}
and
\texttt{NEMO-3} \cite{Arnold:2018tmo}
experiments.
The strongest bounds to date on the half-life of the different isotopes are:
$\Tbb(^{76}{\rm Ge})>8.0\e{25}$~yr from \texttt{Gerda} \cite{Agostini:2018tnm},
$\Tbb(^{82}{\rm Se})>2.4\e{24}$~yr from \texttt{CUPID-0} \cite{Azzolini:2018dyb},
$\Tbb(^{130}{\rm Te})>1.5\e{25}$~yr from \texttt{CUORE} \cite{Alduino:2017ehq}
and $\Tbb(^{136}{\rm Xe})>1.07\e{26}$~yr from \texttt{KamLAND-Zen} \cite{KamLAND-Zen:2016pfg}.

The effective Majorana mass reads as:
\begin{equation}\label{eq:mbb}
 \mbb=
%  \left|
 \sum_{k=1}^N e^{i\alpha_k}|U_{ek}|^2 m_k
%  \right|
 \,,
\end{equation}
where $N$ is the number of neutrino mass eigenstates,
each with its mass $m_k$,
$\alpha_k$ are the Majorana phases
(one of which can be rotated away,
so that there are $N-1$ independent phases),
and
$U_{ek}$ represents the mixing between the electron flavor neutrino
and the $k$-th mass eigenstate.
Notice that the conversion between the half-life of the process
and the effective Majorana mass depends on the NMEs
(see e.g.~\cite{Vergados:2016hso,Engel:2016xgb}),
which are typically difficult to compute.
Several methods can be employed and there is no full agreement
between the results obtained with the different methods.
As a consequence, the quoted limits on \Tbb\ can be translated
into limits on \mbb\ which depend on the NMEs.
If the most conservative values for the NMEs are considered,
none of the current constraints reaches
the level required to start constraining
the inverted ordering in the framework of three neutrinos, see
Figure~\ref{fig:mbb_vs_mlightest}. 

If we compute \mbb\ as a function of the lightest neutrino mass
with the current preferred values of the mixing parameters
and in the context of three neutrinos,
we discover that
the value of \mbb\ depends on the mass ordering only for
$\mlight\lesssim40$~meV, see Figure~\ref{fig:mbb_vs_mlightest}.
For this reason,
neutrinoless double beta experiments can aim to distinguish the mass
ordering only for the smallest values of the lightest neutrino mass.
Please note that this also means that if the lightest neutrino
has a mass above $\sim40$~meV,
perfectly allowed by all the present constraints on the neutrino mass scale,
the two mass orderings will never be distinguished
in the context of neutrinoless double beta decay experiments.

When going to smaller \mlight, the situation changes,
as \mbb\ becomes independent of \mlight.
In the region $\mlight\lesssim10$~meV,
a difference between the two mass orderings appears,
since the effective Majorana mass is constrained
by the mass splittings to be larger than $\sim10$~meV for inverted ordering,
while it must be below $\sim7$~meV for normal ordering.
This means that experiments which can test the region $\mbb<10$~meV
can rule out the inverted scenario.
Note that a positive detection of \Tbb\ in the range that
corresponds to $\mbb\gtrsim10$~meV, on the other hand,
would not give sufficient information to determine the mass
ordering without an independent determination of \mlight.
To resume, in the context of three neutrino mixing,
neutrinoless double beta decay experiments alone
will be able to determine the neutrino mass ordering
only ruling out the inverted scheme,
that is to say if the ordering is normal
and $\mlight\lesssim10$~meV.

In any case, we should remember that if no neutrinoless double beta decay candidate event will ever be observed
we will not have determined the mass ordering univocally:
Dirac neutrinos escape the constraints from this kind of process,
so that it would be still perfectly allowed to have
an inverted ordering scheme and no Majorana fermions in the neutrino sector.
Due to the presence of the Majorana phases in Equation~\eqref{eq:mbb}, unfortunately,
there is a small window for \mlight\ in normal ordering
that can correspond to almost vanishing values of \mbb,
which will possibly never be observable.
As we show in Figure~\ref{fig:mbb_vs_mlightest}, however,
the region of parameter space where this happens has a very small volume
if one considers the phases to vary between 0 and $2\pi$,
so that the credible region for \mbb\ in a Bayesian context
shows that it is rather unlikely to have $\mbb\lesssim2\e{-4}$~eV,
as a significant amount of fine tuning
would be needed in the (completely unknown) Majorana phases.
Our statement, which arises from assuming a logarithmic prior on \mlight,
is in perfect agreement with the results of Ref.~\cite{Agostini:2017jim},
where a linear prior on \mlight\ is assumed.

\begin{figure}
\centering
\begin{tabular}{cc}
\includegraphics[width=0.49\textwidth]{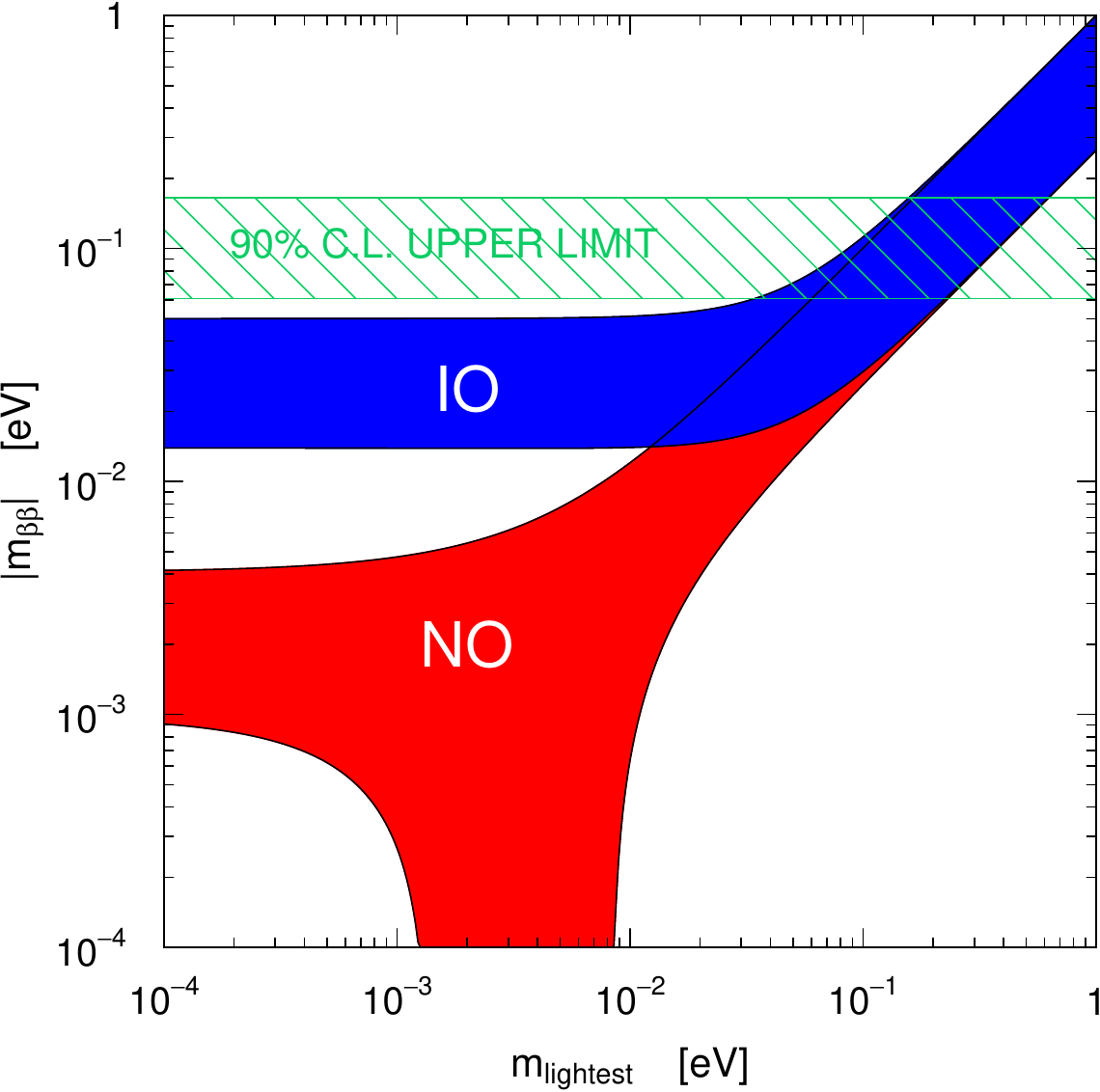}&
\includegraphics[width=0.49\textwidth]{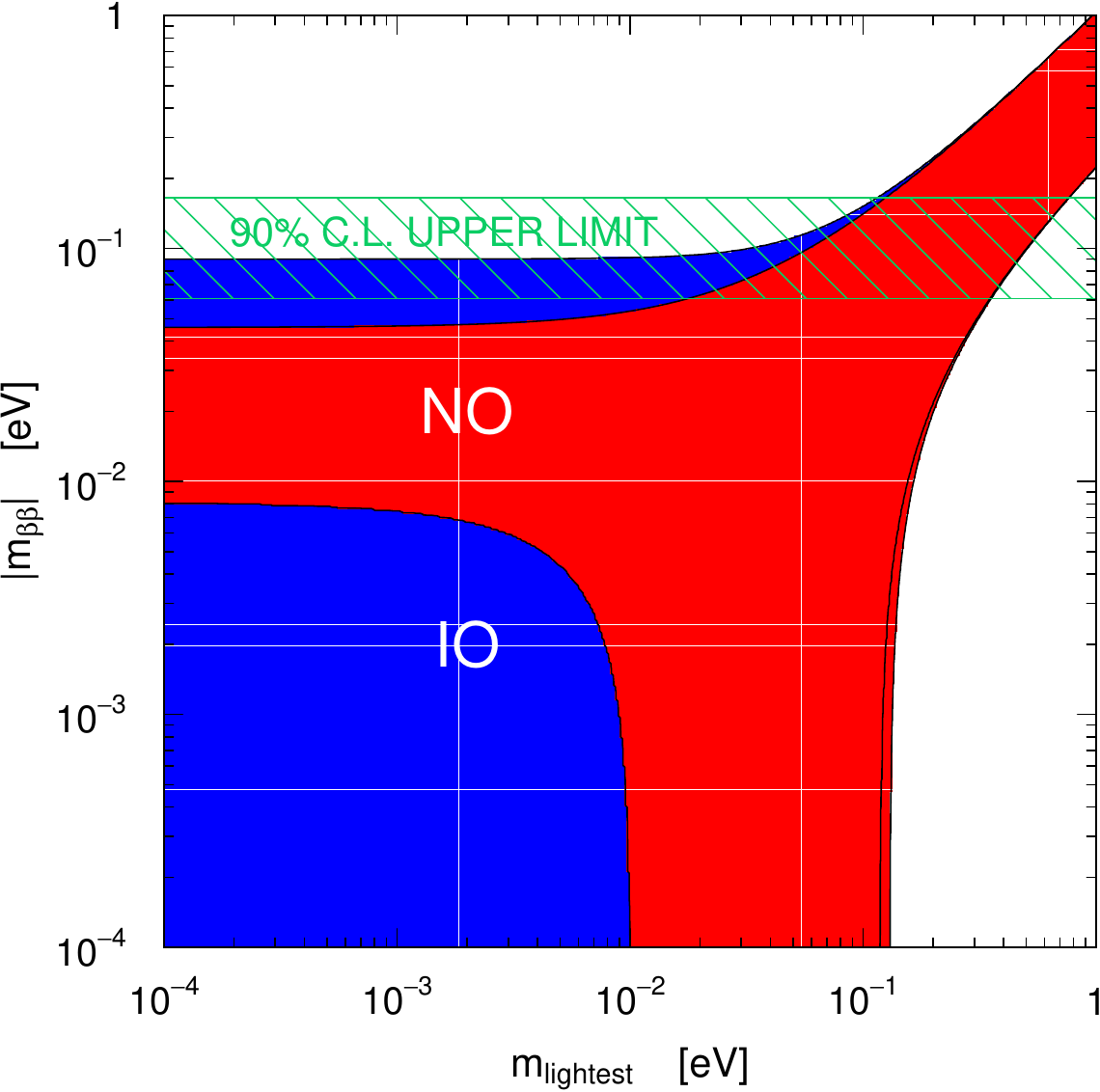}
\end{tabular}
\caption{
\label{fig:mbb_vs_mlightest_3p1}
Effective Majorana mass as a function of the lightest neutrino mass
in the three neutrino (left panel)
and 3+1 neutrino (right panel) scenarios,
at 99.7\%~CL,
comparing normal (red) and inverted (blue) ordering
of the three active neutrinos.
Adapted from Ref.~\cite{Giunti:2017doy}.
The green band represents the 90\%~CL bounds from \texttt{KamLAND-Zen} \cite{KamLAND-Zen:2016pfg},
given the uncertainty on the NME.
}
\end{figure}

Please note that the situation depicted in Figure~\ref{fig:mbb_vs_mlightest}
is only valid if there are only three neutrinos.
If, as the current DANSS \cite{Alekseev:2018efk}
and NEOS \cite{Ko:2016owz} experiments may suggest,
a sterile neutrino with a mass around 1~eV exists
(see e.g.~\cite{Dentler:2017tkw,Gariazzo:2017fdh,Gariazzo:2018mwd,Dentler:2018sju}),
the situation would be significantly different.
The allowed bands for \mbb\ as a function
of the lightest neutrino mass when a light sterile neutrino
is introduced are reported for example in Ref.~\cite{Giunti:2015kza}
(see also Ref.~\cite{Gariazzo:2015rra}).
In this three active plus one sterile neutrino case (3+1),
the contribution of the fourth neutrino mass eigenstate
(mainly mixed with the sterile flavor)
must be added in Equation~\eqref{eq:mbb},
with the consequence that the allowed bands
are located at higher \mbb.
In Figure~\ref{fig:mbb_vs_mlightest_3p1},
adapted from Ref.~\cite{Giunti:2017doy},
we reproduce the dependence of the effective Majorana mass
on the lightest neutrino mass
when one assumes the 3+1 neutrino scenario,
compared with the standard three neutrino case.
As we can see, with the introduction of an extra sterile neutrino state,
\mbb\ is significantly increased for the normal ordering case,
reaching the level of the inverted ordering bands,
which are less shifted towards higher values of \mbb.
Furthermore, in the 3+1 scenario, also in the inverted ordering case
it is possible to have accidental cancellations
due to the three independent Majorana phases in Equation~\eqref{eq:mbb}
(see the detailed discussion of Ref.~\cite{Giunti:2015kza}),
so that a non-detection of the neutrinoless double beta decay process
would never be sufficient to rule out the inverted ordering.
The opposite situation may occur in case the lightest neutrino mass
will be independently constrained to be below $\sim10$~meV
while $\mbb\lesssim10$~meV: in this case, however,
we would rule out normal ordering.
Consequently, if a light sterile neutrino exists, 
neutrinoless double beta experiments will never be able to determine
the mass ordering if the mass ordering is normal,
while some possibility remains if the ordering
of the three active neutrino masses is inverted,
provided that the lightest neutrino is very light
and the Majorana phases are tuned enough.
The \texttt{KamLAND-Zen}, \texttt{Gerda} and \texttt{CUORE} experiments,
using three different materials,
may very soon start probing the inverted ordering region in the case of 3+1 neutrino mixing
for all the possible values of the NMEs,
see Figure~\ref{fig:mbb_vs_mlightest_3p1},
where the current \texttt{KamLAND-Zen} \cite{KamLAND-Zen:2016pfg} constraints are reported.

To conclude and summarize the current status:
neutrinoless double beta decay cannot yet provide constraints
on the neutrino mass ordering.
Depending on the lightest neutrino mass and
on the existence of a fourth (sterile) neutrino,
it would be possible that not even far-future experiments could be able
to reach this goal.

%% file: texs/cosmo.tex
Massive neutrinos affect the cosmological observables in different ways, that we shall summarize in what follows.
For a comprehensive review of the effects of neutrino masses in cosmology,
we refer the reader to the recent work presented in \cite{Lattanzi:2017ubx}.

A very important epoch when discussing the impact of massive neutrinos in the cosmological expansion history
and in the perturbation evolution is the redshift at which neutrinos become non-relativistic.
This redshift is given by
\begin{equation}
1+z_{\rm{nr},i}\simeq 1890 \left(\frac{m_{i}}{ 1\ \textrm{eV}}\right)~,
\end{equation}
with $m_{i}$ referring to the mass of each massive neutrino eigenstate.
Current bounds on neutrino masses imply that at least two out of the three massive eigenstates became non-relativistic in the matter dominated period of the universe.
As stated in the introductory section, and as we shall further illustrate along this section,
cosmological measurements are currently unable to extract individually the masses of the neutrino eigenstates
and the ordering of their mass spectrum and, therefore, concerning current cosmological data,
all the limits on the neutrino mass ordering
will come from the sensitivity to the total neutrino mass $\mnu$.
Consequently, in what follows, we shall mainly concentrate on the effects on the cosmological observables of $\mnu$,
providing additional insights on the sensitivity to the ordering of the individual mass eigenstates whenever relevant.

\subsection{CMB}
\label{sec:cmb}

There are several imprints of neutrino masses on the CMB temperature fluctuations pattern once neutrinos become non-relativistic:
a shift in the matter-radiation equality redshift
or a change in the amount of non-relativistic energy density at late times,
both induced by the evolution of the neutrino background,
that will, respectively, affect the angular location of the acoustic peaks
and the slope of the CMB tail, through the \emph{Late Integrated Sachs Wolfe (ISW) effect}.
The former will mostly modify $\Theta_s$, i.e.\ the angular position of the CMB peaks, which is given by the ratio of the sound horizon and the angular diameter distance, both evaluated at the recombination epoch.
Massive neutrinos enhance the Hubble expansion rate,
with a consequent reduction of the angular diameter distance and an increase of $\Theta_s$,
which would correspond to a shift of the peaks towards larger (smaller) angular scales (multipoles).
The latter, the Late ISW effect, is related to the fact that the gravitational potentials are constant in a matter-dominated universe.
The inclusion of massive neutrinos will delay the dark energy dominated period and
consequently reduce the time variation of the gravitational potential at late times,
suppressing the photon temperature anisotropies in the multipole region $2<\ell<50$.
A very similar effect occurs at early times through the so-called Early ISW effect, which governs the height of the first CMB peak.
Light active neutrino species, indeed, reduce the time variations of the gravitational potential also around the recombination period,
due to the different evolution of these potentials in radiation/matter dominated epochs,
leaving a signature on the CMB photon fluctuations when they become non-relativistic.
Massive neutrinos will therefore decrease the temperature anisotropies by $\Delta C_{\ell}/C_{\ell} \sim \left(m_{\nu,i}/0.1\ \textrm{eV}\right)\,\%$ in the multipole range $20<\ell<500$~\cite{Lesgourgues:2012uu}.

\begin{figure}
\centering
\includegraphics[width=0.6\textwidth]{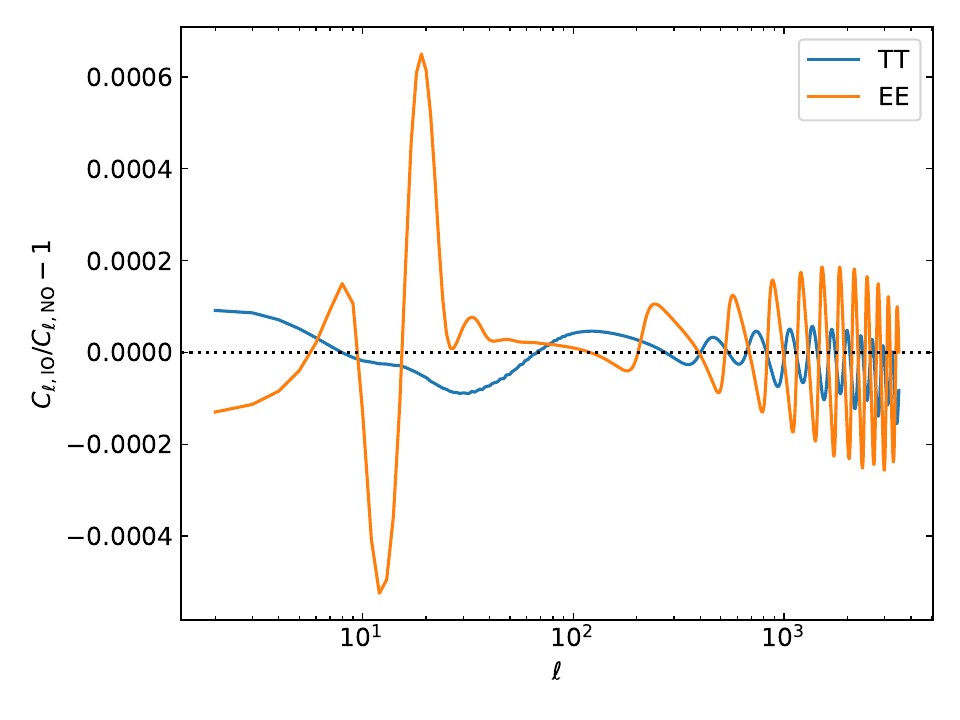}
\caption{\label{fig:Cl}
Relative ratio of the temperature and polarization anisotropies for the inverted over the normal mass orderings, see the text for details.}
\end{figure}

Unfortunately, the Late ISW effect affects the CMB spectrum in a region
where cosmic variance does not allow for very accurate measurements.
From what regards the other two effects, i.e.\ the shift in the location of the acoustic peaks and the Early ISW effect,
they can both easily be compensated varying other parameters which
govern the expansion of the universe.
For example, within the minimal $\Lambda$CDM framework,
the total amount of matter in the universe and the Hubble constant $H_0$
can be tuned in order to compensate the effects of massive neutrinos.
Therefore, CMB primary anisotropies alone can not provide very tight bounds on the neutrino masses, due to the strong parameter degeneracies.
This automatically implies that CMB measurements alone are unable to extract any information concerning the neutrino mass ordering,
as shown in Figure~\ref{fig:Cl}, obtained by means of the publicly available Boltzmann solver Cosmic Linear Anisotropy Solving System (CLASS)~\cite{Lesgourgues:2011re,Blas:2011rf,Lesgourgues:2011rg,Lesgourgues:2013bra}.
In the figure we can notice that the difference between normal and inverted neutrino mass orderings,
for $\mnu=0.12$~eV~\footnote{This is the most constraining $95\%$~CL limit~\cite{Palanque-Delabrouille:2015pga} at present,
excluding combinations of data sets that are in tension, and we have chosen it as the benchmark value in the following discussions throughout this review.}
is almost negligible.
Moreover, the largest differences appear in the multipole range where cosmic variance dominates.

Among the secondary CMB anisotropies,
i.e.\ those generated along the photon line of sight and not produced at recombination,
there are two effects that can notably improve the sensitivity to the total neutrino mass $\mnu$ from CMB observations.
One of them is CMB lensing, that is, a distortion of the photon paths
because of the presence of matter inhomogeneities along the line of sight.
Due to such distortion, the CMB acoustic oscillation features will be smeared out in a scale-dependent way,
mostly due to matter overdensities at $z\lesssim 5$.
By measuring the non-gaussianities of CMB polarization and temperature maps
it is possible to extract the power spectrum of the lensing potential.
This, in turn, contains very useful information on the integrated matter distribution along the line of sight.
Since massive neutrinos behave differently from a pure cold dark matter component, characterized by zero velocities,
the small-scale structure suppression induced by the non-negligible neutrino dispersion velocities
will decrease the CMB lensing signal expected in the absence of neutrino masses~\cite{Kaplinghat:2003bh,Song:2004tg,Lesgourgues:2005yv,Smith:2006nk,dePutter:2009kn,Allison:2015qca},
leaving unchanged the power spectrum of the lensing potential at large scales, and suppressing it at small scales.
Furthermore, since CMB lensing involves high redshifts,
non-linearities do not enter in the calculation of the matter density field.
Therefore, CMB lensing enhances the capabilities to bound the neutrino masses using CMB data.
In the future, this technique may even surpass weak lensing capabilities,
based on statistical analyses of the ellipticity of remote galaxies, see below and section~\ref{sec:futcosmo}.
Indeed, nowadays, measurements from the \pla\ satellite
constrain the neutrino masses dominantly through CMB gravitational lensing.
As stated in Ref.~\cite{Ade:2013zuv}, increasing the neutrino mass implies an increase on the expansion rate at redshifts $z \ge 1$,
corresponding to a suppression of clustering at scales below the size
of the horizon at the non-relativistic transition.
This effect leads to a decrease in CMB lensing that,
at multipoles $\ell=1000$,
is $\sim 10\%$ for $\mnu=0.66$~eV.

On the other hand, we have the reionization process in the late universe,
when the first generation of galaxies emitted ultraviolet (UV) photons that ionized the neutral hydrogen,
leading to the end of the so-called \emph{dark ages}.
Reionization increases the number density of free electrons $n_e$ which can scatter the CMB with a probability
given by a quantity named \emph{reionization optical depth}, $\tau$,
which can be computed as an integral over the line of sight of $n_e$.
The consequence of an increase of $\tau$ on the CMB temperature fluctuations
is the suppression of the acoustic peaks by a factor $\exp(-2\tau)$ at scales
smaller than the Hubble horizon at the reionization epoch.
Even if from the point of view of CMB temperature anisotropies
this effect is highly degenerate
with a change in the amplitude of the primordial power spectrum $A_{s}$,
which governs the overall amplitude of the CMB spectra,
reionization induces linear polarization on the CMB spectrum,
leading to a ``reionization bump'' in the polarization spectra at large scales,
which otherwise would vanish.
Even if the reionization signal is rather weak,
as it amounts to no more than $\sim 10\%$ of the primary polarization signal~\cite{Aghanim:2007bt},
very accurate measurements of the reionization optical depth $\tau$
sharpen considerably the CMB neutrino mass bounds~\cite{Vagnozzi:2017ovm},
as they alleviate the degeneracy between $A_s$ and $\tau$ and consequently the existing one between $\mnu$ and $A_s$.

\subsection{Large scale structure of the universe}
\label{sec:large}
\begin{figure}
\centering
\includegraphics[width=0.53\textwidth]{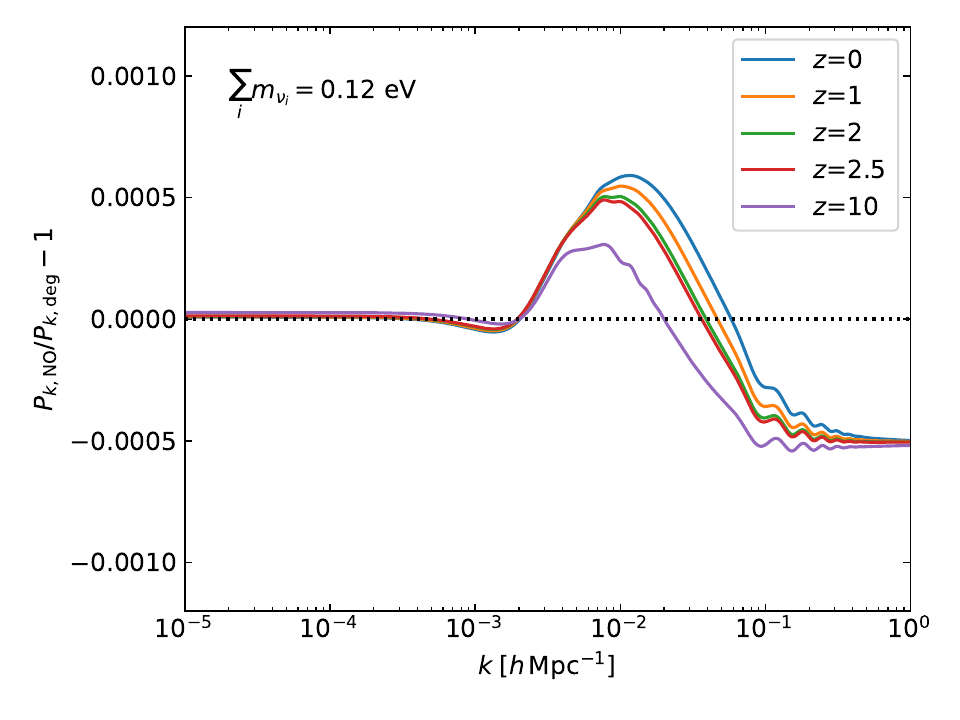}
\includegraphics[width=0.53\textwidth]{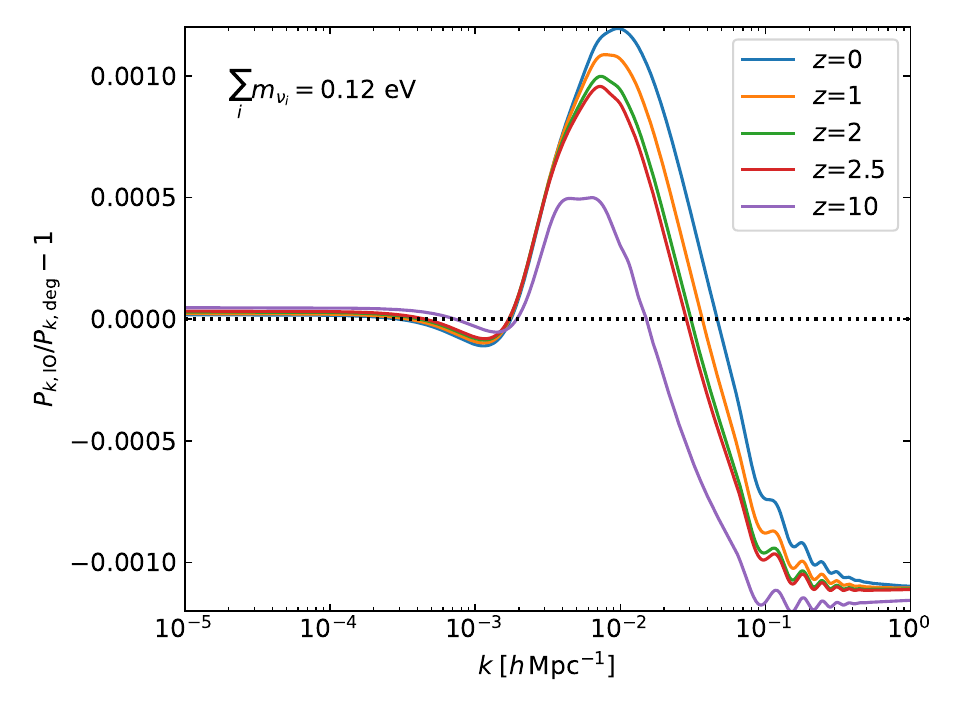}
\includegraphics[width=0.53\textwidth]{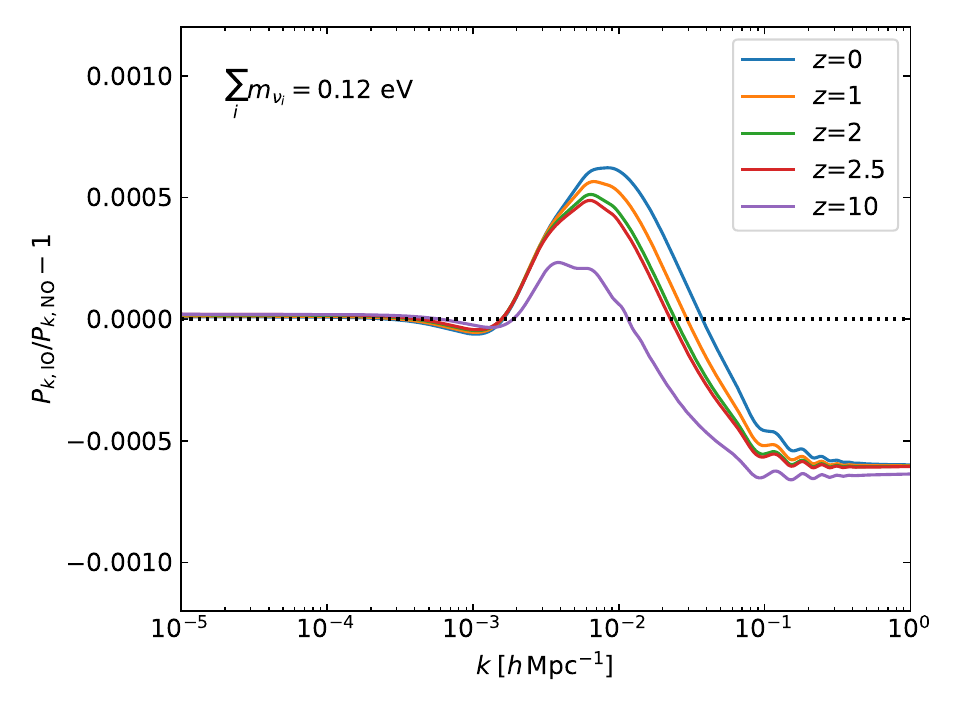}
\caption{The top (middle) panel shows the ratios of the matter power spectrum of normal (inverted) ordering over the degenerate case.
The bottom panel shows the ratio of the matter power spectra for the normal and inverted mass orderings.
See the text for details.}
\label{fig:Ps}
\end{figure}

The largest effect of neutrino masses on the cosmological observables is imprinted in the matter power spectrum~\cite{Bond:1980ha,Hu:1997mj}.
Neutrinos are \emph{hot} dark matter particles and, therefore,
due to the pressure gradient,
at a given redshift $z$,
the non-relativistic neutrino overdensities can only cluster at scales for which the wavenumber of perturbations
is below the neutrino free streaming scale $k_{f_s}$
(i.e.\ at scales $k < k_{f_s}$), with
\begin{equation}
k_{f_s}(z)=\frac{0.677}{(1+z)^{1/2}}\left(\frac{m_\nu}{1\ \textrm{eV}}\right)
\sqrt{\Omega_{m}}\, h \ \textrm{Mpc}^{-1}~,
\end{equation}
being $\Omega_m$ the ratio of the total matter energy density over the critical density at redshift zero.
The free-streaming nature of the neutrino will be directly translated into a suppression of the growth of matter fluctuations at small scales.
One could then conclude that extracting the neutrino relic masses and their ordering is a straightforward task,
once that measurements of the matter power spectrum at the relevant scales are available at a different number of redshifts.
The former statement is incorrect, not only because
it does not consider the existence of degeneracies
with the remaining cosmological parameters,
but also because a number of subtleties must be taken into account,
as we shall explain in what follows.
The decrease of the matter power spectrum due to the total neutrino mass $\mnu$ is, in principle, currently measurable.
Nonetheless, when fixing $\mnu$, the total mass could be splitted differently
among the three neutrino mass eigenstates (i.e.\ $m_1$, $m_2$ and $m_3$),
modifying slightly the relativistic to non-relativistic transition.
This will affect both the background evolution and the perturbation observables~\cite{Lesgourgues:2004ps}:
the different free-streaming scales associated to each of the three neutrino mass eigenstates will be imprinted in the matter power spectrum.
Figure~\ref{fig:Ps} shows the ratios of the matter power spectrum for normal over degenerate, inverted over degenerate,
and inverted over normal neutrino mass spectra for a total neutrino mass of $0.12$~eV.
We illustrate such ratios at different redshifts.
Notice that the differences among the possible neutrino mass schemes are tiny, saturating at the $0.06\%$ level at $k> 0.2h$~Mpc$^{-1}$.
Therefore, only very futuristic means of measuring the matter power spectrum could be directly sensitive to the neutrino mass ordering,
and, eventually, be able to isolate each of the free-streaming scales associated to each individual neutrino mass eigenstate.
We shall comment on these future probes in section~\ref{sec:cosmo21}.

Since, at present, matter power spectrum data constrain exclusively $\mnu$,
it is only via these bounds, combined with CMB or other external data sets,
that nowadays a limit on the neutrino mass ordering can be obtained, see section~\ref{sec:cosmolimits}.
Nevertheless, and as aforementioned, there are a number of problems which may interfere with a proper understanding
of the scale-dependent neutrino mass suppression of clustering.
The first of them is due to the fact that observations measure
the spatial distributions (or their Fourier transforms, the power spectra) of galaxies, clusters, or quasars,
e.g.\ of given tracers,
mapping the large scale structure of the Universe at a number of redshifts,
by measuring the growth of fluctuations at different scales.
However, the matter distribution is not directly measured, i.e.\ it needs to be inferred from the tracers observed.
A simple model of structure formation suggests that
at large scales and, therefore, when the perturbation evolution is still in the linear regime,
the galaxy power spectrum is related to the matter one by a constant named $b$, the light-to-mass bias~\cite{Desjacques:2016bnm}.
The galaxy bias can be determined either separately by independent
methods or to be considered as an additional free parameter to be measured
together with the neutrino mass $\mnu$.
This approach has been followed in many studies in the past~\cite{Cuesta:2015iho,Giusarma:2016phn,Vagnozzi:2017ovm}.
However, when dealing with neutrino masses,
the relationship between the tracers and the underlying matter field may be more complicated,
as neutrinos themselves may induce scale-dependent features in the bias~\cite{LoVerde:2013lta,Castorina:2013wga,Munoz:2018ajr}
due to their free-streaming nature
(see also the recent work of Ref.~\cite{Giusarma:2018jei}
for a new method to extract a scale-dependent bias,
based on the cross-correlation of CMB lensing maps and galaxy samples).

Another additional complication when extracting the neutrino mass
from clustering observations arises from to the presence
of non-linearities at scales
$k\gtrsim k_{\rm NL}^0\equiv0.1-0.2$~$h$~Mpc$^{-1}$ at $z=0$.
The effect of neutrino masses is very-well understood on linear scales,
i.e.\ scales below $k_{\rm NL}$ at $z=0$
(or located at slightly larger values of $k$ but at higher redshifts).
Massive neutrinos induce a suppression in the linear matter power spectrum below their free streaming scale
$\Delta P/P \propto -8 f_\nu$,
with $f_\nu$ the fraction of matter in the form of massive neutrinos~\cite{Hu:1997mj}.
Accurate descriptions of the matter power spectrum in the non-linear regime are therefore mandatory
in order to be sensitive to the full neutrino mass signature.
This is particularly important in the case of galaxy surveys,
in which the information depends on the number of independent modes available,
and where going to smaller scales (i.e.\ larger values of $k$)
has a profound impact on the sensitivity to neutrino masses.
Several approaches have been followed in the literature to account for the effect of massive neutrinos in the non-linear regime,
most of them relying on N-body cosmological simulations,
which have been upgraded to include the effects of neutrino clustering in the evolution of the cosmological structures.
Methods range from perturbative attempts~\cite{Brandbyge:2008js,Saito:2008bp,Shoji:2010hm,AliHaimoud:2012vj,Upadhye:2015lia,Archidiacono:2015ota,Senatore:2017hyk,Dakin:2017idt}
to the fully non-linear inclusion~\cite{Banerjee:2018bxy,Banerjee:2016zaa,Inman:2015pfa,Liu:2017now,Dakin:2017idt,Villaescusa-Navarro:2013pva}
of neutrinos as an extra set of particles.
A conservative alternative consists on using exclusively power spectrum measurements within the linear regime
(i.e., $k<0.1$~$h$~Mpc$^{-1}$).
Some of the cosmological constraints have also been obtained
using the mildly non-linear regime ($k<0.2$~$h$~Mpc$^{-1}$) by means
of the so-called Halofit
formalism~\cite{Smith:2002dz,Takahashi:2012em}.
The Halofit prescription models the non-linear matter power
spectrum, and it has been calibrated against a wide range of
CDM  simulations.
It has also been extended for massive neutrino
cosmologies~\cite{Bird:2011rb}.
Other predictions for the non-linear
matter power spectrum include the Coyote emulator~\cite{Heitmann:2013bra}, which is
based on a set of high-accuracy N-body simulations.

However, there is also another avenue to use large scale structure information,
the \emph{geometrical} approach, which exploits the so-called
Baryon Acoustic Oscillations (BAO)
rather than the measurements of the broad-band \emph{shape} of the galaxy power spectrum.
The BAO signal appears as a peak in the two-point mass correlation function
corresponding to the distance a sound wave can travel in the photon-baryon fluid
from very early in the universe until the drag epoch,
when the baryon optical depth equals one.
The BAO signature provides a standard ruler to measure the distance to various redshifts,
and it can be measured either \emph{along the line of sight},
in which the radial distance is inversely proportional to the Hubble expansion rate $H(z)$,
or \emph{across the line of sight}, in which case the angular distance is proportional to an integral of $H(z)$, the angular diameter distance $d_A(z)$.
To use the BAO method, one must, therefore, extract the acoustic scale
from the clustering of some tracer of the baryon distribution (galaxies, quasars).
This is typically done statistically using the two-point correlation function of the spatial distribution of tracers,
or from its Fourier transform, the power spectrum.
From these functions, it is possible to measure two different quantities
corresponding to the oscillations parallel and perpendicular to the line of sight,
that is $r_s H(z)$ and $d_A(z)/r_s$, with $r_s$ the sound horizon at the drag epoch.
Many of the BAO analyses to date have used spherically averaged clustering statistics,
measuring an effective distance
$D_V \equiv (zd_A(z)^2/H(z))^{\frac{1}{3}}$,
which is the volume-averaged distance.
Some of the most recent BAO extractions by the
Sloan Digital Sky Survey III (\texttt{SDSS-III})~\cite{Eisenstein:2011sa}
Baryon Oscillation Spectroscopic Survey (\texttt{BOSS})~\cite{Dawson:2012va}
have achieved, by measuring the clustering of 1.2 million galaxies with redshifts $0.2<z<0.75$,
$1.8\%$ precision on the radial BAO distance and
$1.1\%$ precision on the transverse distance in the $z<0.75$ redshift region~\cite{Ross:2016gvb,Beutler:2016ixs,Vargas-Magana:2016imr}.
These results improve former determinations from previous data releases of \texttt{BOSS} and \texttt{SDSS}~\cite{Anderson:2013zyy,Anderson:2012sa,Tojeiro:2014eea,Ross:2014qpa,Eisenstein:2005su}
or other galaxy surveys~\cite{Percival:2001hw,Cole:2005sx,Blake:2011en,Kazin:2014qga,Beutler:2011hx},
see also the recent works of Ref.~\cite{Bautista:2017wwp} for a $2.6\%$ measurement of $D_V$
at $2.8\sigma$ significance with the extended Baryon Oscillation Spectroscopic Survey (\texttt{eBOSS}) from \texttt{SDSS-IV}~\cite{Dawson:2015wdb}.
The Dark Energy Survey (\texttt{DES}) has also achieved a $4.4\%$ accuracy on the measurement of $d_A(z)/r_s$ at $z=0.81$~\cite{Abbott:2017wcz}.

Galaxy clustering measurements can also be exploited to constrain, at a number of redshifts,
the product of the linear growth rate $f\times \sigma_8$~%
\footnote{Here, $\sigma_8$ corresponds to the normalization of the matter power spectrum on scales of $8 h^{-1}$~Mpc, see Equation~\eqref{eq:sigma8}.},
by means of the so-called redshift space distortions, caused by galaxy peculiar velocities,
see the recent analyses of Refs.~\cite{Beutler:2016arn,Satpathy:2016tct,Sanchez:2016sas}.

Apart from the spatial distribution of galaxies, there are also other ways of mapping the large scale structure of the universe at different cosmic times.
The Lyman-$\alpha$ forest power spectrum from distant quasars plays a major role for constraining the neutrino masses,
as it is sensitive to smaller scales, where the effect of neutrino masses is more pronounced.
We refer the reader to the seminal works of Refs.~\cite{Croft:1999mm,Seljak:2004xh,Goobar:2006xz,Seljak:2006bg,Gratton:2007tb,Fogli:2008ig,Viel:2010bn}.
In addition, since the redshifts at which Lyman-$\alpha$ forest probes are sensitive to
are higher than those corresponding to galaxy surveys,
a fixed scale $k$ will be closer to the linear regime in the Lyman-$\alpha$ case.
An additional benefit of going to higher redshifts is that uncertainties
related to the evolution of the dark energy fluid will be sub-dominant,
as dark energy effects are expected to be more prominent at very low redshifts.
However, modeling the neutrino mass effect in the Lyman-$\alpha$ forest power spectrum is highly non-trivial
as it may strongly rely on hydrodynamical simulations~\cite{Viel:2010bn}.
These numerical calculations try to properly account for
the late time non-linear evolution of the intergalactic medium (IGM),
including reionization processes~\cite{Viel:2010bn,VillaescusaNavarro:2011me}.
The BAO signature can also be measured in the flux correlation function of the Lyman-$\alpha$ forest of quasars,
first detected at a mean redshift  $z=2.3$ in Ref.~\cite{Busca:2012bu},
see also Refs.~\cite{Slosar:2013fi,Font-Ribera:2013wce,Delubac:2014aqe,Bautista:2017zgn,Bourboux:2017cbm} and Ref.~\cite{Aubourg:2014yra},
in which joint constraints from the BAO signature from galaxies and
quasars have been presented.

Galaxy clusters provide yet another test which allows us to trace the clustering of matter perturbations
and, therefore, to test the suppression due to the presence of a non-zero $\mnu$.
Galaxy clusters are, by far, the largest virialised objects in the universe,
providing a measurement of the so-called cluster number count function $d N/dz$.
This function gives the number of clusters of a certain mass $M$ within a redshift interval (bin) $z+\delta z$ and, for a given survey:
\begin{equation}
{d N\over dz}\Big|_{M>M_{\rm min}}=f_{\rm sky} {dV(z)\over dz}\int_{M_{\rm min}}^\infty dM \,{dn\over dM}(M, z)~.
\end{equation}
The quantity $f_{\rm sky}=\Delta\Omega/4\pi$ refers to the fraction of sky covered by the survey and the unit volume is given by
\begin{equation}
{dV(z)\over dz}={4\pi\over H(z)} \int_0^z dz' {1\over H(z')^2}~.
\end{equation}
While the redshift is relatively easy to measure,
the main uncertainty of this method comes from the cluster mass estimates,
determined through four main available methods:
X-rays, velocity dispersion, Sunyaev-Zeldovich (SZ) effect~%
\footnote{The thermal SZ thermal effect consists on a spectral distortion on CMB photons
which arrive along the line of sight of a cluster.}
and weak lensing.
The overall error in the cluster mass determination is usually around $\Delta M/M\sim 10\%$.
Furthermore, in order to relate the cluster number count function to the underlying cosmological parameters,
one needs as an input a mass function $d n(z, M)/d M$ describing the abundance of virialised objects at a given redshift,
usually obtained by means of $N$-body simulations~\cite{Tinker:2008ff,Costanzi:2013bha}.
This mass function depends on both the matter mass-energy density and on the standard deviation (computed in linear perturbation theory) of the density perturbations:
\begin{equation}
\label{eq:sigma8}
\sigma^2=\frac{1}{2\pi^2}\int^\infty_{0} dk k^2 P(k) W^{2}(kR)~,
\end{equation}
where $P(k)$ is the matter power spectrum,
$W(kR)$ is the top-hat window function,
$R$ is the comoving fluctuation size,
related to the cluster mass $M$ as $R = (3M/4\pi \rho_m)^{1/3}$,
and
\begin{equation}
\label{eq:wf}
W(kR)=\frac{3 \left(\sin(kR) - (kR) \cos(kR) \right)}{(kR)^3}~.
\end{equation}
There are still some degeneracies in the cosmological parameters
probed by cluster surveys, whose results are reported 
by means of a relationship between the matter clustering amplitude $\sigma_8$
(obtained from Equation~\eqref{eq:sigma8}),
and the matter mass-energy density $\Omega_m$ parameters.
More concretely, cluster catalogues measure the so-called cluster normalization condition,
$\sigma_8 \Omega^\gamma_m$,
where $\gamma \sim 0.4$~\cite{Allen:2011zs,Weinberg:2012es}.
Current cluster catalogs include X ray clusters
(see e.g.~\cite{Hilton:2017gal,Sohn:2017oax} and references therein),
the optically detected \texttt{SDSS} photometric \texttt{redMaPPer} cluster catalog~\cite{Rykoff:2013ovv,Rozo:2013vja,Rozo:2014zva}
and the \pla\ SZ galaxy clusters (\texttt{PSZ2})~\cite{Ade:2015gva},
which contains more than a thousand confirmed clusters.
Other SZ cluster catalogs are those detected from the Atacama Cosmology Telescope (\texttt{ACT})~\cite{Hilton:2017gal}
and from the South Pole Telescope (\texttt{SPT})~\cite{deHaan:2016qvy}.

Last but not least, weak lensing surveys are also an additional probe of the large scale structure effects of massive neutrinos~\cite{Cooray:1999rv,Abazajian:2002ck,Hannestad:2006as,Kitching:2008dp,Ichiki:2008ye,Tereno:2008mm,DeBernardis:2009di,Debono:2009bd,Jimenez:2010ev}.
Light rays from distant galaxies are bent by the matter density perturbations between the source galaxies and the observer, thereby inducing distortions in the observed images of the source galaxies, see the reviews~\cite{Munshi:2006fn,Kilbinger:2014cea}.
Commonly, the deformations in the source galaxies are rather weak and to extract the lensing signature one needs a correlation among different galaxy images,
the so-called shear-correlation functions.
By measuring the angular correlation of these distortions, one can probe the clustering statistics of the intervening matter density field along the line of sight,
without relying strongly on bias assumptions, setting therefore independent constraints on the neutrino masses.
Weak lensing surveys usually report their cosmological constraints in terms of the clustering amplitude $\sigma_8$
and the current matter energy density $\Omega_m$.
More specifically, they make use of the combination
$S\equiv \sigma_8 \sqrt{\Omega_m}$
as an accurate description of the amplitude of structure growth in the universe.
The most recent weak lensing cosmological analyses profiting
of weak lensing data from \texttt{DES}
and from the Kilo Degree Survey (\texttt{KiDS}),
consisting of $\sim 450$ deg$^2$ of imaging data, are presented in Refs.~\cite{Abbott:2017wau}
and \cite{Hildebrandt:2016iqg,Joudaki:2016kym,Kohlinger:2017sxk}, respectively.

\subsection{Cosmological bounds on neutrino masses and their ordering}
\label{sec:cosmolimits}
In the following, we shall review the current cosmological bounds on neutrino masses and on their ordering,
firstly in the standard $\Lambda$CDM scenario and then when considering extended cosmological models.

\subsubsection{Constraints within the $\Lambda$CDM universe}
Focusing on bounds exclusively from the \pla\ collaboration,
making use of their CMB temperature anisotropies measurements in the multipole range $\ell\lesssim 2500$ (\plaT)
and of their low-multipole (up to $\ell=29$) polarization data, \texttt{lowP},
a bound of $\mnu <0.72$~eV at $95\%$~CL~\cite{Ade:2015xua} is reported.
When high-multipole (i.e.\ small scale, $\ell>30$) polarization measurements are included in the analyses (\plaTElowP),
the quoted constraint is $\mnu <0.49$~eV at $95\%$~CL.
As the \plaTE\ data combination may still have some systematics due to temperature-to-polarization leakage~\cite{Ade:2015xua},
the bounds including these measurements provide the less conservative approach when exploiting CMB data.
In 2016, the \pla\ collaboration presented a series of new results
based on a new analysis,
in which the modeling and removal of unexplained systematics
in the large angular polarization data were accounted for~\cite{Aghanim:2016yuo}.
The value of the optical depth $\tau$ found in these refined analyses (using the \texttt{SimLow} likelihood)
was smaller than that quoted in previous analyses~\cite{Ade:2015xua}:
while the \texttt{lowP} data was providing $\tau=0.067\pm 0.022$,
the \texttt{SimLow} likelihood results in $\tau=0.055\pm 0.009$.
The most important consequence of this lower value of $\tau$ on the CMB bounds on $\mnu$
is related to the degeneracy between the amplitude of the primordial power spectrum, $A_s$, and $\tau$,
as already introduced in section~\ref{sec:cmb}:
a lower value of $\tau$ will imply a lower value of $A_s$,
thus implying a lower overall normalization of the spectrum,
leading therefore to tighter constraints on neutrino masses.
The $95\%$~CL limits of $\mnu <0.72$~eV and $\mnu <0.49$~eV,
respectively from the \plaTlowP\ and \plaTElowP\ analyses,
are updated to $\mnu <0.59$~eV and $\mnu <0.34$~eV
when using \plaTsl\ and \plaTEsl, respectively.
These constraints are clearly located away from the region
in which a preference for a given mass ordering (normal versus inverted) may show up.
Indeed, the CMB data alone were used by the authors of Ref.~\cite{Gerbino:2016ehw} which,
by means of a novel approach to quantify the neutrino mass ordering,
have shown that the odds favoring normal ordering versus inverted
ordering are $1:1$ and $9:8$ in the case of the \plaTlowP\ and \plaTElowP\ data combinations, respectively.
These results point to an inconclusive strength of evidence, see Table~\ref{tab:jeffreys}.
Based on a full Bayesian comparison analysis, Ref.~\cite{Gariazzo:2018pei} has shown,
using \plaTElowP\ measurements together with global neutrino oscillation data,
that the Bayes factor for such a combination is $\lnBsim{2.5}$ for almost all the possible parameterizations and prior choices considered.
This value of the Bayes factor,
which only points to weak preference for normal ordering,
is entirely due to neutrino oscillation data,
in agreement with the results of Ref.~\cite{Caldwell:2017mqu}.
Therefore, \pla\ temperature and polarization measurements alone can not further improve
our current knowledge of the neutrino mass ordering from global oscillation data.

The CMB limits on neutrino masses can also include the lensing likelihood,
which leads to $\mnu <0.59$~eV at $95\%$~CL for the case of \plaTElowP\ + \lens\ measurements~\cite{Ade:2015xua}.
Notice that the bound with the lensing likelihood is less tight than that obtained
without the lensing potential extraction
($\mnu <0.49$~eV at $95\%$~CL from \plaTElowP).
The reason is related to the fact that,
while the \pla\ CMB power spectra favor a larger lensing amplitude,
the lensing potential reconstructions prefer a lower one.
Since increasing the neutrino masses reduces the lensing amplitude,
the one dimensional posterior distribution of $\mnu$ arising
from the combination of CMB temperature, polarization and lensing data sets
shifts the neutrino mass constraints away from zero,
so that less posterior volume is found near zero
than when constraining $\mnu$ only with CMB temperature and polarization data.

A significant strengthening on the aforementioned limits can be obtained by means of additional data sets,
which help enormously in breaking the degeneracies which are allowed when only CMB data are considered.
Among them, the one existing between $\mnu$ and the Hubble constant $H_0$ (see e.g~\cite{Giusarma:2012ph}).
Large scale structure data from galaxy clustering are of great help in breaking degeneracies.
When exploited in the geometrical (BAO) form,
the \pla\ collaboration quotes $95\%$~CL limits of $\mnu<0.17$~eV from the combination
\plaTEsl\ + \lens\ + BAO data~\cite{Aghanim:2016yuo}~%
\footnote{The BAO measurements exploited by the \pla\ collaboration include the 6dF Galaxy Survey (\texttt{6dFGS})~\cite{Beutler:2011hx},
the \texttt{BOSS LOWZ} BAO extraction of the spherical averaged $D_v/r_s$~\cite{Ross:2014qpa,Anderson:2013zyy}
and the \texttt{BOSS CMASS-DR11} data of Ref.~\cite{Anderson:2013zyy}.}.
Concerning the neutrino mass ordering,
the addition of BAO measurements to CMB \pla\ measurements
leads to odds for the normal versus the inverted ordering
of $4:3$ and of $3:2$,
in the case of the \plaTlowP\ + BAO and \plaTElowP\ + BAO respectively,
suggesting only very mild evidence for the normal ordering case~\cite{Gerbino:2016ehw}.
These results confirmed the previous findings obtained in Ref.~\cite{Hannestad:2016fog}.
The authors of Ref.~\cite{Vagnozzi:2017ovm} reported odds
for the normal versus the inverted ordering of $2.4:1$ from the combination of
\plaTE\ + BAO plus the \texttt{SimLow} prior on the reionization optical depth,
i.e.\ $\tau=0.05\pm 0.009$.
Notice that if data are not informative enough, the choice of prior on \mlight\ will make a difference in the odds ratio~%
\footnote{See e.g.\ the work of the authors of Ref.~\cite{Simpson:2017qvj} and the explanation of their results in \cite{Schwetz:2017fey,Gariazzo:2018pei}.
See also Refs.~\cite{Long:2017dru,Heavens:2018adv,Handley:2018gel} for useful discussions
concerning the prior choice on the neutrino mass ordering extraction.
}.

Another possible avenue to exploit galaxy clustering data
is to use the information contained in the full-shape of the galaxy power spectrum
(see e.g.\ Refs.~\cite{Barger:2003vs,Spergel:2003cb,Hannestad:2003ye,Elgaroy:2003yh,Allen:2003pta,Hannestad:2003xv,Tegmark:2003ud,Crotty:2004gm,Tegmark:2006az,Hamann:2006pf,Fogli:2006yq,Fogli:2008ig,Hamann:2008we,Hamann:2010pw,RiemerSorensen:2011fe,dePutter:2012sh,Zhao:2012xw,Giusarma:2013pmn,Riemer-Sorensen:2013jsa,dePutter:2014hza,Cuesta:2015iho,Giusarma:2016phn,Vagnozzi:2017ovm}).
Notice however that using BAO is currently a more robust method,
as the effects of the galaxy bias and non-linearities are not as severe as in the \emph{shape} approach.
In the minimal $\Lambda$CDM scenario, the BAO \emph{geometrical} approach can supersede the neutrino mass constraints obtained from the \emph{shape} one, see e.g.\ Refs.~\cite{Hamann:2010pw,Giusarma:2012ph}.
Indeed, a dedicated analysis has been devoted in Ref.~\cite{Vagnozzi:2017ovm}
to compare the constraining power of these two different approaches
to large scale structure data with the \texttt{SDSS-III BOSS} measurements.
The conclusions are that,
even if the latest measurements of the galaxy power spectrum map a large volume of our universe, the geometric approach is still more powerful,
at least within the minimal $\Lambda$CDM + $\mnu$ cosmology.
The better performance of BAO measurements
is partly due to the upper cutoff applied in the scale $k$ of the power spectrum
when dealing with shape analyses (mandatory to avoid non-linearities),
and partly due to the fact that two additional nuisance parameters are further required
to relate the galaxy power spectrum to the matter power one~%
\footnote{As stated in Ref.~\cite{Vagnozzi:2017ovm},
in the future, a deeper understanding of the non-linear regime of the galaxy power spectrum with massive neutrinos included,
plus a better understanding of the galaxy bias
could change the constraining power of full-shape analyses versus BAO ones.}.
As an example, the $95\%$~CL bound of $\mnu<0.118$~eV obtained with
\plaTE\ + BAO plus \texttt{SimLow}
is degraded to $\mnu<0.177$~eV
when replacing part of the BAO data
[more concretely, the high redshift \texttt{BOSS CMASS} Data Release 11 (\texttt{DR11}) sample by the full-shape power spectrum measurements from the \texttt{BOSS CMASS} Data Release 12 (\texttt{DR12})].

An alternative tracer to map out the large scale structure in our universe
and improve the CMB-only bounds
on the sum of the three active neutrinos is the Lyman-$\alpha$ forest,
leading to neutrino mass bounds which turn out to be among the most constraining ones.
By means of the one-dimensional Lyman-$\alpha$ forest power spectrum
extracted by Ref.~\cite{Palanque-Delabrouille:2013gaa} and combining these measurements
with \plaTElowP\ + BAO,
the authors of Ref.~\cite{Palanque-Delabrouille:2015pga} find a $95\%$~CL upper limit of $\mnu<0.12$~eV.
It is also remarkable the fact that, even without the addition of CMB data,
the combination of the Lyman-$\alpha$ forest power spectrum of~\cite{Palanque-Delabrouille:2013gaa},
together with those from the \texttt{XQ-100} quasars at $z\simeq 3.5-4.5$
and the high-resolution \texttt{HIRES/MIKE} spectrographs at $z=4.2$ and $z=4.6$~\cite{Viel:2013apy},
is already able to provide a limit of $\mnu <0.8$~eV~\cite{Yeche:2017upn},
showing clearly the enormous potential of small-scale probes to extract the neutrino masses.

The degeneracies among $\mnu$ and the other cosmological parameters
that appear when considering CMB data only
can also be strongly alleviated by the addition
of Supernova Ia luminosity distance data and/or
local measurements of the Hubble parameter~%
\footnote{See Refs.~\cite{Jackson:2007ug,Freedman:2010xv}
for dedicated reviews concerning the different possible local measurements of $H_0$.
Among them, the one based on Cepheid variables.}.
Concerning Supernovae Ia data,
the most complete photometric redshift calibrated sample joins
the SuperNova Legacy Survey (\texttt{SNLS}) and \texttt{SDSS} supernova catalogs.
This \texttt{Joint Light-Curve Analysis (JLA)} catalogue~\cite{Betoule:2012an,Betoule:2014frx,Mosher:2014gyd}
has been used by the \pla\ collaboration and by other analyses to improve the constraints on $\mnu$,
being its impact particularly crucial in non-minimal cosmologies~\cite{Vagnozzi:2018jhn},
as we shall explain towards the end of this section.
Concerning the value of $H_0$,
as there exists a strong anti-correlation between the Hubble constant and $\mnu$ when considering CMB measurements,
larger mean values of $H_0$ will lead to tighter constraints
on the neutrino mass and consequently on the inverted mass ordering.
When performing combined analyses of CMB and $H_0$ data,
the 2015 \pla\ release relies
on the reanalysis \cite{Efstathiou:2013via}
of a former $H_0$ measurement based on the \texttt{Hubble Space Telescope (HST)}
[$H_0=(73.8\pm 2.4)$~km~s$^{-1}$~Mpc$^{-1}$~\cite{Riess:2011yx}],
which was in mild ($2.5\sigma$) tension with the value of the Hubble parameter derived
from 2013 \pla\ CMB data~\cite{Ade:2013zuv}.
This reanalysis \cite{Efstathiou:2013via}
considers the original Cepheid data of Ref.~\cite{Riess:2011yx} and
uses a new geometric maser distance estimate to the active galaxy NGC 4258~\cite{Humphreys:2013eja},
which is used as a distance anchor to
find a value of the Hubble constant $H_0=(70.6\pm 3.3)$~km~s$^{-1}$~Mpc$^{-1}$~%
\footnote{The final result of Ref.~\cite{Efstathiou:2013via}
is however $H_0=(72.5\pm 2.5)$~km~s$^{-1}$~Mpc$^{-1}$,
when the combination of the $H_0$ results obtained
with three different distance estimators is performed.
The value $H_0=(70.6\pm 3.3)$~km~s$^{-1}$~Mpc$^{-1}$ is
the only one of the three which shows a milder tension
with the $H_0$ estimate from \pla.}.
The limit on the sum of the three active neutrino masses
reported by the \pla\ collaboration using this value of $H_0$ is
$\mnu <0.23$~eV at $95\%$~CL,
when combined with \plaTlowP\ + \lens\ + BAO + \texttt{SNIa} data.
Other estimates of the Hubble constant, however, exist.
The $2.4\%$ determination of Ref.~\cite{Riess:2016jrr} profits from new,
near-infrared observations of Cepheid variables,
and it provides the value
$H_0=(73.02\pm 1.79)$~km~s$^{-1}$~Mpc$^{-1}$~\cite{Riess:2016jrr}.
As the former mean $H_0$ value is higher than the one considered by the \pla\ collaboration,
it will lead to tighter limits on $\mnu$.
Indeed, the work of Ref.~\cite{Vagnozzi:2017ovm} quotes
the $95\%$~CL bounds of $\mnu<0.196$~eV and $\mnu<0.132$~eV
when combining with external data sets using the priors
$H_0=(70.6\pm 3.3)$~km~s$^{-1}$~Mpc$^{-1}$ and
$H_0=(73.02\pm 1.79)$~km~s$^{-1}$~Mpc$^{-1}$, respectively.
Focusing on the less conservative choice
$H_0=(73.02\pm 1.79)$~km~s$^{-1}$~Mpc$^{-1}$,
odds for the normal versus the inverted neutrino mass ordering of $3.3:1$ were found
for both the \plaTE\ + BAO + \texttt{SimLow} + \texttt{$H_0$}
and the \plaTE\ + BAO + \texttt{SimLow} + \texttt{$H_0$} + \plaSZ\ data sets~\cite{Vagnozzi:2017ovm}.
The $95\%$~CL upper bounds on the neutrino mass
for these two combinations are
$\mnu <0.094$~eV and $\mnu <0.093$~eV, respectively.
These results indicate, once again,
very mild evidence for the normal mass ordering, even within these more aggressive and less conservative scenarios,
in which the very tight limit on $\mnu$ is mostly due to the tension between CMB
and direct measurements of the Hubble constant $H_0$,
together with the strong degeneracy between $\mnu$ and $H_0$.
Using these results, we stress that having an upper bound $\mnu\lesssim0.1$~eV at $95\%$~CL
is not equivalent to having a 95\%~CL preference for normal ordering:
the probabilities for normal ordering and inverted ordering, as computed from the odds $3.3:1$,
are approximately 77\% and 23\%
(see also section~\ref{ssec:bayesian}).

In general, the combination of data sets that are inconsistent is potentially dangerous.
Apart from the constraining effect on the neutrino mass limits
when considering a particular prior on the Hubble constant $H_0$,
there have been also other related cases in which the neutrino masses
were a tool to accommodate tensions among different data sets.
For instance, in the case of galaxy cluster counts,
a larger neutrino mass could in principle fit both CMB and low-redshift universe constraints
on the power spectrum normalization $\sigma_8$~\cite{Allen:2003pta}.
The effect of combining CMB and BAO observations with clusters and/or shear data is presented in Ref.~\cite{Costanzi:2014tna},
where it is shown that the inclusion of either cluster or shear measurements in the \pla\ + BAO joint analysis
indicates a preference for $\mnu>0$ at more than $2\sigma$.
However, the authors clearly state that these results can not be interpreted
as a claim for a cosmological detection of the neutrino mass,
but rather as a remedy to palliate the existing tension
between clusters/shear data and \pla/BAO observations.

Finally, weak lensing constraints from the
\texttt{Dark Energy Survey} Year 1 results~\cite{Abbott:2017wau} (\texttt{DES Y1}),
have also recently provided bounds on the sum of the total neutrino mass.
Based on $1321$~deg$^2$ imaging data,
\texttt{DES Y1} analyses exploit the galaxy correlation function
(from 650.000 luminous red galaxies divided into five redshift bins)
and the shear correlation function
(from twenty-six million source galaxies from four different redshift bins)
as well as the galaxy-shear cross-correlation.
The $95\%$~CL upper bound reported on the neutrino mass
after combining their measurements with \plaTElowP\ + BAO +\texttt{JLA} is $\mnu<0.29$~eV,
$\sim20\%$ higher than without \texttt{DES} measurements.
The reason for this higher value of $\mnu$ is that the clustering amplitude in the case of \texttt{DES Y1}
is mildly below the one preferred by \pla\ measurements.
Since larger values of the neutrino mass will decrease
the value of the clustering amplitude,
the upper limit on the total neutrino mass is loosened
by $\sim 20\%$ after the \texttt{DES} results are also considered.

\subsubsection{Extensions to the minimal $\Lambda$CDM universe}
\label{sec:mnuw}
So far we have discussed the neutrino mass and neutrino mass ordering sensitivities within the minimal $\Lambda$CDM universe.
However, these limits will change when additional parameters are introduced in the analyses.

The first and most obvious scenario one can consider
is to test the stability of the neutrino mass limits
when new physics is added in the neutrino sector.
As already mentioned in section~\ref{ssec:doublebeta_pres},
short baseline neutrino experiments indicate that a light sterile neutrino at the eV scale may exist.
These extra sterile species will contribute to the
effective number of relativistic degrees of freedom, \neff,
defined by
\begin{equation}
\rho_{\rm{rad}}
=
\left(1+\frac{7}{8}\left(\frac{4}{11}\right)^{4/3} \neff\right)
\rho_\gamma~,
\label{eq:rho_rad}
\end{equation}
where $\rho_{\rm{rad}}$ ($\rho_\gamma$) is
the total radiation (CMB photons) energy density.
In the standard picture $\neff=3.046$ \cite{Mangano:2005cc,deSalas:2016ztq}.
This number accounts for the three active neutrino contribution
and considers effects related to
non-instantaneous neutrino decoupling and QED finite temperature corrections to the plasma evolution~%
\footnote{The work of Ref.~\cite{deSalas:2016ztq},
including three-flavor neutrino oscillations,
has revisited previous calculations including all the proper collision integrals
for both diagonal and off-diagonal elements in the neutrino density matrix
and quotes the value of $\neff=3.045$.}.
Variations in \neff, apart from the light sterile neutrino,
may be related to the existence of additional relativistic particles,
as thermally-produced axions (see below).
Analyses in which both the active neutrino masses
and the number of additional massless or massive species are varied simultaneously
have been extensively carried out in the literature~\cite{Hamann:2007pi,Hamann:2010bk,Giusarma:2011ex,Hamann:2011ge,Giusarma:2012ph,RiemerSorensen:2012ve,Archidiacono:2013fha,DiValentino:2013qma,Archidiacono:2013lva},
showing that the bounds on the active neutrino mass
are relaxed when additional sterile species are added
to the fermion content of the SM of particle physics.
The constraints on the total neutrino mass $\mnu$
are less stringent than in the standard three neutrino case
due to the large degeneracy between $\mnu$ and \neff,
which arises from the fact that a number of
massless or sub-eV sterile neutrino species contributing to the radiation content of the universe
will shift both the matter-radiation equality era
and the location of the CMB acoustic peaks.
This effect could be compensated by enlarging the matter content of the universe,
implying therefore that larger values for the neutrino masses could be allowed.
Consequently, a priori, the constraints on $\mnu$ when \neff\ is also a free parameter in the analyses
are not very competitive.
Fortunately, CMB measurements from the \pla\ collaboration
help enormously in sharpening the measurement of \neff,
especially when considering polarization measurements
at small scales: including data at high multipoles,
one obtains $\Delta\neff<1$ at more than $4 \sigma$ significance
from \pla\ CMB observations alone.
Indeed, the limit on the sum of the three active neutrino species
considering also additional radiation neutrino species
(i.e.\ massless sterile neutrino species) is
$\mnu<0.178$~eV at $95\%$~CL from \plaTElowP\ + BAO data,
very similar to the bound $\mnu<0.168$~eV at $95\%$~CL
arising from the very same dataset
within the minimal $\Lambda$CDM scenario with three active massive neutrinos.
Another possible way of relaxing (or even avoiding)
the cosmological neutrino mass limits is via the addition
of non-standard interactions in the active neutrino sector~\cite{Beacom:2004yd,Bell:2005dr,Hannestad:2004qu,Fardon:2003eh,Afshordi:2005ym,Brookfield:2005bz,Brookfield:2005td,Bjaelde:2007ki,Mota:2008nj,Ichiki:2008rh,Boehm:2012gr,Archidiacono:2013dua,Dvali:2016uhn,DiValentino:2017oaw}.

Furthermore, additional relics different from sterile neutrinos,
as thermal axions~\cite{Peccei:1977hh,Peccei:1977ur,Weinberg:1977ma,Wilczek:1977pj},
contributing to both \neff\ at early times and to the hot dark matter component in the late-time universe,
suppress small-scale structure formation
and show effects very similar to those induced
by the (active) three massive neutrino species.
Therefore, the cosmological bounds on the three active neutrino masses are modified in scenarios with thermal axions,
see Refs.~\cite{Melchiorri:2007cd,Hannestad:2007dd,Hannestad:2008js,Hannestad:2010yi,Archidiacono:2013cha,Giusarma:2014zza,DiValentino:2015zta,DiValentino:2015wba,DiValentino:2016ikp},
as these two species have to share the allowed amount of dark matter.
Nonetheless, there are non-negligible differences among neutrinos and thermal axions:
\textit{(a)} axions are colder than neutrinos, as they decouple earlier;
\textit{(b)} since the axion is a scalar particle,
an axion mass larger than the neutrino one is required
in order to make identical contributions to the current mass-energy density of the universe;
\textit{(c)} in the case of axions, the contribution to \neff\
is related to their mass, while for neutrinos this is usually not true.
Consequently, the bounds on the axion mass are always less constraining than for the neutrino,
and \mnu\ is slightly more constrained in scenarios
in which thermal axions are also present.
For instance, Ref.~\cite{DiValentino:2015sam} quotes
$\mnu <0.175$~eV at $95\%$~CL from the \plaTElowP\ + BAO data combinations
when considering only neutrinos,
while the analyses in Ref.~\cite{DiValentino:2015wba},
including massive axions, find
$\mnu <0.159$~eV and $m_a <0.763$~eV, both at $95\%$~CL,
for the very same data combination.

There are also other ways of relaxing the cosmological neutrino mass bounds,
related either to the early or the late-time accelerating periods in the universe.
In the former case one can play with inflationary processes.
There have been a number of studies devoted to explore their degeneracies with the neutrino sector,
see the recent works of Refs.~\cite{Hamann:2006pf,Joudaki:2012fx,Archidiacono:2013lva,dePutter:2014hza,DiValentino:2016ikp,Canac:2016smv,Gerbino:2016sgw}.
The authors of Ref.~\cite{DiValentino:2016ikp} have considered
a non-standard and parametric form for the primordial power spectrum,
parameterized with the
\texttt{PCHIP} (piecewise cubic Hermite interpolating polynomial) formalism
with twelve nodes between $k_1= 5\times10^{-6}$~Mpc$^{-1}$ and $k_2= 10$~Mpc$^{-1}$
and derived the neutrino mass constraints within this more general scenario.
When only \plaTlowP\ measurements were considered,
the $95\%$~CL mass bound of $\mnu<0.75$~eV obtained
with the usual power-law description of the primordial power spectrum
was relaxed to $\mnu<2.16$~eV.
This large value is explained in terms of the strong degeneracy between $\mnu$
and the \texttt{PCHIP} nodes corresponding to the wave-numbers
where the contribution of the Early ISW effect is located,
in such a way that the effect induced by a non-zero neutrino mass
is easily compensated by an increase of the primordial power spectrum at these scales only.
BAO information improves considerably the limits in the \texttt{PCHIP} prescription,
but it is the addition of high-$\ell$ polarization data
what further constrains the effect.
The $95\%$~CL upper limit in the \texttt{PCHIP} scenario
from the \plaTElowP\ + BAO data combination is $\mnu<0.218$~eV,
quite close to the bound found when the usual power-law description is applied ($\mnu<0.175$~eV).
Reference~\cite{Gerbino:2016sgw} deals instead with the robustness of the constraints
on the scalar spectral index $n_s$ under several neutrino physics scenarios.
The authors have explored the shifts induced in the inflationary parameters
for different choices of the neutrino mass ordering,
comparing the approximate massive neutrino case
(one massive eigenstate plus two massless species
when the total mass is close to the minimum allowed value by oscillation data,
and three degenerate massive neutrinos otherwise)
versus the exact case (normal or inverted mass orderings).
While the mass-ordering assumptions are not very significant when $\mnu$ is fixed to its minimum value,
there is a shift in $n_s$ when $\mnu$ is a free parameter,
inherited from the strong degeneracies in the $\mnu$, $H_0$ and $\Omega_m h^2$ parameter space.
Fortunately, BAO measurements revert the $\mnu$-$n_s$
anti-correlation present with CMB data only,
and the shift in the spectral index turns out to be negligible.

The other possibility is to play with the late-time acceleration period and study how the neutrino mass bounds change.
The current accelerated expansion of the universe,
explained in terms of a cosmological constant in the minimal $\Lambda$CDM scenario,
may be due to a dynamical dark energy fluid
with a constant equation of state $w\neq -1$ or
a time-dependent $w(z)$~\cite{Chevallier:2000qy,Linder:2002et},
or to quintessence models, based on the existence of a cosmic scalar field~\cite{Caldwell:1997ii,Zlatev:1998tr,Wang:1999fa,Wetterich:1994bg,Peebles:1987ek,Ratra:1987rm},
which provide a dynamical alternative to the cosmological constant scenario with $w=-1$.
It is naturally expected that the neutrino mass bounds will increase when enlarging the parameter space.
Indeed, when the dark energy equation of state is allowed to vary within the phantom region $w<-1$,
there is a very well-known degeneracy
between the dark energy equation of state $w$
and the sum of the three active neutrino masses,
as first noticed in Ref.~\cite{Hannestad:2005gj}
(see also Refs.~\cite{LaVacca:2009ee,Joudaki:2012fx,Archidiacono:2013lva,Lorenz:2017fgo,Vagnozzi:2018jhn,Sutherland:2018ghu})~%
\footnote{Interacting dark energy models can also change the neutrino mass constraints,
see e.g.~\cite{Gavela:2009cy,Reid:2009nq,Honorez:2009xt,LaVacca:2008mh,Guo:2018gyo}.}.
It has been pointed out that for very high neutrino masses
only dark energy models lying within the phantom region will be allowed.
The reason for that is the following:
a larger $\mnu$ can be compensated by a larger $\Omega_m$,
which in turn can be compensated by a smaller equation of state
of the dark energy component, i.e.\ $w<-1$.
Interestingly, the recent work of Ref.~\cite{Vagnozzi:2018jhn}
shows that the cosmological bounds on $\mnu$ become more restrictive
in the case of a dynamical dark energy component
with $w(z)\ge -1$.
Following the usual dynamical dark energy description,
whose redshift dependence is described by the standard Chevallier-Polarski-Linder (CPL) parametrization~\cite{Chevallier:2000qy,Linder:2002et},
the authors of \cite{Vagnozzi:2018jhn} have shown that
the combination of \plaTE\ + BAO + \texttt{JLA}
plus the \texttt{SimLow} prior on the reionization optical depth provides,
at $95\%$~CL, $\mnu<0.11$~eV in the CPL case
when restricting $w(z)\ge-1$
(within the physical, non-phantom region),
while $\mnu<0.13$~eV in the $\Lambda$CDM case.
When $w(z)$ is also allowed to be in the phantom region ($w(z)<-1$) within the CPL parameterization,
the resulting $95\%$~CL constraint on the three active neutrino masses is $\mnu<0.37$~eV.
These results have a direct impact on the cosmological preference for a given neutrino mass ordering.
Following Refs.~\cite{Hannestad:2016fog,Vagnozzi:2017ovm},
it is found that the normal ordering is mildly preferred over the inverted one,
with posterior odds $3:1$ for the data combination quoted above when $w(z)\ge -1$.
On the contrary, if there is no such a restriction
and $w(z)$ can also take values in the phantom region, the odds are $1:1$.
The odds in the non-phantom dynamical dark energy case
show a mild preference for normal ordering.
Therefore, if neutrino oscillation experiments
or neutrinoless double beta decay searches find
that the neutrino mass ordering is the inverted one,
if the current cosmic acceleration is due to a dynamical dark energy component,
one would require this component to be phantom.

As a final point in this section,
we would like to note that also in scenarios
in which the current accelerated expansion is explained
by means of modifications of gravity at ultra-large length scales,
the cosmological limits on neutrino masses will differ
from those in the standard $\Lambda$CDM model,
see e.g.~\cite{Huterer:2006mva,Baldi:2013iza,Hu:2014sea,Shim:2014uta,Barreira:2014ija,Bellomo:2016xhl,Peirone:2017vcq,Renk:2017rzu,Dirian:2017pwp}.

%% file: texs/analysis.tex
In this section we shall combine the available measurements
that allow us to constrain the neutrino mass ordering,
updating the results presented in Ref.~\cite{Gariazzo:2018pei}.

\subsection{Bayesian model comparison}
\label{ssec:bayesian}
Before performing the analysis, we will briefly summarize the method
we will adopt to compare the two possible orderings.

We will follow a Bayesian approach to model comparison (see previous work suggesting the
Bayesian method  as the most suited one for the mass ordering extraction
in Refs.~\cite{Qian:2012zn} and \cite{Blennow:2013kga})~%
\footnote{We also refer the reader to Ref.~\cite{Blennow:2013oma},
which provides a comprehensive study of the sensitivity reach to the mass ordering in the context of the frequentist approach.},
which makes use of the Bayesian evidence $Z$.
This quantity, which is also known as the marginal likelihood,
is defined as the integral over the entire parameter space
$\Omega_\mathcal{M}$
of the prior $\pi(\theta)\equiv p(\theta|\mathcal{M})$ times
the likelihood $\mathcal{L}(\theta)\equiv p(d|\theta,\mathcal{M})$,
where $\theta$ is the set of parameters that describe the model $\mathcal{M}$
and $d$ represents the available data:
\begin{equation}\label{eq:bayevid}
Z_\mathcal{M}
=
\int_{\Omega_\mathcal{M}}
\mathcal{L}(\theta)\,\pi(\theta)\,d\theta\,.
\end{equation}
The posterior probability of the model $\mathcal{M}$
can be written in terms of its prior probability $\pi(\mathcal{M})$
times the Bayesian evidence $Z_\mathcal{M}$:
\begin{equation}\label{eq:modelposterior}
 p(\mathcal{M}|d)
 \propto
 Z_\mathcal{M}
 \,\pi(\mathcal{M})
 \,,
\end{equation}
where the proportionality constant depends only on the data.
In our case we will be interested in comparing
normal ordering (NO)
and inverted ordering (IO),
which can be considered as two different competing models
$\mathcal{M}_1\equiv\rm{NO}$ and
$\mathcal{M}_2\equiv\rm{IO}$.
The ratio of the posterior probabilities
of the two models can be written as
\begin{equation}\label{eq:post_no_o_io}
 \frac{p({\rm NO} |d)}{p({\rm IO}|d)}
 =
 \Bnoio
 \frac{\pi(\rm{NO})}{\pi(\rm{IO})}
 \,,
\end{equation}
having defined the Bayes factor as
\begin{equation}\label{eq:bayfac_noio}
 \Bnoio
 =
 \Zno/\Zio
 \,.
\end{equation}
Assuming the same prior probabilities for normal and inverted ordering,
the Bayes factor is what determines the odds in favor of one
of the competing models.
In particular we will indicate the results in terms of its natural logarithm
$\ln\Bnoio$, which will be positive when data will prefer normal ordering
and negative otherwise.
Quantitatively, the preference is given in terms of posterior odds,
which are always $|\Bnoio|:1$ in favor of the preferred model.
The strength of the preference can be also translated into an empirical scale,
which in our case is summarized in the third column of Table~\ref{tab:jeffreys}.

\begin{table}
\centering
\begin{tabular}{c|c|c|c}
\hline
$|\ln \Bnoio|$ & Odds & strength of evidence & $N\sigma$ for the mass ordering \\
\hline
$<1.0$         & $\lesssim 3:1$     & inconclusive & $<1.1\sigma$\\
$\in[1.0,2.5]$ & $(3-12):1$         & weak         & $1.1-1.7\sigma$\\
$\in[2.5,5.0]$ & $(12-150):1$       & moderate     & $1.7-2.7\sigma$\\
$\in[5.0,10]$  & $(150-2.2\e4):1$   & strong       & $2.7-4.1\sigma$\\
$\in[10,15]$   & $(2.2\e4-3.3\e6):1$& very strong  & $4.1-5.1\sigma$\\
$>15$          & $>3.3\e6:1$        & decisive     & $>5.1\sigma$\\
\hline
\end{tabular}
\caption{\label{tab:jeffreys}
Jeffreys' scale \cite{Jeffreys:1961a} for estimating the strength of the preference
for one model over the other (adapted from Ref.~\cite{Trotta:2008qt}) when
performing Bayesian model comparison analysis.
The fourth column indicates the approximate correspondence between the quoted Bayes factor levels
and the $N\sigma$ probabilities computed for a Gaussian variable.}
\end{table}

Let us briefly discuss the correspondence of the quoted levels that classify the strength of the preference
in favor of one of the competing models.
In the case of the neutrino mass ordering, we have only two possibilities (normal or inverted),
so that
$
p({\rm NO}|d)+p({\rm IO}|d)
=
\pi(\rm{NO})+\pi(\rm{IO})
=
1$.
If we assign the same prior probability to the two cases, $\pi(\rm{NO})=\pi(\rm{IO})=1/2$,
it is easy to compute the posterior probability for each of the two cases,
which will be
\begin{eqnarray}
p({\rm NO}|d) &=& \Bnoio/(\Bnoio+1)
\,,
\label{eq:post_no}\\
p({\rm IO}|d) &=& 1/(\Bnoio+1)
\label{eq:post_io}\,,
\end{eqnarray}
having used Equations~\eqref{eq:post_no_o_io} and \eqref{eq:bayfac_noio}.
The confidence levels for the rejection of the disfavored (e.g.\ inverted) mass ordering
will then be
$x = 100\times(1 - |\Bnoio|^{-1})\,\%$.
For example, a Bayes factor $\Bnoio=10$ corresponds to a rejection
of the inverted ordering at 90\% CL.
If, instead, we want to reproduce the probability levels $P=\rm{erf}(N/\sqrt{2})$
that are usually associated to the classical $N\sigma$ levels for a Gaussian measurement,
being $\rm{erf}$ the error function and considering, for example, $N\in(1,2,3,4,5)$,
the corresponding Bayes factors $B$ can be computed to be $B=P/(1-P)$,
which gives us $\ln B_{N\sigma} \simeq 0.77,\,3,\,5.9,\,9.7,\,14.37$.
Therefore, our ``strong'', ``very strong'' and ``decisive'' levels
roughly correspond to the $>3\sigma$, $>4\sigma$ and $>5\sigma$ probabilities,
as indicated in the fourth column of Table~\ref{tab:jeffreys}.

\subsection{Parameterization and data}
Our two competing models are described by the same number of parameters,
listed with their priors in Table~\ref{tab:commonParams}:
the three neutrino mixing angles
($\sin^2\theta_{12}$, $\sin^2\theta_{13}$, $\sin^2\theta_{23}$),
the CP violating phase \deltacp\
and the parameters associated with
neutrino masses,
neutrinoless double beta decay (\doublebeta)
and cosmology,
as we shall describe now.

We consider in our analysis the parameterization that uses the two mass splittings
(\dmsq{21} and \dmsq{31}) and the lightest neutrino mass \mlight\ with
logarithmic priors.
This parameterization, strongly motivated by the physical observables, was shown to provide the optimal strategy to
successfully explore the neutrino parameter space, see
Ref.~\cite{Gariazzo:2018pei}~\footnote{As we are making use of logarithmic
priors here, we shall not report the upper limits we obtain on the sum
of the neutrino masses, as they will be much smaller than the usually quoted results
due to the volume effects associated with the use of the logarithmic prior,
that naturally leads to a preference for small neutrino
masses.}.
Within the other possible choice, that is, within the parametrization that uses the three neutrino masses
as free parameters, most of the parameter space at high neutrino
masses is useless for the data fit. Therefore, this second possibility
is penalized by the Occam's razor and we shall not explore it here.

The neutrino mixing parameters are constrained
using the same data we described in section~\ref{sec:osc-current}.
The complete oscillation data set is indicated with the label ``OSC''
in the following.

\begin{table}[t]
\centering
\begin{tabular}{c|c|c|c|c|c}
\multicolumn{2}{c|}{Neutrino mixing and masses} & \multicolumn{2}{c|}{Cosmological} & \multicolumn{2}{c}{\doublebeta{}} \\
\hline
Parameter & Prior & Parameter & Prior & Parameter & Prior \\
\hline
$\sin^2\theta_{12}$ & 0.1 -- 0.6   & $\Omega_bh^2$ & 0.019 -- 0.025 & $\alpha_2$             & 0 -- $2\pi$  \\
$\sin^2\theta_{13}$ & 0.00 -- 0.06 & $\Omega_ch^2$ & 0.095 -- 0.145 & $\alpha_3$             & 0 -- $2\pi$  \\
$\sin^2\theta_{23}$ & 0.25 -- 0.75 & $\Theta_s$    & 1.03 -- 1.05   & $\nme{^{76}{\rm Ge}}$  & 3.3 -- 5.7   \\
$\deltacp/\pi$      & 0 -- 2       & $\tau$        & 0.01 -- 0.4    & $\nme{^{136}{\rm Xe}}$ & 1.5 -- 3.7   \\
$\dmsq{21}/\eVq$    & $5\e{-5}$ -- $10^{-4}$     & $n_s$     & 0.885 -- 1.04 \\
$\dmsq{31}/\eVq$    & $1.5\e{-3}$ -- $3.5\e{-3}$ & $\logA$   & 2.5 -- 3.7    \\
$\log_{10}(\mlight/\rm{eV})$  & -5 -- 0\\
\end{tabular}
\caption{\label{tab:commonParams}
Neutrino, cosmological and \doublebeta\ parameters used in the analysis, with the adopted priors.
All the priors are linear in the corresponding quantity.}
\end{table}

For the cosmological part,
we will describe the universe using the $\Lambda$CDM model
and its six parameters:
the baryon and cold dark matter densities, $\Omega_bh^2$ and $\Omega_ch^2$;
the optical depth to reionization, $\tau$;
the angular scale of the acoustic peaks through $\Theta_s$ and
the amplitude $\logA$
and tilt $n_s$
of the power spectrum of initial curvature perturbations.
In addition, we add the effect of the three massive neutrinos
computing the evolution of the cosmological observables assuming three
independent mass eigenstates, which, in terms of the parameters involved in
our analyses, read as
$m_1=\mlight$ $\left(m_1=\sqrt{\mlight^2+ |\dmsq{31}|}\right)$,
$m_2=\sqrt{\mlight^2 + \dmsq{21}}$ $\left(m_2=\sqrt{\mlight^2+ |\dmsq{31}| + \dmsq{21}}\right)$
and
$m_3=\sqrt{\mlight^2 + \dmsq{31}}$ $\left(m_3=\mlight\right)$
for normal (inverted) neutrino mass orderings.

When considering cosmological data,
we will focus on the \pla\ measurements of the CMB spectrum
and on the most recent results from BAO observations.
For the former we consider the 2015 \pla\ release~\cite{Adam:2015rua,Ade:2015xua}
of the high-$\ell$ likelihood~\cite{Aghanim:2015xee},
together with a prior on $\tau$ as obtained
in the 2016 intermediate results~\cite{Aghanim:2016yuo}.
For the purposes of our analyses,
this will be sufficient to mimic the final \pla\ release
which is expected within the next few months.
Complementary to the CMB, we include in our calculations
the final constraints from the \texttt{SDSS BOSS} experiment,
the \texttt{DR12} release,
in the form denoted as ``final consensus''~\cite{Alam:2016hwk},
which provides constraints from observing
1.2 million massive galaxies in three separate bands
at effective redshifts 0.38, 0.51 and 0.61,
plus results from the \texttt{6DF} survey at $z=0.106$~\cite{Beutler:2011hx} and
from the \texttt{SDSS DR7 MGS} survey at $z=0.15$~\cite{Ross:2014qpa}.
The combined dataset including the mentioned CMB and BAO data
will be denoted as ``Cosmo''.

In addition, we shall impose a prior on the Hubble parameter
as obtained in the recent results from Ref.~\cite{Riess:2016jrr}:
$H_0=(73.24\pm1.74)\mbox{ km s}^{-1}\mbox{ Mpc}^{-1}$.
We will denote the data combinations including this prior with
the label ``$H_0$''.

Finally, concerning neutrinoless double beta decay,
we vary the two Majorana phases in the entire available range (0--$2\pi$)
and the NMEs according to the range allowed by recent theoretical calculations.
We revised the NME ranges adopted in \cite{Gariazzo:2018pei},
which were the ones suggested in \cite{Giuliani:2012zu}.
Here we use these new ranges:
[3.3 -- 5.7] for $^{76}$Ge
and
[1.5 -- 3.7] for $^{136}$Xe,
following the $1\sigma$ range proposed in \cite{Vergados:2016hso}.

We use \doublebeta\ data from the $^{136}$Xe experiments
\texttt{KamLAND-Zen} \cite{KamLAND-Zen:2016pfg} and
\texttt{EXO-200} \cite{Albert:2014awa}
and from the $^{76}$Ge experiment
\texttt{Gerda}, for which we use the results in Ref.~\cite{Agostini:2017iyd},
since the latest publication \cite{Agostini:2018tnm} does not contain enough
information that allows us to parameterize a likelihood function.
The most stringent bounds, anyways, still come from \texttt{KamLAND-Zen},
so that not including the new \texttt{Gerda} results does not affect significantly our results.
For the very same reason we do not include the results of \texttt{CUORE}~\cite{Alduino:2017ehq},
for which the uncertainty on the NME of $^{130}$Te is very large and
the constraints corresponding to most of the values
of $\nme{^{130}{\rm Te}}$ are much looser than the ones from \texttt{KamLAND-Zen},
and of \texttt{CUPID-0} \cite{Azzolini:2018dyb}, which establishes a much less stringent limit on the $^{82}$Se half-life.
The complete neutrinoless double beta set of data will be denoted as ``\doublebeta''.

All the previously listed data are coded as likelihood terms
in a full Bayesian analysis.
We compute the cosmological quantities using the
Boltzmann solver \texttt{CAMB} \cite{Lewis:1999bs},
the likelihoods using the interface provided by
\texttt{CosmoMC} \cite{Lewis:2002ah},
modified in order to take into account the oscillation and
neutrinoless double beta decay data,
while the calculation of the Bayesian evidence is committed
to \texttt{PolyChord} \cite{Handley:2015fda,Handley:2015aa}.

\subsection{Constraints on the mass orderings}

\begin{figure}[tp]
  \centering
  \includegraphics[width=0.5\textwidth]{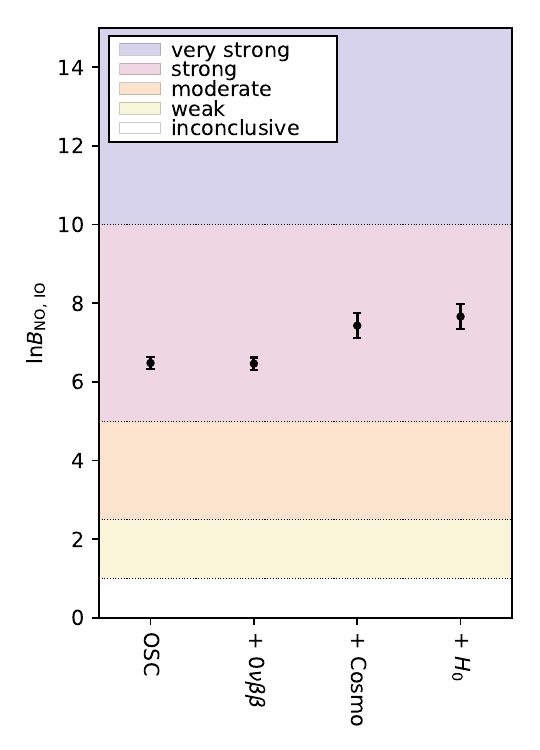}
  \caption{\label{fig:bayesfactors}
  Graphical visualization of the Bayesian factors comparing normal and inverted ordering.
  }
\end{figure}

The main results are depicted in Figure~\ref{fig:bayesfactors}.
The first data point corresponds to the Bayesian evidence from oscillation data only.
Notice that the Bayes factor
[$\ln(\Bnoio)=\input{results/B_osc_novsio.tex}$ for concreteness]
indicates \emph{strong} evidence for the normal
mass ordering \emph{from oscillation data only}.
This Bayes factor is translated into a $\sim \input{results/s_osc_novsio.tex}\sigma$ evidence favoring normal mass ordering.
This result was expected in light of the results presented
in section~\ref{sec:osc-current}, arising from the frequentist joint analysis.
There it was reported a $\Delta\chi^2=11.7$ in favor of the normal mass ordering from the
combination of all long baseline, reactor and atmospheric data, which
corresponds, roughly, to $\sim 3.4\sigma$.
Adding information from neutrinoless double beta decay searches does not affect the Bayesian analysis,
as shown by the second data point in Figure~\ref{fig:bayesfactors},
and as expected from previous work~\cite{Gariazzo:2018pei}.

Once CMB and BAO measurements are also added in the Bayesian analysis,
$\ln(\Bnoio)=\input{results/B_osc_0n2b_cmb_novsio.tex}$ is obtained
(see the third point in Figure~\ref{fig:bayesfactors}),
improving the significance of the preference for normal ordering
from $\sim \input{results/s_osc_novsio.tex}\sigma$ to $\sim \input{results/s_osc_0n2b_cmb_novsio.tex}\sigma$.
Notice that, even if the preference for the normal
neutrino mass ordering is mostly driven by oscillation data,
the information provided by cosmological observations is more powerful than that in the analysis carried out in Ref.~\cite{Gariazzo:2018pei},
as the Bayesian analyses here also include BAO measurements, together with CMB data.
Indeed,
from the two Bayes factors obtained
considering oscillation data only
[$\ln(\Bnoio)=\input{results/B_osc_novsio.tex}$]
and oscillation plus cosmological measurements
[$\ln(\Bnoio)=\input{results/B_osc_0n2b_cmb_novsio.tex}$],
it is straightforward to infer the probability odds for normal ordering arising exclusively from cosmology.
By doing so,
one obtains odds of $2.7:1$ for the normal ordering against the inverted one,
in perfect agreement with the analyses of Ref.~\cite{Vagnozzi:2017ovm},
where odds of $2.4:1$ with cosmological data only were reported
when considering the very same data sets adopted here (albeit the odds
were derived with an alternative method).

Finally, the addition of the prior on the Hubble constant raises the
evidence for the normal ordering to $\ln(\Bnoio)=\input{results/B_osc_0n2b_cmb_H0_novsio.tex}$
(i.e.\ $\sim \input{results/s_osc_0n2b_cmb_H0_novsio.tex}\sigma$).
This improvement is expected,
as previously explained in section~\ref{sec:cosmo},
since a prior on the Hubble constant breaks the degeneracy
between $\mnu$ and $H_0$ and,
therefore, sharpens the neutrino mass bounds from cosmology.
By performing a similar exercise to the one previously quoted,
one finds that the odds for normal versus inverted ordering
from cosmology data only are $3.3:1$ for the combination
of CMB, BAO plus the $H_0$ prior,
again in excellent agreement with the results obtained in Ref.~\cite{Vagnozzi:2017ovm}.

%% file: results/B_osc_novsio.tex
6.5 \pm 0.2

%% file: results/s_osc_novsio.tex
3.2

%% file: results/B_osc_0n2b_cmb_novsio.tex
7.4 \pm 0.3

%% file: results/s_osc_0n2b_cmb_novsio.tex
3.4

%% file: results/B_osc_0n2b_cmb_H0_novsio.tex
7.7 \pm 0.3

%% file: texs/fut_osc.tex
As we have seen in section~\ref{sec:osc-current},
the combination of all current neutrino experiments leads
to a preference for normal ordering of $3.4\sigma$,
within the context of the latest frequentists global data analyses.
The Bayesian analysis described in the previous section confirms these results,
as we have reported a $\input{results/s_osc_novsio.tex}\sigma$ evidence for normal mass ordering.
In principle, one may expect to achieve further sensitivity
on the neutrino mass ordering from more precise data
by the current long-baseline and atmospheric neutrino experiments,
since these experiments will still run for some time before the new experiments will take over.
However, it is not easy to predict the final results of current experiments,
since the sensitivity to the mass ordering is highly correlated to the true value of the CP phase $\deltacp$.
The NO$\nu$A experiment alone expects a 3$\sigma$ sensitivity for 30-50\% of the values of $\deltacp$ by 2024~\cite{nova-himmel}.
If $\deltacp=3\pi/2$, the expected sensitivity would be higher than that and, then,
a very strong result could be obtained by 2024~\cite{nova-himmel}.
Note, however, that the NO$\nu$A sensitivity analysis considers a fixed value of $\theta_{13}$
and does not marginalize over $\Delta m_{31}^2$.
Upgrading T2K to T2K-II will improve the sensitivity substantially,
since the experiment should gather around $20\times10^{21}$~POT by 2026,
which would be roughly 6 times the current amount of data\footnote{We are not aware of any study showing the T2K or SK expectations to the mass ordering in the next few years.}.
Combining beam data from T2K with atmospheric data from SK can improve
the sensitivity even further, as shown in Ref.~\cite{Abe:2017aap}.
Performing a combined fit of T2K, NO$\nu$A and eventually SK
could bring the sensitivity to the $5\sigma$ level within a few years.
In any case, apart from the combinations of different experiments, a very robust determination of the neutrino mass ordering from a single current experiment seems rather unlikely.
Indeed, one of the main goals of the next-generation neutrino oscillation experiments,
including new long-baseline, reactor, and atmospheric neutrino detectors,
will be to perform the determination of the mass ordering by a single experiment.
The upcoming facilities will be able to measure the neutrino mass ordering with astonishing precision.
In this section we briefly discuss some of the proposed projects and their physics potential.

\subsubsection*{Long-baseline experiments}

The Deep Underground Neutrino Experiment
(\texttt{DUNE})~\cite{Acciarri:2016crz,Acciarri:2015uup,Strait:2016mof,Acciarri:2016ooe}
will be a new long-baseline accelerator experiment,
with a small near detector and a huge far detector
with a fiducial mass of 40~kton located 1300~km
away from the neutrino source at Fermilab.
With its powerful 1.1~MW beam,
it will be exposed to around $15\e{20}$~POTs (protons on target) per year,
which will lead to a huge number of events and therefore
to high precision measurements of the neutrino oscillation parameters.
As explained in section~\ref{sec:osc-current},
the presence of matter affects differently
the neutrino appearance probabilities for normal and inverted mass orderings.
\texttt{DUNE}, with the longest baseline ever for an accelerator neutrino experiment,
will be able to measure the neutrino mass ordering
with a significance above $5\sigma$ for any set
of the oscillation parameters $(\theta_{23},\deltacp)$
after 7 years of data taking.
Note that this sensitivity could be further increased
by using an improved energy reconstruction method,
as shown in Ref.~\cite{DeRomeri:2016qwo}.
On the other hand, the sensitivities could also be biased
by the potential presence of new physics beyond the SM,
such as non-standard neutrino interactions~\cite{Farzan:2017xzy,Masud:2016gcl,Deepthi:2016erc,Bakhti:2016gic,Coloma:2015kiu,deGouvea:2015ndi,Forero:2016ghr,Forero:2016cmb,Coloma:2017egw,Miranda:2004nb,Coloma:2016gei},
deviations from unitarity~\cite{Escrihuela:2016ube,Blennow:2016jkn,Dutta:2016czj}
or the presence of light-sterile neutrinos~\cite{Coloma:2017ptb,Agarwalla:2016xxa,Berryman:2016szd,Berryman:2015nua}.
Indeed, besides providing very precise information
about the neutrino oscillation mechanism,
the \texttt{DUNE} experiment will also be very useful
to test different models for neutrino masses and
mixings~\cite{Srivastava:2017sno,Srivastava:2018ser,Chatterjee:2017ilf,Chakraborty:2018dew,Agarwalla:2017wct,Chatterjee:2017xkb,Pasquini:2016kwk}
as well as to check for various effects of new physics such as the ones mentioned above,
neutrino decay scenarios~\cite{Coloma:2017zpg,Choubey:2017dyu,Ascencio-Sosa:2018lbk},
quantum decoherence~\cite{Gomes:2018inp}
or even CPT invariance~\cite{Barenboim:2017ewj,Barenboim:2018lpo,deGouvea:2017yvn}
and Lorentz invariance~\cite{Barenboim:2018ctx,Jurkovich:2018rif}.

There are also plans to build a larger version of the \texttt{Super-Kamiokande} detector,
\texttt{Hyper-Kamiokande}~\cite{Abe:2018uyc},
that will be very similar to its predecessor
but with a fiducial mass of 560~kton,
25 times larger than \texttt{Super-Kamiokande}.
The \texttt{Hyper-Kamiokande} detector will be a requirement for
the upgrade of \texttt{T2K},
the \texttt{T2HK} (Tokai-to-Hyper-Kamiokande) experiment~\cite{Abe:2015zbg}.
The very massive detector together with the upgraded neutrino beam
from J-PARC will guarantee a huge number of neutrino events
and therefore larger statistics.
As a consequence, \texttt{T2HK} will be able to determine
the neutrino mass ordering after few years of running time
with very high significance,
as well as to explore new physics scenarios,
see for instance Refs.~\cite{Abe:2018uyc, Abe:2017jit,Agarwalla:2018nlx}.
In combination with atmospheric data from \texttt{Hyper-Kamiokande},
a $3\sigma$ rejection of the wrong mass ordering
would be expected after five years of data taking.
A third project has been proposed as an extension
of \texttt{T2HK} to Korea,
the \texttt{T2HKK} (Tokai-to-Hyper-Kamiokande-and-Korea) experiment~\cite{Abe:2016ero}.
This proposal includes a second far detector facility for the
J-PARC neutrino beam, located at 1000-1300~km from the source.
The longer path traveled within the Earth by the neutrinos
detected in \texttt{T2HKK} will result
in an enhanced sensitivity to the neutrino mass ordering
if compared to \texttt{T2HK} alone.

The synergies and complementarities among
the three long-baseline proposals above,
\texttt{DUNE}, \texttt{T2HK} and \texttt{T2HKK},
have been discussed in Ref.~\cite{Ballett:2016daj}.
It is found that the combination of their experimental results
may significantly mitigate the limitations of a given
experiment, improving the precision in both the determination
of the mass ordering and the measurement of CP violation.

Note that, although here we have focused
on the long-baseline side of \texttt{DUNE} and \texttt{Hyper-Kamiokande},
they are actually designed as multi-purpose experiments,
with a rich physics program aiming to study
the neutrino oscillations with accelerator,
atmospheric and solar neutrinos as well as to detect neutrinos
from astrophysical sources and proton decay.

\subsubsection*{Atmospheric experiments}
\begin{figure}
\centering
\includegraphics[width=0.49\textwidth]{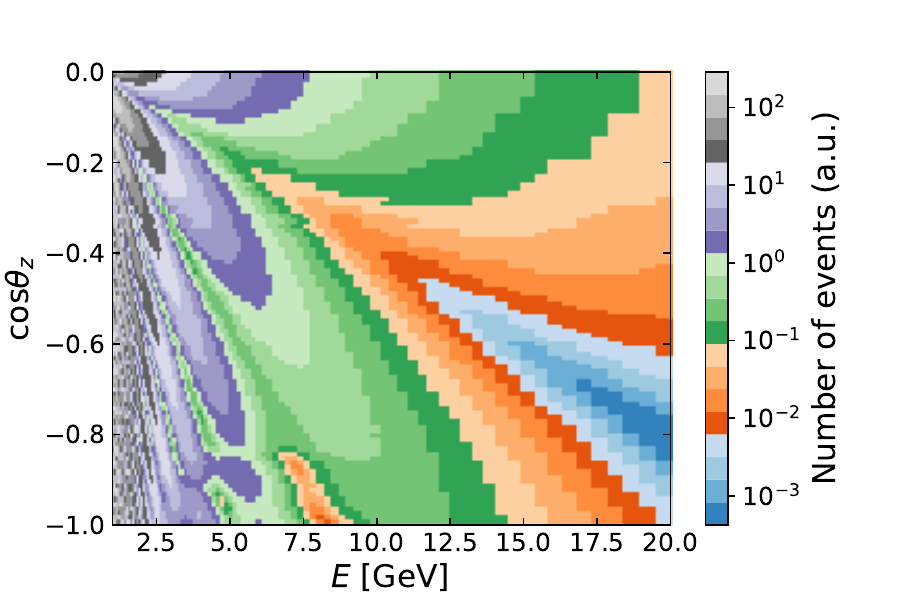}
\includegraphics[width=0.49\textwidth]{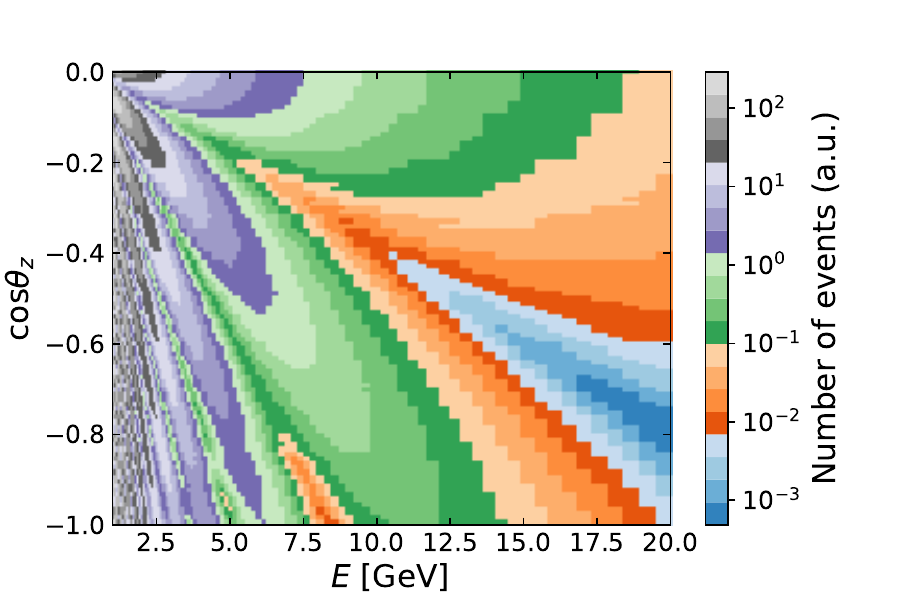}
\caption{\label{fig:Nev_mu_mubar}
Expected number of events (arbitrary units) for an hypothetical atmospheric neutrino
detector with perfect energy resolution 
as a function of the reconstructed neutrino energy $E$
and the cosine of the zenith angle $\cos\theta_z$,
for normal (left) and inverted (right) ordering.
}
\end{figure}

In atmospheric neutrino experiments,
the sensitivity to the mass ordering comes from the matter effects
that distort the pattern of neutrino oscillations inside Earth,
see Equation~\eqref{eq:mixmatter2}.
Based on the oscillatory pattern that depends on the reconstructed
neutrino energy and zenith angle,
an ideal experiment would observe a given number of events
in each energy and zenith angle bin as shown in Figure~\ref{fig:Nev_mu_mubar}.
Comparing the observed two-dimensional histograms
with the theoretical ones for normal (left panel)
or inverted ordering (right panel)
allows to determine the true mass ordering that is realized in nature.
In the following we list some of the future projects with this aim.

% ORCA
The Oscillation Research with Cosmics in the Abyss (\texttt{ORCA}) experiment~\cite{Adrian-Martinez:2016fdl}
will be a large neutrino telescope placed deep inside the Mediterranean sea.
It will detect the Cherenkov light emitted
by the muons and electrons created by the interactions
of atmospheric neutrinos in the sea
and that propagate into water.
Unlike its precursor, \texttt{ANTARES},
with 12 lines and a separation of 70 meters between neighbouring optical modules,
\texttt{ORCA} will have 60 lines with modules separated by 9 meters.
Due to the matter effects on the propagation of atmospheric neutrinos,
the \texttt{ORCA} experiment will be able to measure
the neutrino mass ordering with very good precision.
In particular, a 3$\sigma$ determination of the mass ordering
can be expected after only three years of data taking,
with even higher significance for the case in which nature has chosen 
normal ordering and the upper octant for the atmospheric mixing angle.
Several studies have been performed in order to analyze
the sensitivity of \texttt{ORCA} to the standard oscillation parameters~\cite{Yanez:2015uta,Ribordy:2013set}.
Its potential to determine the Earth matter density
through neutrino oscillation tomography~\cite{Winter:2015zwx}
or to test new physics scenarios~\cite{Ge:2017poy,Capozzi:2018bps}
have also been extensively discussed.

%PINGU
\texttt{PINGU} (Precision IceCube Next Generation Upgrade) \cite{Aartsen:2014oha}
is a planned upgrade of the \texttt{IceCube DeepCore} detector,
an ice-Cherenkov neutrino telescope which uses
the antarctic ice as a detection medium.
The \texttt{IceCube} design aims at the detection
of very high energy neutrinos, with an energy threshold
above the relevant energy range for neutrino oscillations.
However, the denser instrumented region \texttt{DeepCore}
allows \texttt{IceCube} to decrease its energy threshold
down to $E_{\rm th} = 6.3\,\mathrm{GeV}$.
A further improvement with an even denser zone,
\texttt{PINGU}, could lower $E_{\rm th}$ to only a few GeV.
With this very low-energy threshold, one of the main purposes
of \texttt{PINGU} is the determination
of the neutrino mass ordering, with expected sensitivities
similar to the \texttt{ORCA} experiment~%
\footnote{The effect of statistic and systematic uncertainties
on the \texttt{PINGU} sensitivity to the mass ordering
has been presented in Ref.~\cite{Capozzi:2015bxa}.}.
Besides that, \texttt{PINGU} is expected to have
the best sensitivity to $\nu_\tau$ appearance and
to determine accurately the octant of the atmospheric mixing angle.
The \texttt{PINGU} capabilities to detect high-energy supernova neutrinos~\cite{Murase:2017pfe},
and to investigate scenarios beyond the Standard Model,
such as non-standard interactions~\cite{Choubey:2014iia}
or dark matter self-interactions~\cite{Chen:2014oaa,Robertson:2017hdw}
have been also analyzed in the literature.

%INO-ICAL
The India-based Neutrino Observatory (\texttt{INO})
is a very ambitious project,
aiming to detect atmospheric neutrinos
with a 50~kton magnetized iron
calorimeter (\texttt{ICAL})~\cite{Kumar:2017sdq}.
The most outstanding feature of the \texttt{INO} experiment
will be its capability to distinguish neutrinos
from antineutrinos in an event by event basis.
As a result, the identification of the matter effects
in the neutrino and antineutrino propagation will be much cleaner
in comparison with the sea water/ice Cherenkov detectors.
Indeed, one of the main scientific goals of \texttt{INO}
will be the determination of the neutrino mass ordering~\cite{Ghosh:2012px}.
According to the Physics White Paper of the ICAL (INO) Collaboration~\cite{Kumar:2017sdq},
after 10 years run, \texttt{INO} will be able to identify the correct neutrino mass ordering
with a significance larger than 3$\sigma$.
As the experiments discussed above,
the atmospheric neutrino results from \texttt{INO}
can also be used to test the presence of new physics
beyond the SM, such as
CPT- or Lorentz violation~\cite{Chatterjee:2014oda},
sterile neutrinos~\cite{Behera:2016kwr,Thakore:2018lgn},
dark matter related studies~\cite{Dash:2014sza,Choubey:2017vpr},
non-standard neutrino interactions~\cite{Choubey:2015xha}
or decaying neutrinos~\cite{Choubey:2017eyg}.

\subsubsection*{Medium-baseline reactor experiments}
\begin{figure}[t]
\centering
\includegraphics[width=0.5\textwidth]{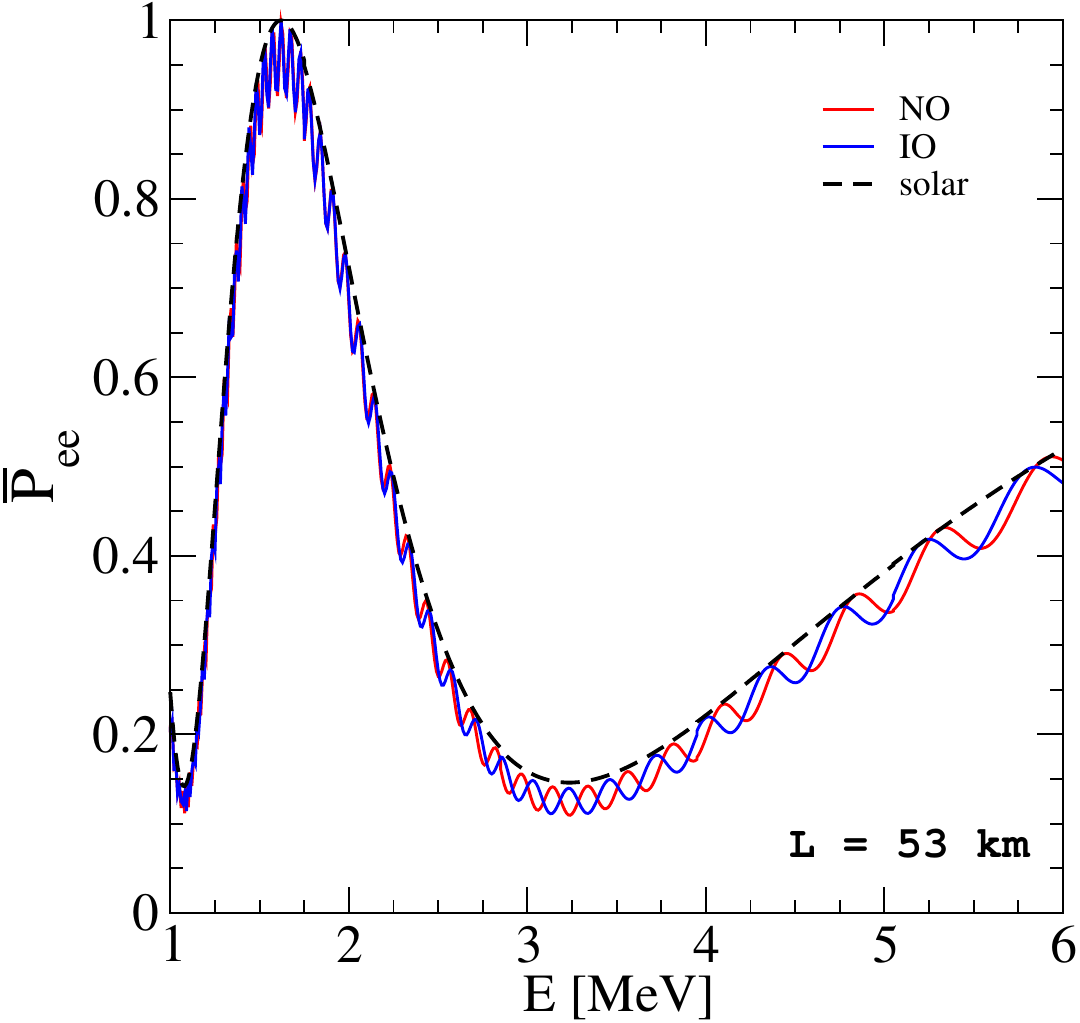}
\caption{
\label{fig:juno}
Electron antineutrino survival probabilities in a medium-baseline reactor experiment with $L=53$~km.
The red (blue) line corresponds to normal (inverted) mass ordering
using the best-fit values from Table~\ref{tab:oscillation_summary},
while the black line contains the main term in the survival probability,
given by the solar mass splitting frequency by setting $\Delta m_{31}^2 = 0$.}
\end{figure}

We have focused so far on extracting the neutrino mass ordering from matter effects
in the neutrino propagation through the Earth interior.
An alternative technique is that provided by medium-baseline reactor neutrino experiments~\cite{Petcov:2001sy}.
For baselines of the order of 50~km,
the survival probability
for reactor antineutrinos exhibits a pattern
that may allow the discrimination
between normal and inverted mass orderings.
Indeed, for such distances, the electron antineutrino
survival probability is given by the following expression:
\begin{eqnarray}
\label{eq:pee-juno}
P_{\overline{\nu}_e\rightarrow\overline{\nu}_e}
&=&
1 -\cos^4\theta_{13}\sin^22\theta_{12}\sin^2\Delta_{21}
\nonumber \\
&&-\sin^2 2\theta_{13}
\left[ \sin^2 \Delta_{31}
+ \sin^2\theta_{12}\sin^2\Delta_{21}\cos 2\Delta_{31}
\mp \frac{\sin^2\theta_{12}}{2}
\sin 2\Delta_{21}\sin 2|\Delta_{31}|\right],
\end{eqnarray}
where $\Delta_{ij}=\frac{\Delta m_{ij}^2 L}{4E}$
and the minus (plus) sign in the last term corresponds
to normal (inverted) mass ordering.
This probability contains a main oscillatory term
with a frequency given by the solar neutrino mass splitting $\Delta m^2_{21}$,
plus an additional term whose frequency
depends on the sign of the atmospheric splitting $\Delta m^2_{31}$,
i.e.\ on the neutrino mass ordering.
The effect of the ordering over the neutrino survival probability
in a medium-baseline reactor experiment is illustrated in Figure~\ref{fig:juno}.
There, we depict in black the oscillatory term corresponding to the solar splitting frequency.
The red (blue) line corresponds to the full neutrino survival probability for normal (inverted) mass ordering.
Note that this plot was obtained using the best-fit values from Table~\ref{tab:oscillation_summary} for each ordering.

% JUNO
The Jiangmen Underground Neutrino Observatory (\texttt{JUNO})~\cite{An:2015jdp}
is a 20~kton multi-purpose underground liquid scintillator detector.
The site of the experiment, located 53~km away
from the Yangjiang and Taishan nuclear power plants in China,
was chosen to optimize its sensitivity to the neutrino mass ordering,
one of its main physics goals.
Like any other reactor neutrino experiment,
\texttt{JUNO} will be sensitive to the disappearance of electron antineutrinos,
with about $10^5$ events expected after six years of run time.
From this high-statistics data sample,
\texttt{JUNO} will try to reconstruct
with extremely good precision the neutrino oscillation spectrum
and to discriminate the different high-frequency behavior
for normal and inverted mass ordering,
as illustrated in Equation~\eqref{eq:pee-juno} and Figure~\ref{fig:juno}.
For a projected energy resolution of 3\% at 1 MeV,
\texttt{JUNO} will be able to establish the neutrino mass ordering
at the level of 3-4$\sigma$ in 6 years.
Its combination with the
\texttt{PINGU} facility could lead to a high significance improvement
of the individual capabilities of these two experiments, see \cite{Blennow:2013vta}.

Apart from the mass ordering,
\texttt{JUNO} will also provide precision measurements
of the solar oscillation parameters,
$\theta_{12}$ and $\Delta m^2_{21}$,
with an accuracy of around 1\%.
In this sense, \texttt{JUNO} might help to solve
the observed disagreement between the mass splitting measured
at solar experiments and
at the reactor experiment \texttt{KamLAND}.
If the discrepancy persists after new measurements
by \texttt{JUNO} and future solar results
by \texttt{Super-Kamiokande},
it could be considered as an indication of new physics~\cite{Farzan:2017xzy}.
Moreover, \texttt{JUNO} will be sensitive to different types
of new physics scenarios beyond the SM,
as studied in
Refs.~\cite{Liao:2017awz,Bakhti:2014pva,Ohlsson:2013nna,Zhao:2016brs,Abrahao:2015rba,Li:2014rya,Girardi:2014wea,Krnjaic:2017zlz,Khan:2013hva}.

% RENO-50
In parallel to \texttt{JUNO}, there is a proposal to extend
the already existing experiment \texttt{RENO}
with a third medium-baseline detector located at a distance of 47 km.
This new project is known as \texttt{RENO-50}~\cite{Kim:2014rfa},
given its location, at approximately 50~km
from the Hanbit power plant, in South Korea.
The detector would consist of
a 18~kton ultra-low-radioactive liquid scintillator instrumented
with 15000 high quantum efficiency photomultiplier tubes.
Using the same technique described above,
\texttt{RENO-50} will be able to determine the neutrino mass ordering
as well as the solar oscillation parameters
with extremely good precision.
Conceived as multi-purpose detectors,
\texttt{JUNO} and \texttt{RENO-50} will have a wide physics program,
including not only the observation of reactor and solar neutrinos,
but also neutrinos from supernova bursts,
the diffuse supernova neutrino background,
atmospheric neutrinos and geoneutrinos.

%% file: texs/fut_dec.tex
\subsection{Prospects from beta-decay experiments}
\label{sec:futbeta}

As already mentioned in section~\ref{sec:beta}, the determination of the mass ordering through the observation
of the energy spectrum near the endpoint of $\beta$-decay or similar will be extremely challenging,
because an impressive energy resolution is required to distinguish the kink due to the second and third
mass eigenstates in the spectrum.
We list here the main projects that aim at detecting the neutrino mass in the future and comment their perspectives
for the mass ordering determination.

The first experiment we will comment on is \texttt{KATRIN},
which has recently started operations and aims at a detection of the effective electron antineutrino mass
with a sensitivity of 0.2~eV \cite{Angrik:2005ep,SejersenRiis:2011sj}.
The first results from \texttt{KATRIN} are expected in early 2019,
but the final target statistics will be reached after 3~yr of data taking.
Thanks to the detailed study of the detector systematics which has been carried out,
it is possible that the final mass determination
will reach a better sensitivity than the nominal one of 0.2~eV,
eventually reaching something closer to 0.1~eV \cite{Parno_neutrino18}.
Even with the more optimistic sensitivity, however,
it will be impossible for \texttt{KATRIN} to determine the mass ordering.

Other tritium experiments exploiting different technologies include
the \texttt{Project-8} \cite{Doe:2013jfe,Asner:2014cwa,Esfahani:2017dmu} experiment, which
will use the Cyclotron Radiation Emission Spectroscopy (CRES)
in order to determine the mass of the electron antineutrino.
The technique consists in measuring the frequency of cyclotron radiation emitted by
the electrons released during tritium decay and spiralling into a magnetic field.
The frequency can then be related with the electron energy and consequently the energy spectrum can be determined.
At the moment, \texttt{Project-8} is in the calibration phase (phase-II) \cite{Rybka_neutrino18}
for a small prototype which will not have enough sensitivity to be competitive in the determination of
the neutrino mass.
Next phases include a large volume system using molecular tritium (phase-III),
starting in 2020,
which will be competitive in determining the neutrino mass
and will serve as an intermediate step before moving to phase-IV,
which will use atomic tritium, required in order to avoid
uncertainties related to the existence
of excited molecular tritium states.
\texttt{Project-8} in its atomic tritium phase is expected to reach the sensitivity
$m_{\bar\nu_e}\lesssim40$~meV with an exposure of $10-100$~m$^3$~yr,
sufficient to probe the values of $m_{\bar\nu_e}$ allowed
in the context of inverted ordering \cite{Esfahani:2017dmu},
so that in case of no observation we will know that the ordering of neutrino masses must be normal.

Another interesting class of the experiments includes the \texttt{HOLMES} \cite{Alpert:2014lfa,Giachero:2016xnn}
and \texttt{ECHo} \cite{Eliseev:2015pda} experiments,
which both aim at the determination of the electron neutrino mass through observations
of the endpoint of the electron capture decay of $^{163}$Ho,
which practically proceeds through the measurement of de-excitation transitions of the Dy atoms,
which are produced in the process $^{163}\mbox{Ho}+e^-\rightarrow^{163}\mbox{Dy}^*+\nu_e$ \cite{DeRujula:1982qt}.
As for the tritium $\beta$-decay, also the endpoint of the $^{163}$Ho electron capture spectrum
depends on the value of the neutrino masses and, in principle, it would be possible
to determine the mass ordering in this way.
Besides the experimental and theoretical problems that the \texttt{HOLMES} and \texttt{ECHo} collaborations must face, however,
it seems that the current technology is not yet at the level of precision required for
the mass ordering determination.
The \texttt{HOLMES} demonstrator, currently running,
should reach a sensitivity of $m_{\nu_e}\lesssim10$~eV by the end of 2018,
while the full-scale experiment, possibly starting in 2019, has a target sensitivity of $m_{\nu_e}\lesssim1$~eV
\cite{Gastaldo_neutrino18}.
\texttt{ECHo}, on the other hand, is running a first phase (\texttt{ECHo-1k}) which has also a target of $m_{\nu_e}\lesssim10$~eV in 1~yr,
while the full scale \texttt{ECHo-100k} will reach $m_{\nu_e}\lesssim1.5$~eV in 3~yr of data taking,
expected to start in 2019 \cite{Gastaldo_neutrino18}.
Both results are impressive when compared with the current upper limit on the electron neutrino mass using the same isotope,
which is 225~eV \cite{Springer:1987zz}, more than two orders of magnitude larger.

Finally, to conclude this subsection we want to mention that the \texttt{PTOLEMY} proposal \cite{Betts:2013uya,Baracchini:2018wwj},
aiming at the detection of the relic neutrino background and recently approved by the Scientific Committee of the Laboratori Nazionali del Gran Sasso (LNGS),
will be able to study and possibly determine the mass ordering through the observation of the $\beta$-spectrum of tritium decay.
\texttt{PTOLEMY} will be discussed later in section~\ref{sec:relic}.

\subsection{Prospects from neutrinoless double beta decay}
\label{sec:futdoublebeta}

We list here the future perspectives for neutrinoless double beta decay experiments
in terms of sensitivity to the half-life for the processes of interest (where possible).
As we already commented in section~\ref{ssec:doublebeta_pres},
the conversion between the half-life \Tbb\ and the effective Majorana mass \mbb\
depends on the NME and the phase space factor of the process of interest, see Equation~\eqref{eq:0n2b_lifetime}.
In order to exclude the inverted ordering allowed range for \mbb\ (in case there is no sterile neutrino),
one would need to constrain $\mbb\lesssim10$~meV,
which corresponds to $\Tbb\simeq1\e{28}$~yr,
with some dependence on the material (phase space and NME).
This means that none of the current generation experiments will be able to reach the required sensitivity,
and we will have to wait for next-generation upgrades and new projects.
Many of the information listed in the following has been taken from
Refs.~\cite{Agostini:2017jim,Giuliani_neutrino18}.

\subsubsection*{Current generation experiments}
Let us firstly address the current generation of experiments, which at most will be able to start exploring
the three-neutrino inverted mass ordering regime or to probe the upper range for \mbb\ allowed within the 3+1 neutrino scenario.
The experiments will be listed in alphabetical order.

\texttt{AMoRE}~\cite{Alenkov:2015dic} is an experiment devoted to determine the life-time of $^{100}$Mo.
After a first pilot run,
the current status (\texttt{AMoRE}-I) is to test the technology with a $^{100}$Mo mass of 5-6~kg,
in order to demonstrate the scalability before moving to the full scale (\texttt{AMoRE}-II) detector,
which will use 200~kg of material and is expected to start around 2020,
with a final target sensitivity of $\Tbb\simeq5\e{26}$~yr.

\texttt{CUORE}~\cite{Artusa:2014lgv,Alduino:2016vjd,Alduino:2018skz}, already mentioned in section~\ref{ssec:doublebeta_pres},
works with $^{130}$Te and is already taking data with the full scale detector,
which will have as ultimate sensitivity $\Tbb\simeq9\e{25}$~yr after 5~yr of data taking~\cite{Ouellet_neutrino18,Adams:2018nek}.

The \texttt{KamLAND-Zen} experiment~\cite{KamLAND-Zen:2016pfg,Gando_neutrino18}, after the previous successful data taking period,
is now upgrading the detector for a new observation run with approximately 750~kg of $^{136}$Xe
and a new balloon inside the \texttt{KamLAND} detector.
The target sensitivity for the upcoming phase is around $\Tbb\simeq5\e{26}$~yr,
a factor of five larger than the current limit \cite{KamLAND-Zen:2016pfg}.

A smaller experiment is \texttt{NEXT}~\cite{Martin-Albo:2015rhw},
which is running background studies in the Canfranc laboratories in Spain.
\texttt{NEXT} will use high pressure $^{136}$Xe TPCs, which will allow
an impressive tracking of the emitted particles through scintillation and electroluminescence.
A prototype with 10~kg of natural Xenon will start data taking this year to demonstrate
that the expected background control and particle tracking have been achieved.
In 2019 \texttt{NEXT} is expected to start a new phase with 100~kg of $^{136}$Xe,
which will reach $\Tbb\simeq1\e{26}$~yr with 5~yr of data.

A similar project is called \texttt{Panda-X-III}~\cite{Chen:2016qcd},
which is based in the Jinping underground laboratories in China.
\texttt{Panda-X-III}  will run the first phase using 200kg of $^{136}$Xe to reach
$\Tbb\simeq1\e{26}$~yr in 3~yr.

Going to a different concept, \texttt{SNO+} \cite{Andringa:2015tza}
will feature a detector of 760~ton of ultra-pure liquid scintillator.
\texttt{SNO+} will be a multipurpose detector, as it will be capable of studying
reactor, solar, supernova and geoneutrinos, and also to probe proton decay \cite{Gann_neutrino18}.
After the background studies will be completed,
a 0.5\% loading will be performed, inserting $^{130}$Te in the
detector to measure double beta decay processes.
The target sensitivity after 5~yr is $\Tbb\simeq2\e{26}$~yr.
Future plans for the \texttt{SNO+} experiment
include the further $^{130}$Te loading to 1\%, or even more, of the detector mass,
with the advantage that increasing the $^{130}$Te amount will not influence the backgrounds but only the signal.
The final target for this second phase is to reach $\Tbb\simeq1\e{27}$~yr,
thus starting to cover the inverted ordering allowed range.

Let us finally comment the
\texttt{SuperNEMO} experiment \cite{Patrick:2017eso,Arnold:2010tu}, which uses $^{82}$Se for its study.
\texttt{SuperNEMO} is particularly interesting because it will be able to perform a full topological reconstruction
of the events, which is extremely important in case of detection
because it opens the possibility to directly test the mechanism that underlies neutrinoless double beta decay
and, in principle, to determine the lepton-number violating process.
A first demonstrator of about 7~kg is expected to start in 2018 and to reach
$\Tbb\simeq6\e{24}$~yr with 2.5~yr of data.
The subsequent plans include an extension with a $\sim100$~kg scale detector with 20 modules,
which will be able to probe \Tbb\ up to $1\e{26}$~yr,
and the possibility to use the $^{150}$Nd isotope, for two reasons:
to have a more favorable phase space when converting \Tbb\ to \mbb\
and to get rid of the Rn background which affects the $^{82}$Se measurements~\cite{Giuliani_neutrino18}.

As a summary, some of the current generation experiments will be able to probe the inverted ordering range
of \mbb\ within the standard three neutrino framework and assuming an exchange of light Majorana
neutrinos.
However, none of them will be able to rule out completely the inverted mass ordering,
because of the uncertainty related to the NMEs.

\subsubsection*{Next generation experiments}
The situation will be different for the following generation of experiments,
which are mostly the natural evolution of current experiments to the ton-scale of decaying material.
With the increased amount of material, a larger statistics will be achieved and stronger bounds, of the order
of $\Tbb\simeq1\e{28}$~yr, will be feasible.
We briefly discuss here the main current proposals for the next 10-20 years.
The time schedules for these projects will be necessarily vague,
as they will depend on the results of the present ones.

Let us start with \texttt{CUPID} (\texttt{CUORE} Upgrade with Particle ID)~\cite{Wang:2015raa,Azzolini:2018dyb},
which will be the evolution of the previously discussed \texttt{CUORE} experiment.
The goal of \texttt{CUPID} is to use particle tracking in order to have a better discrimination of background
and ultimately allow a background-free experiment: the target is $<0.1$ counts/(ton yr)~\cite{Ouellet_neutrino18}.
A first demonstrator, named \texttt{CUPID-0}~\cite{Azzolini:2018dyb},
is already running with about 5~kg of $^{82}$Se,
and already obtained the strongest-to-date constraint on the life-time on this isotope.
In order to reach the target sensitivity $\Tbb\gtrsim1\e{27}$~yr,
however, further improvement in the crystals quality and radio-purity is required.
A full development plan for \texttt{CUPID} is currently under discussion.

Although not specifically designed for neutrinoless double beta decay searches,
the \texttt{DARWIN} (DARk matter WImp search with liquid xenoN) experiment \cite{Aalbers:2016jon}
will have sensitivity to a number of rare decay phenomena.
The primary target of \texttt{DARWIN} is to perform direct detection of dark matter
in a wide mass-range of the experimentally accessible parameter space
for Weakly Interacting Massive Particles (WIMPs),
to the level at which neutrino interactions with the target
become an irreducible background (the so-called neutrino floor).
The core of the detector will be
a multi-ton liquid xenon time projection chamber.
Having a large mass,
low-energy threshold and ultra-low background level,
\texttt{DARWIN} will also search
for solar axions or galactic axion-like particles,
measure the low-energy solar neutrino flux with $<1\%$ precision,
observe coherent neutrino-nucleus interactions,
detect galactic supernovae neutrinos
and study the double beta decay of $^{136}$Xe \cite{Aalbers:2016jon}.
Even if it will be build using natural Xenon without isotope enrichment,
\texttt{DARWIN} will contain 3.5~t of $^{136}$Xe.
If the target energy resolution of $1-2\%$ at 2.5~MeV will be achieved,
the sensitivity of \texttt{DARWIN} will be
$\Tbb\simeq5.6\e{26}$~yr with an exposure of 30~t~yr \cite{Aalbers:2016jon}.
The estimated ultimate sensitivity,
which will be achieved only with a complete mitigation
of the material background and 140~t~yr of exposure,
is claimed to be $\Tbb\simeq8.5\e{27}$~yr \cite{Aalbers:2016jon}.

The successor of \texttt{KamLAND-Zen}, \texttt{KamLAND2-Zen}~\cite{Shirai:2017jyz,Gando_neutrino18,Giuliani_neutrino18}
will benefit the upgrades of \texttt{KamLAND} into \texttt{KamLAND2},
including the improved light collection and better energy resolution guaranteed by the new photomultipliers,
together with an increased amount of $^{136}$Xe, to reach at least 1~ton of material.
These upgrades will be performed after the completion of \texttt{KamLAND-Zen 800},
expected to start this year.
The target sensitivity after 5~yr will be $\mbb\lesssim20$~meV%
\footnote{The collaboration does not report the sensitivity in terms of the half-life of the decay.},
sufficient for ``fully covering the inverted ordering region''~\cite{Shirai:2017jyz}.
Future studies will also test the possibility to accommodate scintillating crystals inside the detector
and run a multi-isotope experiment.

Back to $^{76}$Ge-based experiments, the efforts of the \texttt{Gerda} and \texttt{Majorana} collaborations
will join to work on the
\texttt{LEGEND} (Large Enriched Germanium Experiment for Neutrinoless Double beta decay) experiment.
Learning from both its precursors, \texttt{LEGEND} will need further background rejection
and will be built in different phases.
The first module, \texttt{LEGEND-200}, made of 200~kg of Germanium and expected to start in 2021,
will be built on top of the existing \texttt{Gerda} infrastructures and
will have a target sensitivity $\Tbb\simeq1\e{27}$~yr in 5~yr.
The full scale detector, \texttt{LEGEND-1000},
consisting in several modules summing up to a total of 1~ton of material,
will have as an ultimate goal $\Tbb\simeq1\e{28}$~yr in 10~yr~\cite{Abgrall:2017syy},
giving a full coverage of the inverted mass ordering region.

Even larger in size,
\texttt{nEXO}~\cite{Albert:2017hjq,Kharusi:2018eqi}
will replace the \texttt{EXO-200} experiment after its completion, expected this year.
The new detector will use 5~ton of Xenon in order to reach
$\Tbb\simeq1\e{27}$~yr with just 1~yr of data and $\Tbb\simeq1\e{28}$~yr with the full statistics, after 10~yr.

After the completion of the upcoming phase, \texttt{NEXT-100} will be possibly upgraded
into \texttt{NEXT 2.0}, which will need a 1.5~ton of Xenon to obtain the statistics for achieving
$\Tbb\simeq1\e{27}$~yr after 5~yr of running~\cite{Agostini:2017jim,Giuliani_neutrino18}.

In the same way, the \texttt{Panda-X-III} collaboration is also planning a 1~ton scale phase II
with a target of $\Tbb\simeq1\e{27}$~yr~\cite{Chen:2016qcd}.

The last comment regards another interesting possibility related to the \texttt{SNO+} experiment.
The \texttt{THEIA} proposal~\cite{Gann:2015fba}
is a concept study for a gigantic detector of something around 30-100~kton of target material
which will use water-based liquid scintillator.
Such target allows to track both Cherenkov and delayed scintillation light,
thus enabling high light yield and low-threshold detection with attenuation close to that of
pure water.
The result is that such a detector would be able to achieve excellent background rejection
thanks to directionality, event topology, and particle ID, with very large statistics.
Loading of metallic ions which can undergo neutrinoless double beta decay would enable to use the \texttt{THEIA}
detector for studying the Dirac/Majorana nature of neutrinos.
Given the size of the detector, a 0.5\% loading will allow to store several tons of decaying material,
which naturally result in huge statistics when compared with current experiments.
A 3\% loading with natural (not enriched) Tellurium will be sufficient to reach,
assuming $\mbb\simeq15$~meV, a $3\sigma$ discovery
in 10~yr~\cite{Alonso:2014fwf,Giuliani_neutrino18}.

%% file: texs/fut_cosmo.tex
There are a number of studies in the literature
focused on forecasting the expected sensitivity
from both future CMB and large scale structure surveys
to the total neutrino mass $\mnu$~\cite{dePutter:2009kn,Carbone:2010ik,Carbone:2011by,Abazajian:2011dt,Hamann:2012fe,Basse:2013zua,Font-Ribera:2013rwa,Allison:2015qca,Amendola:2016saw,DiValentino:2016foa,Archidiacono:2016lnv,Sprenger:2018tdb}.

Awaiting for very futuristic measurements which may allow
for the extraction of each of the individual masses associated
to the neutrino mass eigenstates (see section~\ref{sec:cosmo}),
the extraction of the neutrino mass ordering strongly relies
on the error achieved on $\mnu$ for a chosen \emph{fiducial} value of the neutrino mass.

A complete, updated and useful summary is provided in Table~II of Ref.~\cite{Lattanzi:2017ubx},
which shows the expected sensitivity [$\sigma(\mnu)$] from different future cosmological probes,
assuming the fiducial value $\mnu=0.06$~eV.
Nevertheless, the authors of Ref.~\cite{Gerbino:2016ehw}
considered different fiducial values for the total neutrino mass
and computed the odds for the normal versus the inverted ordering
for possible combinations of future cosmological probes
including the current information from oscillation experiments.
We shall comment on these results towards the end of this section.

\subsubsection{CMB prospects}
Two main missions are expected to lead the next decade generation of CMB experiments,
albeit a number of other experiments are in progress between now and then.
The latter list includes ground-based observatories as the
\texttt{ACT} (Atacama Cosmology Telescope)~\cite{DeBernardis:2016uxo},
the \texttt{SPT-3G} (South Pole Telescope-3G)~\cite{Benson:2014qhw},
the \texttt{Simons Array}~\cite{Suzuki:2015zzg},
\texttt{CLASS}~\cite{Essinger-Hileman:2014pja},
\texttt{BICEP3}~\cite{Ahmed:2014ixy}
and the \texttt{Simons Observatory}~%
\footnote{For a detailed study on the prospects
from pre- and post-2020 CMB experiments on the extraction of cosmological parameters,
including the total neutrino mass $\mnu$,
see also Ref.~\cite{Errard:2015cxa}.}.
The two main missions are expected to be the
\texttt{CMB-Stage IV} project~\cite{Abazajian:2016yjj}
and \texttt{CORE} (Cosmic Origin Explorer)~\cite{Delabrouille:2017rct}.
The former, the \texttt{CMB-Stage IV} project~\cite{Abazajian:2016yjj},
expected to be \emph{the definitive ground-based CMB experiment},
aims at $250000$ detectors operating for four years,
covering a $40\%$ fraction of the sky.
Depending on the beam size and on the effective noise temperature,
\texttt{CMB Stage IV} could reach sensitivities of $\sigma(\mnu)=0.073-0.11$~eV,
assuming $\mnu=0.058$~eV as the fiducial model and
an external prior on the reionization optical depth of $\tau=0.06\pm 0.01$,
see Ref.~\cite{Abazajian:2016yjj} for the precise configuration details.
The latter, \texttt{CORE},
a medium-size space mission proposed to the European Space Agency (ESA)~\cite{Delabrouille:2017rct},
is expected to have an one order of magnitude larger number of frequency channels
and a twice better angular resolution than \pla.
With these improved capabilities,
\texttt{CORE} could achieve a sensitivity of $\sigma(\mnu)=0.044$~eV~\cite{DiValentino:2016foa,Lattanzi:2017ubx},
for a fiducial total neutrino mass of $0.06$~eV.
As it is evident from these estimates,
future CMB experiments alone will not be able to determine the neutrino mass ordering.

\subsubsection{Large scale structure prospects}

From the large scale structure perspective, in analogy to the future CMB probes,
there are also two main surveys,
\texttt{DESI} (Dark Energy Spectroscopic Instrument)~\cite{Levi:2013gra,Aghamousa:2016zmz},
a ground-based telescope which will improve the SDSS-III and IV legacies (\texttt{BOSS}~\cite{Dawson:2012va} and \texttt{eBOSS} galaxy surveys~\cite{Dawson:2015wdb}),
and the \texttt{Euclid} space mission~\cite{Amendola:2016saw}.
The baseline design of \texttt{DESI} assumes that it will run over five years,
covering 14000 deg$^2$ of the sky targeting four different tracers:
Bright, Luminous Red and Emission Line Galaxies
plus quasars in the redshift interval ($0.05<z<1.85$),
and a Lyman-$\alpha$ survey in the $1.9<z<4$ redshift interval.
The expected error in $\mnu$ from \texttt{DESI} and \pla\ data is $0.02$~eV.
This number corresponds, approximately, to a $2\sigma$ determination
of the neutrino mass ordering in case neutrinos have
the minimal mass within the normal ordering scenario~\cite{Aghamousa:2016zmz}.
The authors of Ref.~\cite{Font-Ribera:2013rwa} have also explored
a number of possible combinations of \texttt{DESI} with other surveys.
Namely, combining \texttt{DESI} measurements with the final results from \texttt{DES},
an error of $0.017$~eV in $\mnu$ could be achieved.
Their most constraining result, $\sigma (\mnu)=0.011$~eV, however,
arises from an extension of the \texttt{DESI} survey,
together with data from \texttt{Euclid}
and \texttt{LSST} (Large Synoptic Survey Telescope)~\cite{Ivezic:2008fe,Abell:2009aa} (see below).
In case this small error is achieved,
the neutrino mass ordering can be determined with a high accuracy,
again assuming a massless lightest neutrino and normal ordering.
Other analyses have also reduced the nominal
$\sigma (\mnu)=0.02$~eV expected from the \texttt{DESI} survey
replacing the \pla\ CMB information with that expected from
the future \texttt{CMB Stage IV}~\cite{Abazajian:2016yjj}
or \texttt{CORE}~\cite{DiValentino:2016foa} probes.

\texttt{Euclid}, an ESA mission expected to be launched early in the upcoming decade,
mapping $\sim 15000$~deg$^2$ of the sky,
has also been shown to provide excellent capabilities to test the neutrino properties~\cite{Amendola:2016saw}.
\texttt{Euclid} will focus on both galaxy clustering and weak lensing measurements,
which, combined with \pla\ CMB data, will provide errors on the sum of the neutrino masses
of $\sigma (\mnu)=0.04$~eV~\cite{Carbone:2010ik}
and $\sigma (\mnu)=0.05$~eV~\cite{Kitching:2008dp}, respectively,
albeit exploiting the mildly non-linear regime could highly reduce these errors~\cite{Audren:2012vy}.
While these errors are large to extract useful information concerning the neutrino mass ordering,
the weak gravitational lensing abilities from \texttt{Euclid}
have also been considered to extract the neutrino mass ordering
when it lies far enough from the degenerate region,
see e.g.\ Ref.~\cite{Amendola:2016saw}.
The addition of future CMB measurements, as those from \texttt{CORE},
could notably improve the expected \texttt{Euclid} sensitivity.
The authors of \cite{Archidiacono:2016lnv} have shown that CMB measurements from \texttt{CORE},
combined with full shape measurements of the galaxy power spectrum
and weak lensing data from \texttt{Euclid},
could reach $\sigma (\mnu)=0.014$~eV.
This result clearly states the complementarity of cosmic shear and galaxy clustering probes,
crucial to test the neutrino mass ordering.
Further improved measurements of the reionization optical depth $\tau$ could strengthen this bound
and consequently the sensitivity to the ordering of the neutrino masses~\cite{Archidiacono:2016lnv,Sprenger:2018tdb,Liu:2015txa}, see the following section.
Other future large scale structure surveys are
the aforementioned \texttt{LSST} and \texttt{WFIRST}~\cite{Spergel:2013tha,Spergel:2015sza},
that will lead as well to accurate measurements of the total neutrino mass.
Their combination with e.g.\ \texttt{Euclid} could provide an error of a few meV on the total neutrino mass,
$\sigma(\mnu)\lesssim 0.008$~eV~\cite{Jain:2015cpa}.

The above neutrino mass (neutrino mass ordering) projected errors (sensitivities),
even if strongly constraining, are highly dependent on the fiducial value of $\mnu$,
in the sense that the majority of the forecasts
\textit{(a)} are usually carried out assuming the minimal neutrino mass allowed within the normal ordering scheme,
i.e.\ $\mnu\simeq 0.06$~eV~%
\footnote{The authors of Ref.~\cite{Amendola:2016saw} have nonetheless presented constraints for different fiducial models.};
\textit{(b)} the quoted sensitivities in the neutrino mass ordering
are computed via an extrapolation of the error on the sum of neutrino masses
rather than from proper Bayesian comparison tools.
The authors of Ref.~\cite{Gerbino:2016ehw} found that a future CMB \texttt{CORE}-like satellite mission,
even combined with a $1\%$ measurement of the Hubble constant $H_0$
and with the future \texttt{DESI} survey~\cite{Aghamousa:2016zmz,Font-Ribera:2013rwa}
can not extract the ordering
if nature has chosen a value for the neutrino masses of $\mnu=0.1$~eV.
Odds for the normal versus the inverted ordering of
$1:1$ were reported~\cite{Gerbino:2016ehw}.
When considering the minimum allowed value for the total neutrino mass set by neutrino oscillation experiments,
i.e.\ $\mnu=0.06$~eV, they quote odds of $3:2$ ($9:1$) for the case in which \texttt{CORE}
and the prior on $H_0$ without (with) \texttt{DESI} measurements are considered~%
\footnote{For the \texttt{CORE} CMB mission,
data were generated following Refs~\cite{Bond:1998qg,Bond:1997wr}.
For \texttt{DESI},
mock $r_s H(z)$ and $d_A(z)/r_s$ data were generated for the three \texttt{DESI} tracers
in the $0.15<z<1.85$ redshift range, accordingly to Ref.~\cite{Font-Ribera:2013rwa}.}.
Therefore, the next generation of CMB and large scale structure surveys
will be sensitive to the mass ordering only if it is normal and the
lightest neutrino mass is close to zero.
The significance of such a measurement will crucially depend
on how far $\mnu$ lies from its minimum allowed value from oscillation probes.

%% file: texs/cosmo21.tex
Cosmological measurements of the redshifted 21~cm hydrogen line
provide a unique test of the Epoch of Reionization (EoR) and the ``dark ages'',
the period before the first stars formed.
The 21~cm line is due to spin-flip transitions in neutral hydrogen
between the more energetic triplet state and the ground singlet state,
and its intensity depends on the ratio of the populations
of these two neutral hydrogen hyperfine levels.
At a given observed frequency $\nu$,
the 21~cm signal can be measured in emission or in absorption against the CMB.
The so-called differential brightness temperature $\delta T_b$
therefore refers to the contrast between the temperature
of the hydrogen clouds and that of the CMB, which,
for small frequencies and up to first order in perturbation theory,
reads as~\cite{Madau:1996cs, Furlanetto:2006jb, Pritchard:2011xb, Furlanetto:2015apc}
\begin{equation}
\delta T_b(\nu)
\simeq
27 \, x_\textrm{HI} \, (1 + \delta_b)
\left( 1 - \frac{T_\textrm{CMB}}{T_S}\right)
\left( \frac{1}{1+H^{-1} \partial v_r / \partial r} \right) \,
\left( \frac{1+z}{10}\right)^{1/2}
\left(\frac{0.15}{\Omega_m h^2} \right)^{1/2}
\left( \frac{\Omega_b h^2}{0.023}\right)
\,\textrm{mK} ~,
\label{eq:Tbdev}
\end{equation}
where $x_\textrm{HI}$ is the fraction of neutral hydrogen,
$\delta_b$ is the baryon overdensity,
$\Omega_b h^2$ and $\Omega_m h^2$ the present baryon and matter contributions to the mass-energy budget of the Universe,
$H(z)$ the Hubble parameter
and $\partial v_r / \partial r$ the comoving peculiar velocity gradient along the line of sight.
Therefore, 21~cm cosmology aims to trace the baryon overdensities via transitions in neutral hydrogen.

There are a number of current and future experimental setups devoted
to detect the 21~cm global signal averaged over all directions in the sky,
as \texttt{EDGES} (Experiment to Detect the Reionization Step)~\cite{Bowman:2012hf},
the future \texttt{LEDA} (Large Aperture Experiment to Detect the Dark
Ages)~\cite{Greenhill:2012mn} or
\texttt{DARE} (Moon space observatory Dark Ages Radio Experiment)~\cite{Burns:2011wf}.
The \texttt{EDGES} experiment has quoted the observation
of an absorption profile located at a frequency of $78 \pm 1$~MHz,
corresponding to a redshift of $z \sim 17$,
with an amplitude of about a factor of two larger
than the maximum expected in the canonical $\Lambda$CDM framework~\cite{Bowman:2018yin}.
This recent claim has led to a number of studies
aiming either to explain the effect
or to constrain some non-standard scenarios~\cite{Munoz:2018pzp,McGaugh:2018ysb,Barkana:2018lgd,Barkana:2018qrx,Fraser:2018acy,Kang:2018qhi,Yang:2018gjd,Pospelov:2018kdh,Costa:2018aoy,Slatyer:2018aqg,Falkowski:2018qdj,Munoz:2018jwq,Fialkov:2018xre,Berlin:2018sjs,DAmico:2018sxd,Safarzadeh:2018hhg,Hill:2018lfx,Clark:2018ghm,Cheung:2018vww,Hektor:2018qqw,Liu:2018uzy,Hirano:2018alc,Mitridate:2018iag,Mahdawi:2018euy,Feng:2018rje,Ewall-Wice:2018bzf,Witte:2018itc}.

Fluctuations in the redshifted 21~cm signal
can be used to compute
the power spectrum of the differential brightness temperature.
This is the major goal of experiments as
\texttt{GMRT} (Giant Metrewave Radio Telescope)~\cite{Ananthakrishnan:1995, Paciga:2010yy},
\texttt{LOFAR} (LOw Frequency ARray)~\cite{vanHaarlem:2013dsa},
\texttt{MWA} (Murchison Widefield Array)~\cite{Tingay:2012ps} and
\texttt{PAPER} (Precision Array for Probing the Epoch of Reionization)~\cite{Pober:2015ema,Ali:2015uua,Parsons:2009in},
targeting statistical power-spectrum measurements
of the 21~cm signal employing large radio interferometers.
Even if current experiments have not yet detected the 21~cm cosmological signature,
the \texttt{PAPER} collaboration has recently improved
the previous upper limits at $z=8.4$~\cite{Ali:2015uua}.
Next decade, high-redshift 21~cm experiments include
the \texttt{SKA} (Square Kilometre Array)~\cite{Mellema:2012ht}
and \texttt{HERA} (Hydrogen Epoch of Reionization Array)~\cite{Beardsley:2014bea}.
A three-dimensional map of the 21~cm signal
could also be obtained by means of the so-called intensity mapping technique,
which measures the collective emission from neutral hydrogen in dense clumps,
targeting large regions without resolving individual galaxies
in the post-reionization era ($z\lesssim 3$)~\cite{Wyithe:2007rq, Chang:2007xk,Loeb:2008hg,Villaescusa-Navarro:2014cma}.
The experimental efforts for this technique include
the \texttt{GBT-HIM} project,
with the \texttt{GBT} (Green Bank Telescope)~\cite{Chang:2016npo},
\texttt{CHIME} (Canadian Hydrogen Intensity Mapping Experiment)~\cite{CHIME},
the \texttt{Tianlai} project~\cite{Chen:2015oga}
and \texttt{SKA-mid} frequency~\cite{SKA_baseline},
see e.g.~\cite{Bull:2014rha}.

Despite the fact that the primary task of future 21~cm experiments
is to improve our current knowledge of the reionization history,
they provide as well an additional tool for fundamental cosmology~\cite{Scott:1990,Tozzi:1999zh,Iliev:2002gj,Barkana:2005xu,Barkana:2004zy,Bowman:2005hj,McQuinn:2005hk,Santos:2006fp,Mao:2008ug,Visbal:2008rg,Clesse:2012th,Liu:2015txa,Liu:2015gaa},
complementary to CMB missions and galaxy surveys.
Indeed, 21~cm cosmological observations will play a very important role concerning neutrino physics.
As previously stated, there are two types of experiments.
First of all, we will have observations focused on the pre-reionization and EoR periods,
that can probe very large volumes (where the non-linear scale is small).
Remember that the largest signal from relic neutrino masses
and their ordering appears at scales which,
at the redshifts attainable at galaxy clustering surveys,
lie within the mildly non-linear regime.
Therefore one needs to rely on either N-body simulations
or on analytical approximations for the matter power spectrum
to simulate the massive neutrino signature.
EoR 21~cm experiments will achieve the scales required
to observe the neutrino signature within the linear regime,
avoiding the simulation problems described in section~\ref{sec:large}.
In this regard, these probes may widely surpass the constraints
on neutrino masses expected from even very large galaxy surveys~\cite{Tegmark:2008au,Mao:2008ug,McQuinn:2005hk,Pritchard:2008wy,Abazajian:2011dt,Oyama:2012tq,Shimabukuro:2014ava,Oyama:2015gma}.
Furthermore, the neutrino constraints will be largely independent
of the uncertainties in the dark energy fluid, which,
as we have seen in section~\ref{sec:mnuw},
have instead a non-negligible impact in lower redshift, galaxy survey measurements.
This is a byproduct of using the 21~cm line to trace the matter overdensities:
at redshifts $z\lesssim 2$, the universe starts to be dominated
by the dark energy fluid and the growth of matter perturbations
is modified depending on the dark energy equation of state $w(z)$,
whose precise time-evolution remains unknown.
Consequently, for a given perturbation in the matter fluid,
a suppression in its structure growth could be either due
to the presence of massive neutrinos or
to an evolving dark energy fluid.
Focusing at higher redshifts, the neutrino mass constraints
from 21~cm probes will be largely independent
of the uncertainties in the dark energy fluid properties.

Expectations from \texttt{MWA}, \texttt{SKA} and \texttt{FFTT} (Fast Fourier Transform Telescope)~\cite{Tegmark:2008au}
were considered in Ref.~\cite{Mao:2008ug}.
Focusing on 4000 hours of observations of two areas in the sky
in a range of $z= 6.8-8.2$ (divided into three redshift bins)
and a value of $k_{\rm{max}}=2$~Mpc$^{-1}$,
the reported errors on $\mnu$ are
0.19 (0.027),
0.056 (0.017),
0.007 (0.003)
for \texttt{MWA}, \texttt{SKA} and \texttt{FFTT}, respectively,
in their middle (optimistic) scenarios~%
\footnote{These scenarios differ in the assumptions concerning the power modeling,
the prior on the reionization history and
the residual foregrounds cutoff scale, among other factors,
see Ref.~\cite{Tegmark:2008au}.}, when combined with \pla\ measurements.
These forecasts were performed for a fiducial
$\Omega_\nu h^2=0.0875$,
which corresponds to a quite high value for the neutrino mass,
lying in the fully degenerate neutrino mass spectrum.

The authors of Ref.~\cite{Oyama:2012tq} devoted
a dedicated analysis to establish the potential for extracting
the neutrino mass ordering combining the \texttt{FFTT}
capabilities with future CMB polarization measurements.
Based exclusively on the induced effect of the neutrino mass ordering
on the cosmic expansion rate, a robust $90\%$~CL neutrino mass ordering extraction was reported if $\mnu <0.1$~eV,
regardless the underlying true ordering (i.e.\ normal or inverted).
In Ref.~\cite{Oyama:2015gma}, the authors propose to combine
ground-based CMB polarization observations, \texttt{SKA} Phase 2
and
BAO measurements from \texttt{DESI}.
With these data sets, a $2\sigma$
extraction of the neutrino mass ordering seems feasible, 
unless the neutrino spectrum is degenerate.
Notice that these results arise from the signature induced by the neutrino mass ordering
in the cosmic expansion rate, as the minimum cutoff
of the wavenumber in the 21~cm observations is
$k_{\rm{min}}=0.06h$~Mpc$^{-1}$,
while the wavenumber corresponding to the neutrino free-streaming
scale is $k_{\rm{min}}\simeq 0.02h$~Mpc$^{-1}$ for a $0.05$~eV massive
neutrino.

More futuristic 21~cm experiments, as \texttt{FFTT},
may open the possibility of going beyond measurements
of the total neutrino mass $\mnu$ and measure the individual neutrino masses,
revealing the uniqueness of such experiments for constraining the neutrino properties.
As shown in Figure~\ref{fig:Ps} in section~\ref{sec:cosmo},
the differences in the power spectra for the two possible mass orderings are tiny.
Therefore, exquisite precision measurements are required to identify such signatures.
Galaxy surveys, already discussed in the previous section, are limited by two facts.
The first one is related to non-linearities,
which will not allow for a measurement of the power spectrum
at scales $k> 0.2h$~Mpc$^{-1}$ at small redshifts, see section~\ref{sec:large}.
Since the non-linear scale at $z=8$ is $k\simeq 3h$~Mpc$^{-1}$,
both \texttt{SKA} and \texttt{FFTT} can measure the entire linear region
and be more sensitive to the scale-dependent suppression,
which is different in the two neutrino mass orderings.
The second one is related to the fact that a galaxy survey
requires a large number density of tracers to ensure a good sensitivity at small scales,
while for 21~cm surveys, tracing the ubiquitous permeating hydrogen,
a high-density antennae distribution
will already warrant excellent small-scale sensitivities.
One drawback of 21~cm probes are foregrounds,
which should be kept under control.

\begin{figure}
\centering
\includegraphics[width=0.6\textwidth]{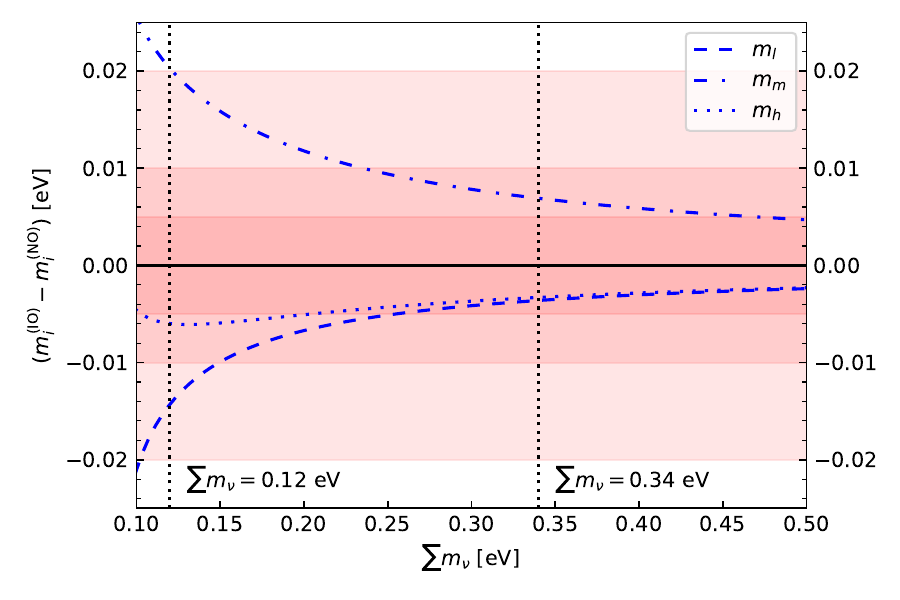}
\caption{\label{fig:diff}
Differences in the masses for three neutrino mass eigenstates
as a function of the total neutrino mass
between inverted and normal orderings.
The vertical dashed lines depict the value $\mnu=0.12$~eV and $\mnu=0.34$~eV,
which are the present most constraining $95\%$~CL limit on $\mnu$~\cite{Palanque-Delabrouille:2015pga}
and the latest $95\%$~CL bound quoted by the \pla\ collaboration~\cite{Aghanim:2016yuo}, respectively.
Different shades of colored bands indicate the possible errors
which could be achieved by future cosmological experiments
on the determination of single neutrino masses:
0.02~eV, 0.01~eV or 0.005~eV.}
\end{figure}

\begin{figure}[t]
\includegraphics[width=0.55\textwidth]{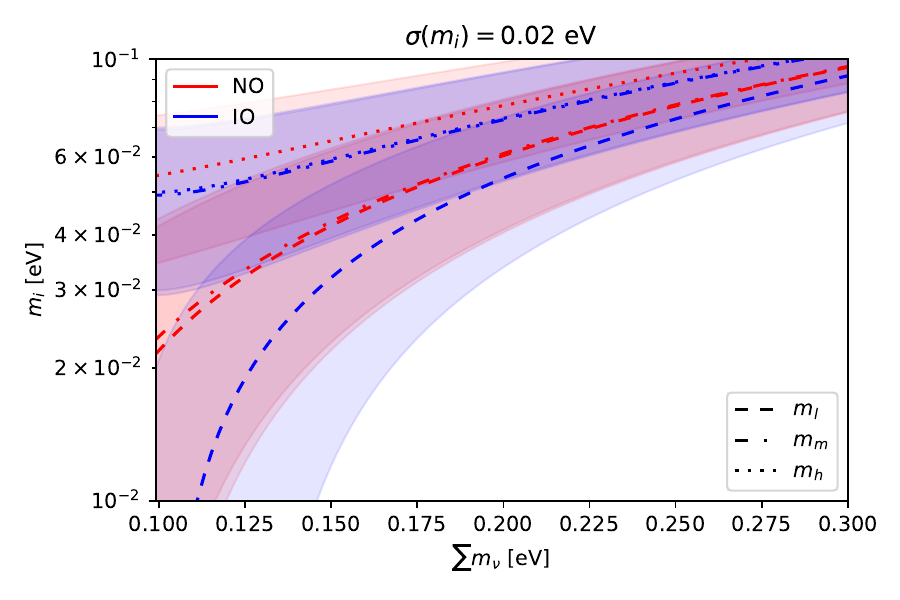}
\includegraphics[width=0.55\textwidth]{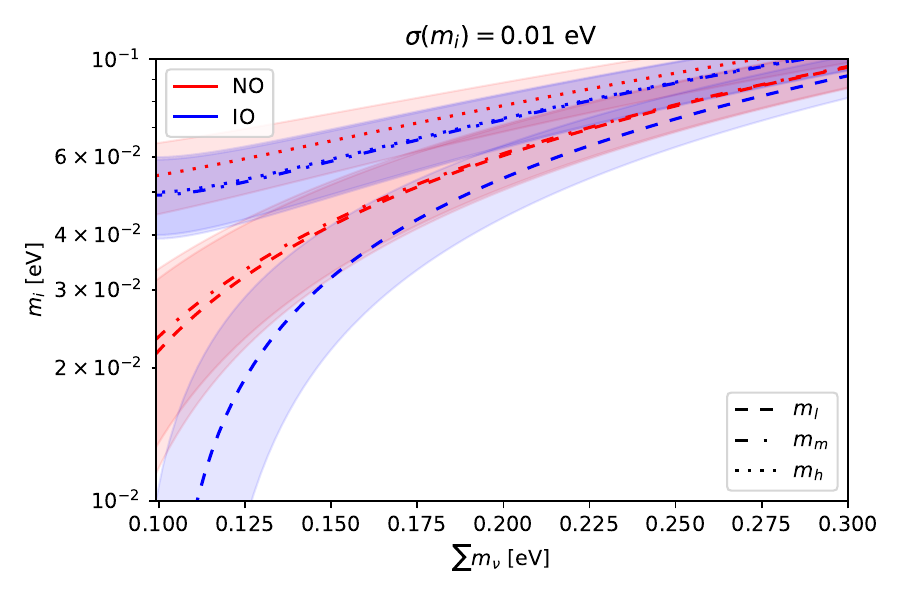}
\includegraphics[width=0.55\textwidth]{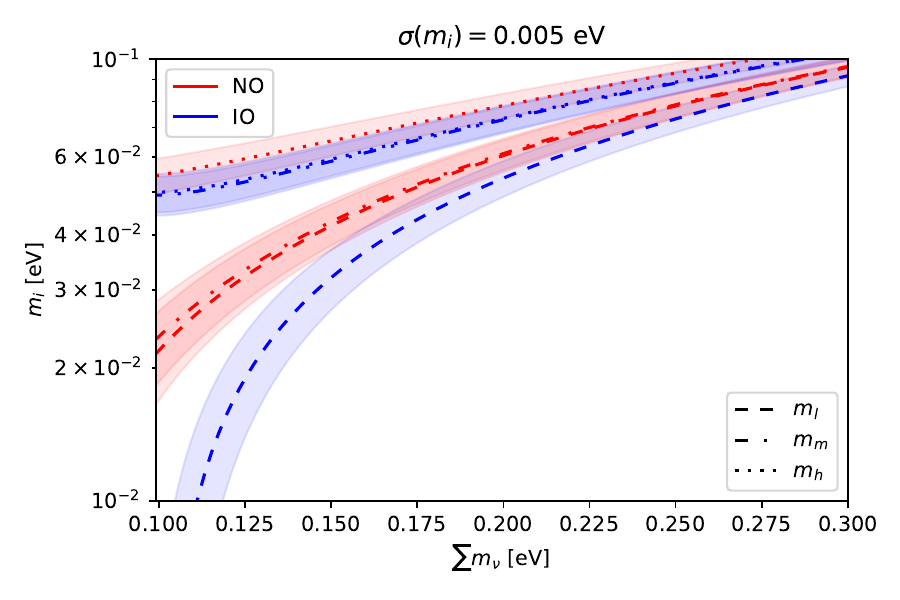}
\caption{\label{fig:mi}
Values of the individual neutrino masses for the heaviest, medium and lightest mass eigenstates
for the normal and inverted orderings
as a function of the total neutrino mass.
The panels, from top to bottom,
depict the error bands $\sigma(m_i)=0.02$~eV, $\sigma(m_i)=0.01$~eV and $\sigma(m_i)=0.005$~eV.}
\end{figure}

The authors of Ref.~\cite{Pritchard:2008wy} have studied the perspectives
for extracting the individual neutrino masses with \texttt{SKA} and \texttt{FFTT},
finding that \texttt{FFTT} could be able to distinguish
all the three neutrino masses from zero at the $3\sigma$ level,
due to its enormous effective volume
(see Figure~3 of Ref.~\cite{Pritchard:2008wy}).
Extracting the neutrino mass ordering directly from the individual masses,
however, was shown to be a very difficult achievement.
Our calculations show that, for the total neutrino mass we use here as a reference, $\mnu=0.12$~eV,
the differences among the lightest ($l$), medium ($m$) and
heaviest ($h$) neutrino mass eigenstates
between the normal and inverted orderings are
$(|\Delta m_{l}|,\ |\Delta m_m|,\ |\Delta m_h|) =
(0.015, 0.0209, 0.0059)$~eV,
which, especially for the case of $|\Delta m_h|=0.0059$~eV,
are tiny and very difficult to resolve,
even with very futuristic 21~cm measurements.
While increasing the exposure of \texttt{FFTT} may improve
its capabilities for this purpose
(the error in the most optimistic \texttt{FFTT} scenario of Ref.~\cite{Mao:2008ug} is $0.003$~eV),
it seems an extremely challenging task.
Figure~\ref{fig:diff} depicts the differences in the values
of the three neutrino masses as a function
of the total neutrino mass between inverted and normal orderings.
We show with a dashed vertical line our representative case $\mnu=0.12$~eV
(the present most constraining $95\%$~CL upper limit)
and another one for $\mnu=0.34$~eV
(the most recent $95\%$~CL bound from the \pla\ collaboration
after the removal of systematics in their polarization data
at high angular scales~\cite{Aghanim:2016yuo}).
Notice that, as expected, the differences between the values of
the three neutrino masses decrease
with the total neutrino mass.
In this regard, the lower the neutrino mass,
the easier it could be to single out the three neutrino mass eigenstates,
because they are more separated.
However, an extraction of the mass ordering
in the non-degenerate region via the values
of the individual neutrino masses seems very difficult.
Indeed, Figure~\ref{fig:mi} illustrates the values
of the individual neutrino masses for the heaviest, medium and lightest states
for the normal and inverted orderings as a function of the total neutrino mass.
The bands, from top to bottom panels,
depict the errors $\sigma(m_i)=0.02$~eV and $\sigma(m_i)=0.01$~eV,
together with the very futuristic \texttt{FFTT} one, $\sigma(m_i)=0.005$~eV.
For an error of $\sigma(m_i)=0.02$~eV,
there is no hope to disentangle the individual neutrino masses,
as the error bands overlap for the heaviest, medium and lightest masses in all the parameter space.
If instead one could achieve $\sigma(m_i)=0.01$~eV,
a measurement of the individual neutrino masses
in the non-degenerate region could be possible at the $1-2\sigma$ level,
but in order to unravel the ordering one would need very extreme conditions
as, for instance, a value of $\mnu$ very close to $0.1$~eV independently determined
with very small errors.
The bottom plot  in Figure~\ref{fig:mi} shows the results if we assume the futuristic value of $\sigma(m_i)=0.005$,
expected to be achieved by \texttt{FFTT}.
In this case, a measurement of the three neutrino masses will be achieved.
Furthermore, in this (very optimistic) situation,
the error bars will be, in principle, sufficiently small
to detect the presence (or the lack) of two massive neutrino states
with masses in the 0.02--0.03~eV range,
required if the ordering is normal to explain $\mnu\simeq0.1$~eV,
which would strongly confirm the normal (or inverted) neutrino mass ordering.
If $\sigma(m_i)=0.005$, the detection of the mass ordering
will still be possible even if $\mnu\lesssim0.1$~eV,
since the error on \mnu\ will allow to exclude the inverted ordering
with great accuracy.

As already mentioned, another possibility is
the so-called 21~cm intensity mapping,
which will focus on low redshifts $z\lesssim 3$ and will measure,
with low angular resolution,
the integrated 21~cm flux emitted from unresolved sources observing large patches of the sky.
The lack of high angular resolution will result in a less precise measurement of non-linear scales.
On the other hand, low angular resolution will imply a much faster survey.
Future planned intensity mapping surveys are developed within the Phase 1 of the \texttt{SKA} experiment,
which will include a wide and deep survey at low redshifts
($z\lesssim 3$, the \texttt{SKA1-MID} array)
and a narrow and deep survey at higher redshift
($3\lesssim z \lesssim 6$, the \texttt{SKA1-LOW} array),
and within the Phase 2 of \texttt{SKA} (\texttt{SKA2}).
Since, in some sense, these intensity mapping probes will be complementary
to future planned optical surveys,
as \texttt{DESI} or \texttt{Euclid},
it makes sense to combine their expected results.
The intensity mapping technique, as galaxy clustering,
is also affected by bias uncertainties and non-linearities at small
scales.
Several studies have been carried out in the literature to unravel
the perspectives of the intensity mapping technique in unveiling the neutrino properties.
Some of them include the combination of the expectations
from future large scale structure and intensity mapping surveys~\cite{Sprenger:2018tdb,Archidiacono:2016lnv,Loeb:2008hg,Visbal:2008rg,Abazajian:2011dt,Villaescusa-Navarro:2015cca}.
Notice that all these studies rely on different assumptions
on the cosmological parameters, on the foregrounds and on the systematic uncertainties, therefore we can not do comparisons among them.
Instead, we quote the most recent findings
and the impact for an eventual future detection of the neutrino mass ordering.

The authors of Ref.~\cite{Villaescusa-Navarro:2015cca} found that,
by combining \texttt{SKA1-LOW} with \pla\ measurements,
the $95\%$~CL error on $\mnu$ could be $\sim 0.089$~eV.
It is remarkable that such a combination could potentially rule out the inverted ordering scenario,
assuming that normal ordering is realized in nature.
These authors also find that,
under identical assumptions in the forecasted analyses,
their combination of intensity mapping surveys (\texttt{SKA1-LOW} and \texttt{MID})
should be regarded as competitive with future spectroscopic surveys
concerning neutrino mass properties.
The authors of Ref.~\cite{Archidiacono:2016lnv} showed
that constraints of the future \texttt{CORE} CMB mission and galaxy redshift/weak lensing
large scale structure surveys (as \texttt{Euclid})
on the neutrino mass can be improved if a prior
on the reionization optical depth from 21~cm probes
as \texttt{HERA} or \texttt{SKA} is also included.
A prior of $\sigma(\tau)=0.001$ will reduce the freedom
in the amplitude of the primordial power spectrum $A_s$,
as CMB measurements mostly constrain the combination $A_s \exp(-2\tau)$, see section~\ref{sec:cmb}.
Therefore, the direct correlation between $\mnu$ and $A_s$,
both modifying the amplitude of the matter power spectrum
(although the change induced by $\mnu$ is, obviously, scale dependent),
is largely affected by the presence of a precise determination of $\tau$.
The $1\sigma$ sensitivity they find for the combination
of \texttt{CORE}, \texttt{Euclid} plus the prior on the optical depth from future 21~cm observations
is $\sigma(\mnu)=0.012$~eV~%
\footnote{More recently, this very important synergy between \texttt{Euclid} and future 21~cm surveys,
concretely with the intensity mapping survey \texttt{SKA1},
has been further assessed in Ref.~\cite{Sprenger:2018tdb}.}.

Nevertheless, as carefully detailed above, even if these tiny errors on $\mnu$ will be reached
and extrapolated to an error on the individual neutrino mass eigenstates,
the possibility of extracting the neutrino mass ordering via singling out the neutrino mass eigenstates
with cosmological observables remains unfeasible,
unless very visionary scenarios,
as \texttt{FFTT} under the most optimistic assumptions, are envisaged. 

%% file: texs/sn.tex
Neutrinos from core-collapse supernovae offer an independent and
complementary way to test neutrino physics.
The existence of these
neutrinos was robustly confirmed by the detection of twenty-five
events from Supernova 1987A in the Large Magellanic Cloud~\cite{Bionta:1987qt,Hirata:1987hu,Alekseev:1987ej},
located at $\sim 50$~kpc from our Milky Way galaxy.
Such a detection allowed to set
very compelling bounds on a number of neutrino
properties~\cite{Schramm:1990pf,Raffelt:1998hw}.
Even if laboratory experiments have surpassed some of these limits,
the eventual detection of supernovae neutrinos will still provide
precious information about the details of the explosion process
(see e.g.~\cite{Scholberg:2017czd,Mirizzi:2015eza,Janka:2012wk} and references therein),
and also of neutrino mixing effects in dense
media, see also Ref.~\cite{Horiuchi:2017sku}.

Neutrino production in core-collapse supernovae occurs
in a number of different stages.
The first one is the \emph{infall}, in which
electron neutrinos are produced, confined, as a result of the process
$e^{-}+p\rightarrow n +\nu_e$.
When electrons are converted, the outwards pressure they generate disappears
and the gravity forces are no more balanced:
the core will start to collapse until its density reaches
that of matter inside atomic nuclei, i.e.\ nuclear densities.
Once these densities are reached, matter becomes incompressible, and a
hydrodynamic shock is formed.
As this shock wave propagates outwards,
it heats up the nuclei and disintegrates them, releasing neutrinos.
This initial neutrino release is commonly known as
\emph{neutronization burst},
and it is mainly composed of $\nu_e$ and may last for a few tens of milliseconds.
After the neutronization burst,
the remnant proto-neutron star may evolve into a
neutron star or collapse to a black hole,
depending on the mass of the progenitor star.
During this phase of \emph{explosion and accretion},
which lasts for one to two seconds,
the $\nu_e$ contribution is still the dominant one,
albeit there is also a contribution from other (anti)neutrino flavors, in particular $\bar{\nu}_e$.
The neutrinos produced in the \emph{cooling} stage
give the main contribution to the total flux,
as it is in this phase when the supernova releases
its energy via all-flavor neutrino-antineutrino pair production,
reaching its final cold state.
This process lasts for about tens of seconds.
The differences in the mean temperature of the neutrino fluxes
of $\nu_e$, $\bar{\nu}_e$ and $\nu_x$ ($\bar{\nu}_x$)
are due to the different medium opacity of each species.
The larger the opacity, the lower the temperature
that the (anti)neutrino will have at decoupling.
The neutrino fluxes read as~\cite{Scholberg:2017czd}
\begin{equation}
\phi(E_\nu)
=
N_0
\frac{(\alpha+1)^{(\alpha+1)}}{\langle E_\nu\rangle\Gamma(\alpha+1)}
\left(\frac{E_\nu}{\langle
E_\nu\rangle}\right)^{\alpha}
\exp\left(-(\alpha+1)\frac{E_\nu}{\langle E_\nu\rangle}\right)~,
\end{equation}
where $N_0$ is the total number of emitted neutrinos, and
both $\alpha$ and the mean energy
$\langle E_\nu\rangle$ are flavor dependent.
The supernova neutrino energy spectra peaks around the $10-20$~MeV region.

The most popular process for
supernova neutrino detection is inverse beta decay on protons
($\bar{\nu}_e+ p \rightarrow n +e^{+}$).
Other possibilities include elastic scattering on
electrons ($\nu+e^{-}\rightarrow \nu+e^{-}$),
whose kinematics may provide information on the supernova location.
Supernova neutrinos can also interact with nuclei
via charged current or neutral current interactions,
giving rise to charged leptons and/or excited nuclei
which may provide flavor tagging.
A very important process on argon nuclei is
$\nu_e+ ^{40}\mathrm{Ar}\rightarrow e^{-} + ^{40}\mathrm{K}^*$,
which allows for electron neutrino tagging.
In practice, water Cherenkov and scintillator detectors
are mostly sensitive to electron antineutrinos via inverse beta decay,
while the liquid argon technique mainly detects electron neutrinos.
While other flavors may also be detected,
the two processes above are the dominant ones.
Large detector volumes (dozens of kilotons) are
required to detect neutrinos from core-collapse supernovae
located at $\sim \mathcal{O}(10)$~kpc.
A convenient way to scale the total number
of supernova neutrino events in a detector
of given effective mass is~\cite{Beacom:1998ya,Mena:2006ym}
\begin{equation}
N
=
N_0
\left(\frac{E_{\rm B}}{3\times 10^{53}\ \rm{erg}}\right)
\left(\frac{10 \ \textrm{kpc}}{D_{\rm{OS}}}\right)^2~.
\end{equation}
In the expression above,
$E_{\rm B}$ is the gravitational binding energy of the collapsing star
and $D_{\rm{OS}}$ the distance between the observer and the supernova.
Assuming sensitivity to all reactions, the
reference rate is $N_0=\mathcal{O} (10^4)$
for the \texttt{Super-Kamiokande} water Cherenkov detector
with $32$~kton and $5$~MeV energy detection threshold.
References~\cite{Scholberg:2017czd,Scholberg:2012id} give
an estimate of the number of neutrino events for a number of ongoing and
future facilities, based on different detection techniques:
water Cherenkov
(including also those with long string photosensors in ice,
as \texttt{Icecube} and \texttt{PINGU}),
liquid argon time projection chambers, and liquid scintillators.
Upcoming neutrino detectors,
already described in section~\ref{sec:futureosc}
and crucial for oscillation physics measurements,
such as the \texttt{JUNO} liquid scintillator~\cite{An:2015jdp},
the liquid argon \texttt{DUNE}~\cite{Acciarri:2016crz,Acciarri:2015uup,Strait:2016mof,Acciarri:2016ooe} and
the water Cherenkov \texttt{Hyper-Kamiokande}~\cite{Abe:2015zbg}
can lead to a number
of $6000$, $3000$ and $75000$ supernova neutrino events respectively,
assuming that the explosion occurs at $10$~kpc from our position.

Flavor transitions inside a supernova have been carefully reviewed in
Refs.~\cite{Scholberg:2017czd,Mirizzi:2015eza}
(see also \cite{Lunardini:2000sw,Lunardini:2001pb,Akhmedov:2002zj,Lunardini:2003eh,Lunardini:2004bj}).
Here we summarize the most relevant results.
As we have seen in section~\ref{sec:osc-current},
when neutrinos propagate through matter
their mixing effects undergo the so-called MSW mechanism,
feeling a matter potential which is proportional
to the electron number density $N_e$.
If the supernova matter density has a profile which varies slowly,
the neutrino matter eigenstates will propagate adiabatically
and their final flavor composition will depend on the neutrino mass ordering,
which will establish whether or not resonant transitions
associated to each neutrino mass squared difference
(solar and atmospheric) take place~%
\footnote{%
In case the matter potential inside the supernova suffers from discontinuities,
the neutrino transitions will be non-adiabatic
and the final flavor composition will depend on the precise matter profile.}.
In the normal ordering case,
the neutrino fluxes will have a significantly transformed spectrum,
while the electron antineutrino one will only be partially transformed
($F^{\rm{final}}_{\nu_e}=F^{\rm{initial}}_{\nu_x}$ and
$F^{\rm{final}}_{\bar{\nu}_e}= \cos^2
\theta_{12}F^{\rm{initial}}_{\bar{\nu}_e}+ \sin^2 \theta_{12}F^{\rm{initial}}_{\bar{\nu}_x}$).
In the inverted ordering case,
the effects on the electron neutrino and antineutrino
fluxes will be approximately the opposite ones
($F^{\rm{final}}_{\nu_e}= \sin^2
\theta_{12}F^{\rm{initial}}_{\nu_e}+ \cos^2
\theta_{12}F^{\rm{initial}}_{\nu_x}$ and
$F^{\rm{final}}_{\bar{\nu}_e}= F^{\rm{initial}}_{\bar{\nu}_x}$).
Once neutrinos exit from supernovae,
they can still undergo flavor transitions if they traverse the Earth.
Their final flavor composition at the detector location
will again depend on the neutrino mass ordering,
as matter effects in Earth depend on it,
see e.g.\ Ref.~\cite{Scholberg:2017czd} and references therein.

Furthermore, \emph{collective effects} from neutrino self-interactions,
due to $\nu_e+\bar{\nu}_e\rightarrow \nu_x+\bar{\nu}_x$ flavor processes,
can lead to departures from the above summarized three-flavor oscillation
picture~\cite{Duan:2010bg,Mirizzi:2015eza,Hannestad:2006nj,Duan:2007bt,Raffelt:2007yz,EstebanPretel:2007ec,EstebanPretel:2008ni}.
The effective potential, proportional to the
difference between the electron antineutrino and
the muon/tau antineutrino fluxes,
and inversely proportional to the supernova radius,
should dominate over the standard matter one, leading to
\emph{spectral swaps or splits}~\cite{Raffelt:2007cb,Raffelt:2007xt,Dasgupta:2009mg,Dasgupta:2010cd}.
In the early stages,
these self-interacting effects are sub-leading for mass ordering
signatures,
albeit we shall comment on possible non-thermal features
in the neutrino or antineutrino spectra which depend
on the mass ordering~\cite{Choubey:2010up}.

In the following,
we shall summarize the most relevant available methods to
extract the neutrino mass ordering using the mentioned fluxes.
For a recent and thorough review
of the mass ordering signatures from supernovae neutrinos,
we refer the reader to Ref.~\cite{Scholberg:2017czd}.
The electron neutrinos produced in the neutronization burst
undergo the MSW effect being fully (only partially) transformed,
i.e.\
$F^{\rm{final}}_{\nu_e}=F^{\rm{initial}}_{\nu_x}$ ($F^{\rm{final}}_{\nu_e}= \sin^2 \theta_{12}F^{\rm{initial}}_{\nu_e}+ \cos^2
\theta_{12}F^{\rm{initial}}_{\nu_x}$) if the mass ordering is normal (inverted), respectively.
Therefore, detectors with good $\nu_e$ tagging,
such as liquid argon or water Cherenkov ones,
will detect a neutronization burst only
in the inverted neutrino mass ordering case.
Concerning the accretion phase,
and once electron antineutrinos are also produced,
as they are almost unchanged in the MSW resonance,
the largest signature is expected to occur
for the normal ordering case for the three type
of aforementioned detector types
(liquid argon, water Cherenkov and scintillator),
although the \texttt{Icecube} detector,
with its excellent capabilities to reconstruct the time dependence of the signal,
could also distinguish between the normal and inverted mass orderings~\cite{Ott:2012jq}.
While a devoted study with precise and accurate mass ordering sensitivities
attainable at these three detector types via supernova neutrinos is,
to our knowledge, missing in the literature,
we exploit the event rates during the accretion phase quoted
for normal and inverted orderings in Ref.~\cite{Scholberg:2017czd}
for a supernova located at $10$~kpc.
For a 40~kton liquid argon detector,
374~kton water Cherenkov and 20~kton scintillator, 
the normal mass ordering could be extracted with $\sim$ $2$, $6$ and
$2$ $\sigma$ significance, respectively, based on a pure statistical-error analysis.

On the other hand, collective effects, which lead to spectral swaps
in the electron (anti)neutrino spectra,
show very sharp features at fixed energy values which depend,
among other factors, on the neutrino mass ordering.
However, these signatures are not as robust as the ones existing
in the neutronization and accretion phases.
Finally, a very significant imprint of the neutrino mass ordering
on the supernovae neutrino fluxes is that due to
their propagation through the Earth interior,
where the standard MSW effect will induce
a few percent-level oscillatory pattern
in the $10-60$~MeV energy range,
in the electron (anti)neutrino spectra in case of (normal) inverted mass ordering.
The detection of these wiggles requires however excellent energy resolution.

%% file: texs/relic.tex
In the early Universe,
neutrinos decoupled from the cosmic plasma during the cool down,
in a process similar to the one leading to the formation of the CMB
but at an earlier time, when the universe was seconds to minutes old.
These neutrinos have been free-streaming for such a long time that they have decohered and are currently propagating as mass eigenstates.
The decoupling of neutrinos occurred just before
$e^\pm$ annihilated and reheated photons,
leading to the following ratio between the photon ($T_\gamma$) and neutrino ($T_\nu$) temperatures,
see Equation~\eqref{eq:rho_rad}:
\begin{equation}
\frac{T_\nu}{T_\gamma} = \left( \frac{4}{11} \right)^{1/3}.
\end{equation}
Today, the temperature of the neutrino background is $T^0_\nu \simeq 1.6\e{-4}$~eV.
Their mean energy is $\langle E_\nu\rangle\simeq3\,T_\nu\simeq5\e{-4}$~eV,
much smaller than the minimal mass of the second-to-lightest neutrino as required by flavor oscillations,
so that at least two out of three neutrinos are non-relativistic today.
The cosmic neutrino background (C$\nu$B) is the only known source of non-relativistic neutrinos and it has never been detected directly.

Apart from the imprints that relic neutrinos leave in the CMB (see section~\ref{sec:cmb}),
which allow to have an indirect probe of their existence through
the determination of \neff,
the direct detection of the C$\nu$B would offer a good opportunity to test neutrino masses and their ordering.
Capturing relic neutrinos is not only rewarding from the point of view of what we can learn about neutrino properties,
but also because it would be a further confirmation of the standard Big Bang cosmological model.
Different ideas on how to achieve such a detection have been proposed~\cite{Weinberg:1962zza,Weiler:1982qy,Weiler:1983xx,Duda:2001hd,Eberle:2004ua,Barenboim:2004di,Gelmini:2004hg,Ringwald:2005zf,Cocco:2007za,Li:2015koa,Vogel:2015vfa,Domcke:2017aqj},
ranging
from absorption dips in the ultra-high-energy (UHE) neutrino fluxes
due to their annihilation with relic neutrinos at the $Z$ boson resonance, 
to forces generated by coherent scattering of the relic bath
on a pendulum and measured by laser interferometers. 
Most of these proposed methods are impractical from the experimental point of view.
The one exploiting UHE neutrinos \cite{Weiler:1982qy,Weiler:1983xx,Eberle:2004ua,Barenboim:2004di} has two problems, one related with the fact that it is difficult to think about a source that produces such UHE neutrinos, of energies
\begin{equation}
E^{\rm res}_\nu = \frac{m_Z^2}{2m_{i}} \simeq 4\cdot 10^{22} \left( \frac{0.1\,\mathrm{eV}}{m_{i}}\right)\,\mathrm{eV},
\end{equation}
and another one regarding the difficulties of detecting a large enough sample of UHE neutrinos in order to resolve the dips.
The method based on interferometers \cite{Domcke:2017aqj} is even more complicated to address.
At interferometers, current sensitivities to accelerations are of the order of $a \simeq 10^{-16}\,\mathrm{cm}/\mathrm{s}^2$,
with an optimistic estimation of $a \simeq 3\cdot 10^{-18}\,\mathrm{cm}/\mathrm{s}^2$~\cite{Domcke:2017aqj}
for the incoming generation.
However, expected accelerations due to relic neutrino interactions are of the order of
$\left( 10^{-27} - 10^{-33}\right) \,\mathrm{cm}/\mathrm{s}^2$~\cite{Duda:2001hd,Domcke:2017aqj}, 
many orders of magnitude below the sensitivity of the next-generation interferometers.

The most promising approach to detect relic neutrinos is to use neutrino capture in a $\beta$-decaying nucleus $A$ 
\begin{equation}
\nua{}_e
+ A \rightarrow
e^\pm + A' ,
\end{equation}
where the signal for a positive detection is a peak located about $2m_\nu$ above the true $\beta$-decay endpoint (see below).
In particular, tritium is considered as the best candidate since it has a high neutrino capture cross section, low Q-value and it is long-lived~\cite{Cocco:2007za,Lazauskas:2007da,Blennow:2008fh,Faessler:2011qj,Long:2014zva}.
The proposal for an experiment chasing this purpose was made in~\cite{Cocco:2007za}.
Currently, efforts are put for such experiment, the PonTecorvo Observatory for Light Early-Universe Massive-Neutrino Yield
(\texttt{PTOLEMY})~\cite{Betts:2013uya,Baracchini:2018wwj},
to be built.
The experiment has recently been approved
by the Scientific Committee of the Italian National Laboratories of Gran Sasso
and, in the following months, the existing prototypes for various components
are expected to be moved from Princeton,
where the R\&D has been performed up to now,
to Gran Sasso.
The idea is to implant the tritium source on graphene layers,
to avoid the problems related to a gaseous source,
then collect and measure the energy of the emitted electrons
using a combination of MAC-E filter, radio-frequency tracking and
micro-calorimetry to obtain a determination
of the $\beta$-decay and neutrino capture spectrum of tritium
with an energy resolution of the order $\Delta\simeq0.05-0.1$~eV.

The total expected event rate from relic neutrino capture
for a \texttt{PTOLEMY}-like experiment,
assuming the estimated tritium mass of $100\,\mathrm{g}$, is
\begin{equation}\label{eq:event-rate-PTOLEMY}
\Gamma_{\mathrm{C}\nu\mathrm{B}} = \left[ n_0 (\nu_{h_{R}}) + n_0 (\nu_{h_L} ) \right] N_T \, \bar{\sigma} \,\sum_{i=1}^{3} \left| U_{ei} \right|^2 f_c (m_i)\,,
\end{equation}
where $n_0 (\nu_{h_{R,L}})$ is the averaged number density of relic neutrinos with right (R) or left (L) helicity,
$N_T = M_T / m({{^3}\mathrm{H}})$ is the approximated number of tritium atoms in the source,
$\bar{\sigma} \simeq 3.834 \times 10^{-45}\,\mathrm{cm}^2$~\cite{Long:2014zva},
and $f_c(m_i)$ is a mass-dependent overdensity factor that accounts for the clustering of relic neutrinos under the gravitational attraction of the matter potential (mostly from the dark matter halo) of our galaxy.
This last factor was originally computed in Refs.~\cite{Singh:2002de,Ringwald:2004np} and later updated in Ref.~\cite{deSalas:2017wtt} (see also Ref.~\cite{Zhang:2017ljh}),
where smaller masses were considered for the neutrinos, and the treatment of the matter potential of the Milky Way was improved.
The values of $f_c(m_\nu)$ range from $1.1-1.2$ for a neutrino with $m_\nu = 60$~meV
to $1.7-2.9$ for $m_\nu = 150$~meV~\cite{deSalas:2017wtt}.

\begin{figure}
\centering
\includegraphics[width=0.5\textwidth]{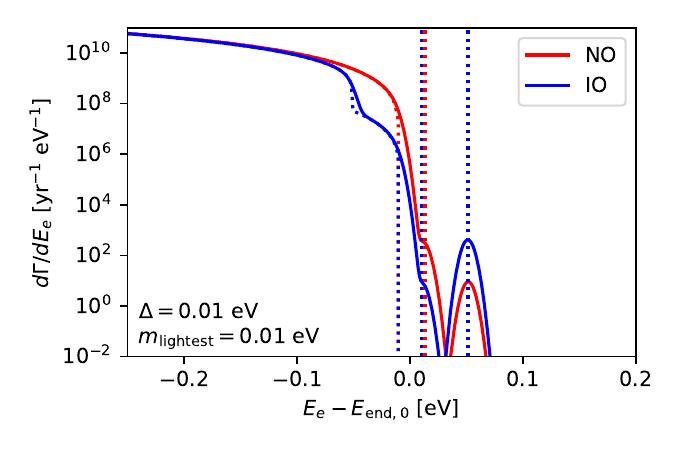}
\includegraphics[width=0.5\textwidth]{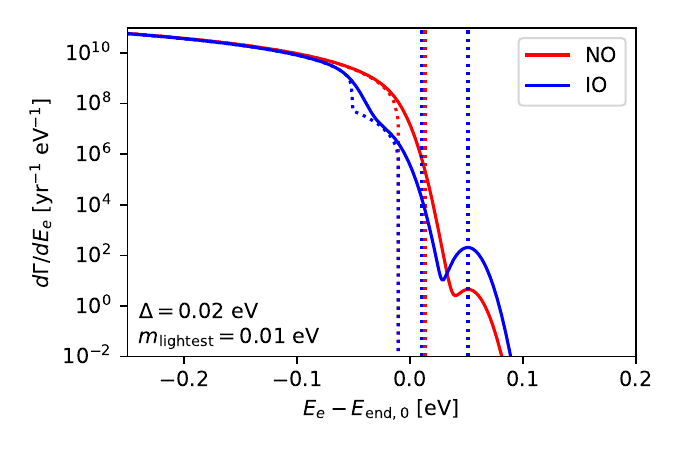}
\includegraphics[width=0.5\textwidth]{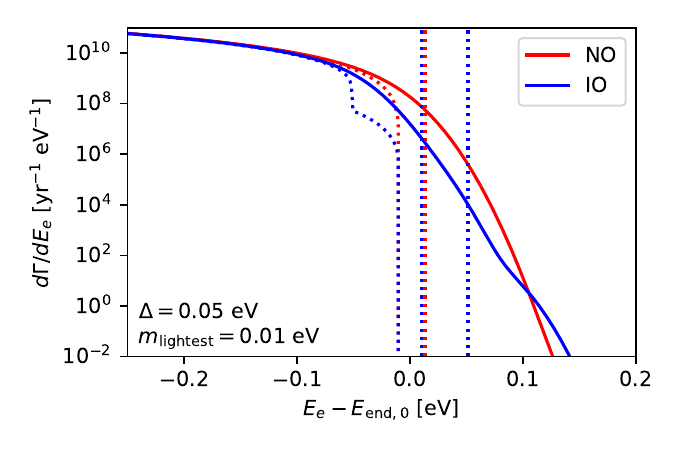}
\caption{
\label{fig:ptolemy_no_io}
Electron spectrum in a \texttt{PTOLEMY}-like experiment,
comparing normal (red) and inverted ordering (blue) with $\mlight=10$~meV
and
three different energy resolutions:
$\Delta=10$~meV (top),
$\Delta=20$~meV (middle),
$\Delta=50$~meV (bottom).
Dashed lines indicate the spectrum as it would be measured by an experiment with perfect energy resolution.
}
\end{figure}

For unclustered neutrinos (i.e.\ $f_c = 1$) and $100$~g of tritium, the expected number of events per year is~\cite{Long:2014zva}
\begin{equation}
\Gamma_{\mathrm{C}\nu\mathrm{B}}^{\rm D} \simeq 4\,\mathrm{yr}^{-1}, \qquad\qquad \Gamma_{\mathrm{C}\nu\mathrm{B}}^{\rm M} = 2\Gamma_{\mathrm{C}\nu\mathrm{B}}^{\rm D} \simeq 8\,\mathrm{yr}^{-1},
\end{equation}
where the upperscripts D and M stand for the possible Dirac and Majorana neutrino character.
If neutrinos are Majorana particles,
the expected number of events is doubled with respect to the Dirac case.
The reason is related to the fact that,
during the transition from ultra-relativistic to non-relativistic particles,
helicity is conserved, but not chirality.
The population of relic neutrinos is then composed
by left- and right-helical neutrinos in the Majorana case,
and only left-helical neutrinos in the Dirac case.
Since the neutrino capture can only occur for left-chiral electron neutrinos,
the fact that in the Majorana case the right-handed neutrinos
can have a left-chiral component leads to a doubled number of possible interactions.
While this means that in principle it is possible to distinguish
the Dirac or Majorana neutrino nature with a precise determination of the event rate, there are two problems.
First of all, even without assuming new physics,
the factor of two coming from the neutrino nature is degenerate with the clustering factor, see Equation~\eqref{eq:event-rate-PTOLEMY},
so that a precise calculation of $f_c$ is required to determine if neutrinos are Dirac or Majorana particles through the direct detection
of relic neutrinos \cite{deSalas:2017wtt}.
Moreover, non-standard interactions can increase the event rate
in the Dirac case by a factor larger than two, canceling the difference with Majorana neutrinos in some scenarios \cite{Arteaga:2017zxg}.

Let us come back to the \texttt{PTOLEMY} proposal.
Instead of considering the total event rate,
for this kind of experiment it is much better to study the energy spectrum,
as the direct detection of relic neutrinos can only be possible if
one can distinguish the signal events due to neutrino capture
from the background events due to the $\beta$-decay of tritium.
A crucial issue for such an experiment,
actually more important than the event rate,
is therefore the energy resolution.
In order to distinguish the peak due to the captured relic neutrinos from the $\beta$-decay background,
a full-width half maximum (FWHM) energy resolution
$\Delta \lesssim 0.7 m_\nu$ is needed~\cite{Long:2014zva}.
If neutrinos are non-degenerate in mass,
the neutrino capture signal has a peak for each of the separate neutrino mass eigenstates.
The full expression of the energy spectrum of neutrino capture,
given an energy resolution $\sigma=\Delta/\sqrt{8\ln 2}$,
can be written as:
\begin{equation}\label{eq:dgamma_nc_de}
\frac{d\widetilde\Gamma_{\mathrm{CNB}}}{dE_e}(E_e)
=
\frac{1}{\sqrt{2\pi}\sigma}
n_{0}
N_T\,
\bar\sigma\,
\sum^{N_\nu}_{i=1}
|U_{ei}|^2\,
f_{c,i}\,
\times
\exp\left\{-\frac{[E_e-(E_{\rm end}+m_i+m_{\rm lightest})]^2}{2\sigma^2}\right\}\,,
\end{equation}
where
$E_{\rm end}$ is the energy of the $\beta$-decay endpoint, $E_{\rm end} = E_{\rm end,0} - m_{\rm lightest}$,
being $E_{\rm end,0}$ the endpoint energy when $\mlight=0$.
If the energy resolution is good enough,
the three peaks coming from the three neutrino mass eigenstates could be resolved,
each of them with an expected number of events modulated by $\left| U_{ei} \right|^2$.
This might lead to a positive detection of the neutrino mass ordering,
since the electron-flavor component of $\nu_1$ is larger than the one of $\nu_2$ and $\nu_3$,
and therefore the furthest peak from the $\beta$-decay endpoint (again if neutrinos are non-degenerate) is enhanced
if the ordering of neutrino masses is inverted.
This can be seen in the three panels of Figure~\ref{fig:ptolemy_no_io},
which also show the effect of changing the mass ordering on the $\beta$-decay spectrum.
Dashed lines represent the spectrum which would be determined
by an experiment capable of measuring the $\beta$ spectrum with zero energy uncertainty,
while solid lines represent the shape of the spectrum
that one would observe in a real experiment.
We plot in red (blue) the spectrum obtained using normal (inverted) ordering, a FWHM resolution 
$\Delta=10$~meV (top),
$\Delta=20$~meV (middle),
$\Delta=50$~meV (bottom)
and a lightest neutrino mass $\mlight=10$~meV.
As we can see from the figure, the kink commented in section~\ref{sec:beta} is clearly visible when one observes
the huge number of events that come from the 100~g of decaying tritium
with a sufficient energy resolution.
While for distinguishing the relic neutrino events
from the $\beta$-decay background and for having
a direct detection of the C$\nu$B
the energy resolution is a crucial requirement,
in principle even a worse energy resolution may allow to determine
the neutrino mass scale and the mass ordering, thanks to the fact that
we expect less events near the endpoint when the ordering is inverted.
A direct observation of the amplitude of all the C$\nu$B peaks,
however, would give a much cleaner signal,
because the peak corresponding to the heaviest neutrino
would be always higher in the inverted ordering case, independently of any other factor.

In summary, the C$\nu$B capture event rate in a \texttt{PTOLEMY}-like experiment (Equation~\eqref{eq:dgamma_nc_de}),
even within SM physics and without considering
non-standard interactions,
depends on several main unknowns:
\textit{i)} the absolute neutrino mass,
\textit{ii)} the matter distribution (especially that of dark matter) in our galaxy,
\textit{iii)} the nature of neutrino masses (whether neutrinos are Dirac or Majorana particles), and
\textit{iv)} the true mass ordering.
This last dependence is encoded in the $\left| U_{ei} \right|^2$ factor in Equation~\eqref{eq:dgamma_nc_de} and it is only accessible if neutrinos are non-degenerate.
A quantitative study on the \texttt{PTOLEMY} capabilities
in determining the mass ordering has not been published yet,
but a new Letter of Intent is in preparation \cite{Baracchini:2018wwj}~%
\footnote{%
We suggest the interested readers to look forward to the publication
of this document,
which will describe in more detail the physics reach and
the technical characteristics of \texttt{PTOLEMY}.}.

%% file: texs/summary.tex
Identifying the neutrino mass ordering is one of the major pending
issues to complete our knowledge of masses and mixings in the lepton sector.
The two possibilities, normal versus inverted,
may result from very different underlying symmetries
and therefore to single out the one realized in nature is a mandatory
step to solve the flavor puzzle, i.e.\ to ensure a full theoretical
understanding of the origin of particle masses and mixings.
We have presented a comprehensive review on the current status and on
future prospects of extracting the neutrino mass ordering via a number
of different ongoing and upcoming observations.
Furthermore, the most updated and complete result
on the preference for a given neutrino mass ordering
from a Bayesian global fit to all 2018 publicly available neutrino
data has also been presented.

Currently, among the three available methods to extract
the neutrino mass ordering
(oscillations, neutrinoless double beta decay searches and cosmological observations),
the leading probe comes from oscillations in matter,
measured at long-baseline accelerator or atmospheric neutrino beams in combination with reactor experiments.
The latest frequentists global data analysis results in a preference
for normal mass ordering with $\Delta\chi^2=11.7$ ($\sim 3.4\sigma$),
mostly arising from the combination of the
long-baseline \texttt{T2K} and \texttt{NO$\nu$A} data
with reactor experiments
(\texttt{Daya Bay}, \texttt{RENO} and \texttt{Double Chooz}),
plus the latest atmospheric neutrino results from \texttt{Super-Kamiokande}.
Similar results for the preference in favor of the normal mass ordering
arise from other global fit analyses~\cite{Capozzi:2018ubv}.

Cosmological measurements are able to set indirect, albeit independent
bounds on the neutrino mass ordering.
Neutrinos affect Cosmic Microwave Background (CMB)
primary anisotropies by changing the gravitational potential
at the recombination period when they become non-relativistic.
However, for sub-eV neutrino masses this effect is tiny
and the most prominent effect on the CMB is via lensing,
as neutrinos, having non-zero velocities,
will reduce the lensing effect at small scales.
Nevertheless, the largest impact of neutrinos in
cosmology gets imprinted in the matter power spectrum.
Once neutrinos become non-relativistic,
their large velocity dispersions will prevent the clustering of matter
inhomogeneities at all scales smaller than their free streaming length.
At present, the cosmological constraints on the neutrino mass ordering
come from the sensitivity to the total neutrino mass $\mnu$
and not via the effects induced in the CMB and matter power spectrum
by each of the individual neutrino masses $m_i$.
Within the context of the minimal $\Lambda$CDM model with massive neutrinos,
current cosmological probes cannot provide odds
stronger than $\sim3:1$ in favor of normal ordering.

Neutrinoless double beta decay searches can also
test the neutrino mass ordering if neutrinos are Majorana particles.
However, present constraints on the so-called effective Majorana mass
do not affect the overall Bayesian analyses.

All in all,
the 2018 Bayesian global analysis,
including all the neutrino oscillation data available before the Neutrino 2018 conference,
results in a $\input{results/s_osc_novsio.tex}\sigma$ preference for the normal neutrino mass ordering
which, in Bayesian words, implies a \emph{strong} preference for such a scenario.
One can then combine the oscillation data with
\doublebeta\ data from \texttt{KamLAND-Zen}, \texttt{EXO-200} and \texttt{Gerda}
and cosmological observations from \pla, \texttt{SDSS BOSS}, \texttt{6DF} and \texttt{SDSS DR7 MGS}.
Using this conservative cosmological data combination,
the aforementioned preference becomes $\input{results/s_osc_0n2b_cmb_novsio.tex}\sigma$,
which raises to $\input{results/s_osc_0n2b_cmb_H0_novsio.tex}\sigma$
if a prior on the Hubble parameter $H_0$ from local measurements is considered in addition.
This clearly states the current power of oscillation results
when dealing with neutrino mass ordering extractions.

While in the very near perspective an improved sensitivity
(i.e.\ above the $\input{results/s_osc_0n2b_cmb_H0_novsio.tex}\sigma$ level)
is expected mostly from more precise measurements of current
long-baseline and atmospheric experiments,
and, to a minor extent,
from cosmological surveys
(\texttt{Planck}, \texttt{DES} and \texttt{eBOSS} among others),
there will be a number of planned experiments
which will be crucial for extracting the neutrino mass ordering in the non-immediate future.

Of particular relevance are the upcoming neutrino oscillation facilities,
as they will be able to measure the neutrino mass ordering
with astonishing precision without relying on combinations of different data sets.
Such is the case of the Deep Underground Neutrino Experiment (\texttt{DUNE}),
that will be able to measure the neutrino mass ordering
with a significance of $5\sigma$ with seven years of data.
Atmospheric neutrino observatories as \texttt{PINGU} or
\texttt{ORCA} will also mainly focus on the mass ordering measurement.
Some of these future devices could also identify the
neutrino mass ordering via the detection of matter effects in the
neutrino fluxes emitted at the eventual explosion
of a supernova in our galaxy or in its neighbourhood.
On the other hand, medium baseline reactor neutrino detectors
such as \texttt{JUNO} or \texttt{RENO} will also be able
to extract the neutrino mass ordering despite matter effects are
negligible for these two experiments.
They will focus instead on an extremely accurate measurement
of the survival electron antineutrino probability.

Improved masses and detection techniques
in neutrinoless double beta decay future searches
could go down the $10$~meV region in the
effective Majorana mass $m_{\beta\beta}$, and
they could be able to discard at some significance level the
inverted mass ordering scenario,
in the absence of a positive signal.
These limits, however, will apply only
in case neutrinos have a Majorana nature.
Moreover, the determination of the neutrino mass ordering
may be complicated by the presence of a light sterile neutrino
at the eV scale, as currently suggested by the NEOS and DANSS results.

Concerning future cosmological projects,
the combination of different probes will still be required.
Near-future CMB and large scale structure surveys will
only be sensitive to the neutrino mass ordering
via their achieved error on $\mnu$.
Furthermore, the accuracy in the extraction of the neutrino mass ordering
will strongly depend on how far $\mnu$ lies
from the minimum allowed value from oscillation probes.
The future CMB mission \texttt{CORE} plus the \texttt{DESI}
galaxy survey could provide odds of $9:1$ for normal neutrino mass
ordering assuming $\mnu=0.056$~eV.
Even if very futuristic surveys,
based on the observation of the $21$~cm redshifted line in neutral hydrogen,
may be able to extract the individual values of the neutrino masses,
their precision on the $m_i$ values may not be enough to guarantee
a direct determination of the neutrino mass ordering by these means,
albeit they can achieve an accurate measurement of the ordering
thanks to their unprecedented precision on $\mnu$.

Last, but not least, relic neutrino capture in tritium
in a \texttt{PTOLEMY}-like experiment could also establish
the neutrino mass ordering via an almost perfect
energy reconstruction of the $\beta$-decay spectrum,
ensured by the extremely large amount of tritium adopted.
The detection is possible both from a kink in the $\beta$-decay spectrum
which only appears if the ordering is inverted
and from the peaks due to neutrino capture just above the endpoint.

All these future probes may either confirm or reject
the current \emph{strong} preference ($\sim\input{results/s_osc_0n2b_cmb_H0_novsio.tex}\sigma$)
in favor of the normal neutrino mass ordering.
Such a preference has kept gaining significance in the recent years,
thanks to the fact that current neutrino oscillation experiments have
enormously improved our knowledge of neutrino flavor physics.